\documentclass[12pt, a4paper]{article}
\pdfoutput=1
 \usepackage[font=small,format=plain,labelfont=bf,up,textfont=normal,up,justification=justified,singlelinecheck=false]{caption}
\usepackage{subcaption}

\usepackage[a4paper, left=2cm,right=2cm]{geometry}
\usepackage[colorlinks=true,linkcolor=black,citecolor=teal,urlcolor=MidnightBlue,filecolor=black]{hyperref}
\usepackage{amsfonts}
\usepackage{amsmath,amssymb}
\usepackage{setspace}
\usepackage[dvipsnames]{xcolor}
\definecolor{SchoolColor}{rgb}{0.6471, 0.1098, 0.1882} 

\usepackage[utf8,applemac]{inputenc}
\usepackage{tensor}
\usepackage{cite}
\usepackage{tikz}
\usepackage{graphicx}
\usepackage{bm} 

\usepackage[utf8,applemac]{inputenc}
\usepackage{tensor}
\usepackage{cite}
\usepackage{tikz}
\usepackage{graphicx}
\graphicspath{{figure/}}
\usepackage{array}
\usepackage{booktabs} 
\usepackage{multirow}

\usepackage{dcolumn}
\usepackage{bm}

\usepackage{verbatim}

\usepackage{textcomp} 
\usepackage{graphicx} 

\numberwithin{equation}{section}
\newcommand{\bea}{\begin{eqnarray}}
\newcommand{\eea}{\end{eqnarray}}
\newcommand{\be}{\begin{equation}}
\newcommand{\ee}{\end{equation}}
\def\nn{\nonumber}

\newcommand{\beqs}{\begin{eqnarray}}
\newcommand{\eeqs}{\end{eqnarray}}
\numberwithin{equation}{section}

\newcommand{\cubicroot}[1]{{}^{\hspace{3pt}\tiny 3 \hspace{-5pt}}\sqrt{#1}}

\newcommand{\simneq}{\includegraphics[scale=0.28]{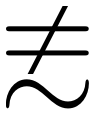}}

\begin{document}
\begin{titlepage}

\begin{flushright}\vspace{-3cm}
{\small
\today }\end{flushright}
\vspace{0.5cm}
\begin{center}
	{{ \LARGE{\bf{Classification of radial Kerr geodesic motion}}}\vspace{30pt}}

	\centerline{\large{\bf{Geoffrey Comp\`ere$^\clubsuit$\footnote{geoffrey.compere@ulb.be}, Yan Liu$^\clubsuit$\footnote{yan.liu@ulb.be}, Jiang Long$^\diamondsuit$\footnote{e-mail:
					 longjiang@hust.edu.cn}}}}
	\vspace{2mm}
	\normalsize
	\bigskip\medskip
	\textit{$^\clubsuit$ Universit\'e Libre de Bruxelles and International Solvay Institutes, Campus Plaine CP231, B-1050 Brussels, Belgium}\\
\vspace{10pt}
	\textit{$^\diamondsuit$ School of Physics, Huazhong University of Science and Technology, \\Wuhan, Hubei 430074, China
	}
	
	\vspace{10mm}
	
	\begin{abstract}
		\noindent
		We classify radial timelike geodesic motion of the exterior nonextremal Kerr spacetime by performing a taxonomy of inequivalent root structures of the first-order radial geodesic equation, using a novel compact notation and by implementing the constraints from polar, time, and azimuthal motion. Four generic root structures with only simple roots give rise to eight nongeneric root structures when either one root becomes coincident with the horizon, one root vanishes, or two roots becomes coincident. We derive the explicit phase space of all such  root systems in the basis of energy, angular momentum, and Carter's constant and classify whether each corresponding radial geodesic motion is allowed or disallowed from the existence of polar, time, and azimuthal motion. The classification of radial motion within the ergoregion for both positive and negative energies leads to six distinguished values of the Kerr angular momentum. The classification of null radial motion and near-horizon extremal Kerr radial motion are obtained as limiting cases  and compared with the literature. We explicitly parametrize the separatrix describing root systems with double roots as the union of the following three regions that are described by the same quartic respectively obtained when (1) the pericenter of bound motion becomes a double root, (2) the eccentricity of bound motion becomes zero, (3) the turning point of unbound motion becomes a double root.
		\end{abstract}
	

\end{center}

\end{titlepage}

\tableofcontents

\section{Introduction}

The direct observation of gravitational waves from binary black hole mergers \cite{Abbott:2016blz} and the prospects of future observatories, such as LISA \cite{Audley:2017drz}, the Einstein Telescope \cite{Maggiore:2019uih} , TianQin \cite{Luo:2015ght,Mei:2020lrl} or Taiji \cite{Hu:2017mde,Guo:2018npi}, strongly encourage the development of more accurate waveform models within general relativity. In particular, self-force methods \cite{Barack:2018yvs,Pound:2021qin} model binaries for small (or not that small \cite{vandeMeent:2020xgc}) mass ratios in terms of perturbed timelike Kerr geodesics.  Besides, timelike Kerr geodesics are directly relevant for the study of dark matter spikes around Kerr black holes \cite{Ferrer:2017xwm}. The phase space of negative energy geodesics is also relevant to estimate the energy released from the ergoregion from the Penrose process \cite{1969NCimR...1..252P,1971NPhS..229..177P}  in the approximation where the electromagnetic field and gravitational backreaction can be neglected. The direct imaging of the supermassive black hole M87$^*$ by the Event Horizon Telescope\cite{Akiyama:2019cqa} and future black hole imaging prospects also encourage the comprehensive description of null Kerr geodesics. Furthermore, recent interest in two-body scattering \cite{Damour:2017zjx} motivates an inclusive study of unbounded timelike Kerr geodesics. 

The study of Kerr geodesics has a long history. The Kerr solution found in 1963 \cite{PhysRevLett.11.237} describes the stationary axially symmetric solution of the vacuum Einstein equations, describing a spinning black hole. In 1968, Carter discussed the global structure of the Kerr spacetime \cite{PhysRev.174.1559} and found a nontrivial Killing tensor, which implies the existence of a third conserved quantity, the Carter constant $Q$, along geodesic orbits besides the energy $E$, and the (component along the Kerr axis of the) angular momentum $\ell$. In 1972, Wilkins studied the bound geodesics in Kerr spacetime\cite{PhysRevD.5.814} and described them in terms of their azimuthal, radial and polar frequencies, which were later given in explicit form by Schmidt \cite{Schmidt:2002qk}, Drasco and Hughes \cite{Drasco:2003ky} and Fujita and Hikida \cite{Fujita:2009bp}. In 1973, Bardeen \cite{Bardeen:1973tla} initiated the study of equatorial timelike geodesics and general null geodesics, which were further analyzed in \cite{DEFELICE1979307,1983grg1.conf....6C,1994ApJ...421...46R,Neill:1995aa,Vazquez:2003zm,2003GReGr..35.1909T,Slezakova:2006aa,Hackmann:2010aa,Hackmann:2015vla,Hackmann:2015ewa}.  Negative energy geodesics within the ergoregion were studied in  \cite{1984GReGr..16...43C,Grib:2013hxa,Vertogradov:2015hza} where it was established that only trapped orbits (i.e., emerging from the white hole and plunging into the black hole) are allowed. The decoupling of radial and polar motion was accomplished by Mino using what is now called Mino time \cite{Mino:2003yg}. The geodesics in the near-horizon region of high-spin Kerr were analyzed in \cite{AlZahrani:2010qb,Hadar:2014dpa,Hadar:2015xpa,Gralla:2015rpa,Hadar:2016vmk,Porfyriadis:2016gwb,Compere:2017hsi,Hod:2017uof,Gralla:2017ufe,Lupsasca:2017exc,Kapec:2019hro,Compere:2020eat,Li:2020val,Hou:2021okc}. Part of the complete separatrix, as defined below, namely the separatrix between plunging and bounded orbits,  was reduced to a fourth-order polynomial in terms of semilatus rectum and eccentricity \cite{Glampedakis:2002ya,OShaughnessy:2002tbu} and was further described in \cite{Glampedakis:2005hs,Levin:2008yp,PerezGiz:2008yq}.  Algorithmic codes implementing (a subset of the) Kerr geodesics are publicly available \cite{Dexter:2009fg,BHPToolkit}.

In the last two years, several novel analytic results on Kerr geodesics were achieved \cite{Kapec:2019hro,Rana:2019bsn,vandeMeent:2019cam,Gralla:2019ceu,Stein:2019buj,Himwich:2020msm,Gralla:2020yvo,Gralla:2020srx,Gates:2020sdh,Hadar:2020fda,Compere:2020eat,Teo:2020sey,Cunha:2020azh,Tavlayan:2020cso}. Explicitly real, fully explicit, ``initial data-dependent'' analytical solutions, in terms of elliptic functions, were given for (i) radial and polar motion for timelike bounded orbits \cite{vandeMeent:2019cam}, (ii) generic (i.e., excluding zero measure sets) polar motion for null or timelike orbits \cite{Kapec:2019hro,Compere:2020eat}, (iii) generic radial motion for null orbits \cite{Gralla:2019ceu}, and (iv) general (i.e., including zero measure sets) radial near-horizon motion in the high-spin limit \cite{Kapec:2019hro,Compere:2020eat}. The only missing piece of information, in order to complete such a state-of-the-art analytical standard for all Kerr orbits, is the nongeneric polar motion and general (generic and nongeneric) radial motion for timelike geodesics, which is the main interest of this paper.

A necessary condition to obtain such analytic formulae for radial  motion is to first classify the possible classes of radial motion and derive their domain of existence in the phase space of parameters. In order to describe all geodesic classes, a relevant basis of the phase space is simply the set of conserved quantities $(E,Q,\ell)$. The classification of the roots of the radial potential is nontrivial since its  discriminant is a quintic in $Q$, a polynomial of degree 10 in $\ell$, and of degree 12 in $E$, which admits \emph{a priori} no analytic solution in radials. Following different routes, partial results in this endeavor were recently obtained. Constant radial motion (i.e., spherical orbits) was comprehensively analyzed by Teo \cite{Teo:2020sey} based on earlier results \cite{Hughes:1999bq,Hughes:2001jr,Fayos:2007ks,Grib:2014afa,Grib:2015rea,Rana:2019bsn}, and the resulting phase space was partially implicitly derived using $Q$ as the main parameter, even though  more information is required to derive the full phase space, namely the bound on $Q$ implied by the existence of polar motion for $\vert E \vert >1$ \cite{PhysRev.174.1559}, the bound on $\ell$ from orbits threading the ergosphere \cite{ 1984GReGr..16...43C}, and the classes of orbits with a root coincident with the horizon. It was also independently shown by Stein and Warburton \cite{Stein:2019buj} that the subset of unstable spherical orbits with $\vert E \vert < 1$ that describes the separatrix between bounded and plunging orbits is described by a twelveth-order polynomial in the semilatus rectum and eccentricity. 

Building upon this earlier work, we classify in this paper the radial motion of timelike geodesics of the exterior nonextremal Kerr spacetime, and we describe, in particular, the complete separatrix, i.e., the codimension 1 region in phase space containing spherical orbits. We will achieve this goal by first classifying the roots of the quartic potential controlling the radial motion as a function of the conserved geodesic quantities $(E,Q,\ell)$ for all nongeneric root systems, taking into account the existence of polar motion, thereby inferring the generic cases as the codimension 0 domains bounded by the codimension 1 (and codimension 2) nongeneric cases. Secondly, we will use the bounds on radial motion implied by the existence of time and azimuthal motion within the ergoregion to infer the allowed radial geodesic classes for each generic or nongeneric root system. We will use the energy $E$ as our main parameter for our classification, and we will treat both non-negative and negative energies. 

This paper is organized as follows. In Section \ref{sec:bounds} we first review the bound on Carter's constant $Q$ inferred from the existence of polar motion, and we derive the bounds on $\ell$ inferred from the existence of time and azimuthal motion within the ergoregion for both signs of the energy. In Section \ref{sec2}, we introduce a novel convenient notation for labelling the qualitatively distinct root structures of the radial geodesic potential. We first derive the list and properties of root structures in particular subcases: large $E$, $Q$, or $\ell$ charges, the case where one root coincides with the outer horizon, the double root case where spherical orbits occur, the marginal case $E=1$ where one root disappears due to the lowering of the polynomial order of the radial potential and, finally, the generic case. We conclude with the null case obtained as a limit of infinite energy. In Section \ref{sec:ergo}, we introduce the position of the ergosphere and discuss the radial root systems and allowed radial motion within the ergoregion, first on the equator and then generically. We also obtain the classification of radial motion within the near-horizon region of near-extremal Kerr black holes. In Section \ref{sec3}, we obtain an explicit parametrization of the complete separatrix, and we finally conclude in Section \ref{sec:ccl}. Several useful reviews are relegated to appendices. In Appendix \ref{rootsec}, we review the theory of discriminants of a polynomial. In Appendix \ref{sec:schw}, we review the classification of geodesic orbits of Schwarzschild in our notation. 

\section{Bounds on the constants of motion}
\label{sec:bounds}

The Kerr geodesics are essentially determined by the radial and polar potentials,
\bea
\hspace{-6pt}\!\!\!\!\!\!R(r)&\!\!\!\!=\!\!\!\!&(E^2-\mu^2)r^4+2M \mu^2 r^3+(a^2(E^2-\mu^2)-Q-\ell^2)r^2+2M((a E-\ell)^2+Q)r-a^2 Q,\label{R}\\\hspace{-6pt}
\!\!\!\!\!\!V(u)&\!\!\!\!=\!\!\!\!&a^2(\mu^2-E^2)u^4-(a^2(\mu^2-E^2)+Q+\ell^2)u^2+Q,\label{V}
\eea 
with $u=\cos \theta$. Here $E$, $\ell$, and $Q$ are the conserved energy, angular momentum, and  Carter constant, associated with the two Killing vectors and the nontrivial Killing tensor; $M$ and $a$ are the mass and dimensionless spin of the Kerr black hole; $\mu$ is the mass of the test object. 

The constants of motion $(E,\ell,Q)$ are constrained by the polar motion and, for the orbits entering the ergoregion, by the time and azimuthal motion within the  ergoregion. We derive these constraints in the following. 

\subsection{Polar motion: Bound on Carter's constant}
\label{sec:bound}

The well-known bound on Carter's constant is $Q \geq -(a E -\ell )^2$. Let us discuss the stronger bound on $Q$ imposed by the reality of polar motion in Kerr, $a \neq 0$. Such bound was first discussed in \cite{PhysRev.174.1559,PhysRevD.5.814}. In this section we allow the energy $E$ to be of any sign. 

The potential $V(u)$ defined in Eq. \eqref{V} has the property 
\be
V(0)=Q,\quad V(1)=-\ell^2.
\ee
Therefore, if $Q\ge0$, there is always one root $u=u_0$, $V(u_0)=0$. Then, there is a range of $u$ around $u_0$, for which motion exists for $Q\ge0$. 

In order to discuss $Q <0$, we first rewrite the potential as a quadratic in $z \equiv u^2$ as 
\be
V(z)=V_0-a^2(E^2-\mu^2)(z-z_0)^2 ,
\ee 
where 
\bea
z_0 &\equiv & \frac{1}{2}-\frac{\ell^2+Q}{2a^2(E^2-\mu^2)}, \\
V_0 & \equiv &\frac{Q-\ell^2}{2}+\frac{1}{4}a^2(E^2-\mu^2)+\frac{(Q+\ell^2)^2}{4a^2(E^2-\mu^2)}. 
\eea 

For $E^2 < \mu^2$ and $\ell \neq 0$, the parabola has positive curvature, $V''(z)>0$ but is negative at both $z=0$ and $z=1$. Therefore, it is negative in the range $0 \leq z \leq 1$, and there is  no possible motion. The only exception is $\ell=0$ and $u^2=1$, for which $V=0$. The non-negative Carter constant is $k = Q+a^2 E^2$. Polar motion on the north or south pole is therefore allowed for $E^2 < \mu^2$ when $\ell = 0$ and $Q \geq -a^2 E^2$. 

For $E^2 > \mu^2$, the parabola has negative curvature, $V''(z)<0$, and its endpoints at $z=0,1$ have $V \leq 0$. The existence of motion requires that the maximum of $V(z)$ be non-negative in the range $0 < z \leq 1$. The bound $z_0=1$ is reached only for $\ell =0$, for which $Q = Q_0 \equiv -a^2(E^2-\mu^2) \geq -a^2E^2$ and $V_0 >0$. Motion on the north and south pole is therefore allowed in that range.

For $E^2 > \mu^2$ and $\ell \neq 0$, this implies $V_0 \geq 0$ and $0 < z_0 < 1$. The first inequality implies either 
\begin{equation}
Q\ge -(|\ell|-a \sqrt{E^2-\mu^2})^2\quad \label{Q1cond}
\text{or} \quad Q\le -(|\ell|+a\sqrt{E^2-\mu^2})^2.
\end{equation}
The second set of inequalities is equivalent to 
\be
-a^2(E^2-\mu^2)-\ell^2 <Q< a^2(E^2-\mu^2)-\ell^2.\label{Q2cond}
\ee 
Now the second condition of \eqref{Q1cond} is incompatible with \eqref{Q2cond}. Therefore, for $Q<0$, $\ell \neq 0$, we can only consider 
\be
Q\ge -(|\ell|-a \sqrt{E^2-\mu^2})^2.
\ee 

For $E^2 = \mu^2$, the potential reduces to $V(u)=-(\ell^2+Q)u^2+Q$. Its roots are $u_0^2=\frac{Q}{Q+\ell^2}$. Existence of motion requires $0\le u_0^2\le 1$. This is equivalent to either $Q \geq 0$ or $Q<0$ with $\ell=0$. For $Q\ge 0$, the roots are $u_0=\pm \sqrt{\frac{Q}{Q+\ell^2}}=\pm u_c$. In this case, the orbit librates between $\theta_0=\arccos u_c$ and $\pi-\theta_0$. The angular becomes largest for $\ell=0$. In this case, $0\le \theta\le \pi$. For $Q<0$ and $\ell=0$, we find $V(u)=Q(1-u^2)$ is negative except for $u=\pm 1$. This corresponds to the north pole $\theta=0$ or $\theta=\pi$. We still have $Q \geq -a^2 E^2$.

 Therefore, so far, we have the bounds for $E^2>\mu^2$,
\bea
Q \geq \left\{ \begin{array}{ll} 
-a^2E^2 & \ell =0, \\
-(\vert \ell\vert - a \sqrt{E^2-\mu^2})^2 & 0< \vert \ell \vert < a \sqrt{E^2-\mu^2} , \\ 
0 & \vert \ell \vert > a\sqrt{E^2-\mu^2}. \end{array} \right. \label{boundEmmutmp1}
\eea
For $E^2\le\mu^2$, we have the bounds 
\bea
Q\geq\left\{\begin{array}{ll}
-a^2E^2 & \ell=0,\\
0 & \ell\not=0.\end{array}\right.\label{boundEmmutmp}
\eea 

Now, there is another bound from the definition of $Q$. One can easily check that for $\theta \neq 0,\pi$, 
\be
Q= v_{\theta}^2+\cos^2\theta \left( a^2(\mu^2-E^2)+\frac{\ell^2}{\sin^2\theta} \right),
\ee 
where $v_\theta=g_{\theta\theta} d\theta/d\tau=\Sigma d\theta/d\tau$ is the velocity along the polar coordinate and $\Sigma=r^2+a^2\cos^2\theta$. When $0\le \vert E \vert \le \mu$, we find $Q\ge 0$.  This tightens the bound \eqref{boundEmmutmp} for $\ell=0$. The equality $Q=0$ is only reached for equatorial geodesics,  $\theta=\frac{\pi}{2}$. When $\vert E\vert >\mu$, we find the lower bound 
\bea
Q\ge - a^2(E^2-\mu^2).\label{Q4}
\eea 
The equality is asymptotically reached for $\ell=0$, $v_\theta=0$, and $\theta \mapsto 0,\pi$. For $\theta=0,\pi$ exactly, $V=0$, $v_\theta=0$, and $Q$ is strictly unconstrained, but we constrain it as \eqref{Q4} by continuity. This tightens the bound \eqref{boundEmmutmp1} for $\ell=0$ since $- a^2(E^2-\mu^2)>-a^2 E^2$. In summary, we have the bounds
\bea 
\boxed{Q\ge\left\{\begin{array}{cc}0 & 0\le \vert E\vert \le \mu\, , \\ 0&\vert E \vert >\mu \; \& \; |\ell | \ge a\sqrt{E^2-\mu^2}\, ,\\
-\left( |\ell|-a\sqrt{E^2-\mu^2}\right)^2& \vert E\vert >\mu \;\& \; 0\le  |\ell| < a\sqrt{E^2-\mu^2}.\end{array}\right.}\label{boundQ}
\eea 
The lowest bound for $\vert E \vert >\mu$ and $0\le |\ell|<a\sqrt{E^2-\mu^2}$ is only reached for $\theta$ constant and  $|\ell|=a\sqrt{E^2-\mu^2}\sin^2\theta$. We will refer to these bounds as $Q \geq Q_B(E,\ell)$. 
For the null geodesic case $\mu = 0$ and $\ell \neq 0$, the bound reduces to the one stated in Eq. (24) of \cite{Gralla:2019ceu}. 

\subsection{Time and azimuthal motion: Constraints from the ergoregion}
\label{sec:boundergo}

From now on, we set $M=\mu=1$\footnote{These quantities can be restored by noting that $E\sim \mu$, $r \sim a \sim M$, $Q \sim \mu^2 M^2$, $\ell \sim \mu M$.}.  Kerr spacetime is characterized by an ergosphere with radial range 
\begin{equation}
r_+ \leq r \leq r_{\text{ergo}}(\theta ; a) \equiv  1+\sqrt{1-a^2 \cos^2 \theta}. \label{ergo}
\end{equation}
The region between the horizon and the ergosphere is called the ergoregion. Since $\partial_t$ is spacelike in the ergosphere, negative energy $E<0$ is allowed within the ergoregion.  This ergoregion leads to constraints on geodesic motion in the $t,\phi$ directions for both signs of the energy which, in turn, restrict radial motion. We will derive a complete set of such constraints in this section. We consider $a >0$ since the ergoregion disappears when $a=0$. 

A feature of the ergoregion is that $\nabla^\mu t$ is past directed timelike and $g_{t\phi}<0$, which implies that $\frac{dt}{d\tau} >0$ and $\frac{d\phi}{d\tau}>0$ strictly inside the ergoregion as reviewed e.g., in \cite{Wald:1984rg}. Since $\frac{d\phi}{d\tau}$ can have either sign outside the ergoregion, we may have $\frac{d\phi}{d\tau}=0$ on the ergosphere. Timelike geodesics are therefore moving forward in coordinate time $t$ and corotating along the spin direction of the black hole. This gives the explicit two conditions valid for any $E$
\begin{eqnarray}
a \ell - a^2 E\sin^2\theta + (r^2+a^2)\frac{E(r^2+a^2)-a \ell}{\Delta}>0, \label{eqergo1} \\
-a E +\frac{\ell}{\sin^2\theta}+a \frac{E(r^2+a^2)-a \ell}{\Delta} \geq 0,\label{eqergo2}
\end{eqnarray}
where $\Delta=r^2-2r+a^2$ and $r$ belongs to the region \eqref{ergo}. The condition \eqref{eqergo2} is explicitly violated at the ergosphere for $E<0$ but is obeyed for $E \geq 0$. It implies that no motion with $E<0$ is allowed to reach the ergosphere, i.e., all negative energy motion takes place strictly within the ergoregion. 

We note that the constraints \eqref{eqergo1}-\eqref{eqergo2} are odd under the flip $(E,\ell) \mapsto -(E,\ell)$. It implies that if one motion is allowed for a given radial range and given values of $(E,\ell)$, it will be disallowed for the same radial range and opposite values $-(E,\ell)$, and vice-versa. There is therefore a central symmetry breaking in the phase space of radial motion in the $(E,\ell)$ plane: each radial motion is either allowed or disallowed for either $(E,\ell)$ or $-(E,\ell)$.

Orbits that reach the horizon $r \rightarrow r_+$ have special properties. Since $\Delta \rightarrow 0$, the second term in \eqref{eqergo1} or \eqref{eqergo2} dominates. This implies $E (r_+^2+a^2)-a \ell \geq 0$ or, equivalently, 
\begin{equation}
\ell \leq \ell_+(E) \equiv \frac{E}{\Omega_+},\label{boundT} 
\end{equation}
for any $E \in \mathbb R$. This bound coincides with the first and second laws of black hole thermodynamics $T\delta S = \delta M-\Omega_+ \delta J \geq 0$ upon substituting the variations of the parameters of the black hole with the plunging probe energy and angular momentum $\delta M=E$ and $\delta J=\ell$. The thermodynamic bound \eqref{boundT} therefore applies for any plunging orbit. Moreover, by contraposition, if an orbit has $\ell > \ell_+(E)$, it cannot reach the horizon. In particular, positive energy trapped orbits with $\ell > \ell_+(E)$ are disallowed.

Negative energy geodesics have necessarily $\ell<0$. Indeed, using $\ell=g_{t\phi}u^t+g_{\phi\phi}u^\phi$, $E=-g_{tt}u^t-g_{t\phi}u^\phi<0$ we find $\ell < \frac{g_{\phi\phi}g_{tt}-g_{t\phi}^2}{-g_{t\phi}} u^t< 0 $. Since $\sin^2\theta \leq 1$, we then have $-a^2 E \geq -a^2 E\sin^2\theta$ and $\ell \geq \ell/\sin^2\theta$. Therefore, for any negative energy orbit, the bounds \eqref{eqergo1}-\eqref{eqergo2} imply the same bounds at the equator $\theta=\pi/2$.
In turn, the inequalities  \eqref{eqergo1}-\eqref{eqergo2} evaluated at $\theta=\pi/2$ are equivalent for $E<0$ to 
\begin{equation}
\ell\le \frac{2a E}{2-r}.\label{ellE}
\end{equation}
Since $r_+\le r<2$, it implies in particular for all orbits with $E<0$ that 
\be 
\ell\le \frac{2a E}{2-r}\le \frac{2aE}{2-r_+}=\frac{E}{\Omega_+}. \label{thermoc}
\ee 
The thermodynamic bound \eqref{boundT} is therefore obeyed for all orbits with $E<0$. The upper bound $\ell=\ell_+(E)$ corresponds to root structures with one root at the horizon, see Section \ref{sec:horizontouching}. 

Further detailed constraints on negative energy geodesics will be discussed in Section \ref{sec:details}.

\section{Classification of radial geodesic motion in Kerr}
\label{sec2}

\begin{table}[!tbh]    \centering
\begin{tabular}{|c|c|c|c|c|}\cline{1-2}\cline{4-5}
\rule{0pt}{13pt}\textbf{Notation} & \textbf{Denotes} & & \textbf{Notation} & \textbf{Denotes}\\\cline{1-2}\cline{4-5}
$\vert$ &  outer horizon & & $\bullet$ & simple roots (turning points) \\\cline{1-2}\cline{4-5}
$+$ & allowed region & &  $\bullet \hspace{-2pt}\bullet$ & double roots (spherical orbits)\\\cline{1-2}\cline{4-5}
$-$ & disallowed region & & $\bullet\hspace{-4pt}\bullet\hspace{-4pt}\bullet$ & triple roots (ISSO)\\\cline{1-2}\cline{4-5}
$\rangle$ & radial infinity & & ${\vert \hspace{-5pt} \bullet }$ & roots touching the horizon\\\cline{1-2}\cline{4-5}
\end{tabular}\caption{Notations for the root structures.}\label{table:notations}
\end{table}

In this section we classify the categories of radial motion of Kerr geodesics by classifying the distinct root structures of the quartic radial potential $R(r)$ defined in Eq.\eqref{R}. We will concentrate, for the sake of simplicity, on $E \geq 0$ geodesics. The negative energy geodesics will be studied in detail in Section \ref{sec:ergo}. For our purposes, we introduce the following convenient notation, see Table \ref{table:notations}. The four symbols $\vert$, $+$, $-$ and $\rangle$ label respectively the black hole outer horizon, a region where motion is allowed ($R>0$), a region where motion is disallowed ($R<0$) and radial infinity. The four symbols $\bullet$, $\bullet \hspace{-2pt}\bullet$ , $\bullet\hspace{-4pt}\bullet\hspace{-4pt}\bullet$, ${\vert \hspace{-5pt} \bullet }$ label respectively the distinct root degeneracies: simple, double or triple root, and root touching the outer horizon. The triple root is physically associated with the Innermost Stable Spherical Orbit (ISSO). 

Because of the bounds in Eqs. \eqref{eqergo1}-\eqref{eqergo2}, root structures which admit a positive $R(r)>0$ region might be disallowed. In that case, we will denote the $+$ and $\bullet$ symbols within the root structure in red color as ${\color{red} +}$ and ${\color{red} \bullet}$. In this section, the disallowed region will be assigned to $E \geq 0$ orbits, see Section \ref{sec:ergo} for the case of $E <0$ orbits.

Our final classification to be proven in this section is given in Tables \ref{table:GeoClasses1}  and \ref{table:GeoClasses2}. Generic root structures occur in codimension 0 regions of phase space. Imposing one constraint leads to the root structures of codimension 1 while imposing two constraints leads to the root structures of codimension 2. Each root structure may correspond to distinct geodesic classes: for each allowed radial region of motion $+$ there is a corresponding class (which can be further refined by the initial sign of the radial velocity), and for each double or triple root there are, in addition, spherical orbits. Simple roots correspond to turning points of motion where the velocity vanishes but not the acceleration. Double or triple roots correspond to either spherical orbits or ``whirling'' orbits that asymptotically approach or leave the corresponding radial location. We define a generic geodesic class as a geodesic class where both endpoints are either a simple root, the horizon or infinity. A nongeneric geodesic class is defined as geodesic class such that at least one endpoint differs from a simple root, the horizon, or infinity. 

\begin{table}[!tbh]    \centering
\begin{tabular}{|c|c|c|}\hline
\rule{0pt}{13pt}\textbf{$E \geq 1$ or Null} & \textbf{Root structure} & \textbf{Number of geodesic classes} \\\hline
\textbf{Generic}& $\vert + \rangle$ & 1\\\cline{2-3}
& ${\vert + \bullet-\bullet\hspace{2pt}+\rangle}$ & 2 \\ \hline
\textbf{Codimension 1}& $\vert +\bullet \hspace{-4pt}\bullet + \rangle$ & 3\\\cline{2-3}
& ${\vert \hspace{-7pt}\bullet - \bullet +\rangle}$ & 1 \\\hline
\end{tabular}\caption{The 2 generic and 2 nongeneric inequivalent root structures and their associated 6 distinct geodesic classes of $E\ge 1$ timelike/null geodesics outside the horizon (of which 3 are continuous with $0 \leq E < 1$ geodesic classes).}\label{table:GeoClasses1}
\end{table}

\begin{table}[!tbh]    \centering
\begin{tabular}{|c|c|c|}\hline
\rule{0pt}{13pt}\textbf{$0 \leq E < 1$} & \textbf{Root structure}  & \textbf{Number of geodesic classes}\\\hline
\textbf{Generic}& $\vert + \bullet \hspace{2pt} - \rangle$  & 1 \\\cline{2-3} 
& $\vert + \bullet - \bullet + \bullet \hspace{2pt}- \rangle$ & 2 \\\hline
\textbf{Codimension 1}&  $\vert \hspace{-7pt}\bullet - \rangle$ & 0 \\\cline{2-3}
& $\vert \hspace{-7pt}\bullet - \bullet + \bullet-\rangle$ & 1\\\cline{2-3} 
&  $\vert + \bullet-\bullet \hspace{-4pt}\bullet- \rangle$  & 2 \\\cline{2-3}
& $\vert +\bullet \hspace{-4pt}\bullet + \bullet- \rangle$ & 3 \\\hline 
\textbf{Codimension 2}&$\vert \hspace{-7pt}\bullet - \bullet \hspace{-4pt}\bullet \hspace{2pt} - \rangle$ & 1 \\\cline{2-3} & $\vert +\bullet \hspace{-4pt}\bullet \hspace{-4pt}\bullet \hspace{2pt} - \rangle$ & 2  \\\hline
\end{tabular}\caption{The 2 generic and 6 nongeneric inequivalent root structures and their associated 8 distinct geodesic classes of $0\le E< 1$ timelike geodesics outside the horizon (of which 3 are continuous with $E \geq 1$ geodesic classes).  }\label{table:GeoClasses2}
\end{table}

In Section \ref{sec:largecharges} we discuss the root structures with large charges. In Sections \ref{sec:horizontouching} and \ref{sec:marginal} we investigate the two special cases where the orbits touch the horizon with zero velocity and the so-called marginal orbits with $E=1$. In Section \ref{sec:nonmarginal} we finally classify the generic nonmarginal orbits in the phase space, taking into account the bound on Carter's constant.

\subsection{Root structures for large charges}
\label{sec:largecharges}

In this section, we detail the root structure of Kerr orbits for large values of either the angular momentum $\ell$, Carter's constant $Q$ or the energy $E$. 

\subsubsection{Large $\ell$ limit}
In the limit $\ell \rightarrow \infty$, we consider the following cases.
\begin{enumerate}
    \item $E>1$. There are four real roots:  
    \bea
    r_1&=&-\frac{\ell}{\sqrt{E^2-1}}+O(\ell^0),\qquad  
    r_2=\frac{Q a^2}{2\ell^2}+O(\ell^{-3}),\nn\\
    r_3&=&2-\frac{4aE}{\ell}+O(\ell^{-2}),\qquad 
    r_4=\frac{\ell}{\sqrt{E^2-1}}+O(\ell^0), \nn
\eea
which obey $r_1<0<r_2<r_+<r_3<r_4$. This gives the root structure ${\vert + \bullet - \bullet\hspace{2pt} + \rangle}$.

\item $E=1$. There are three real roots: 
\bea
r_1=\frac{Q a^2}{2\ell^2}+O(\ell^{-3}),\qquad
r_2=2-\frac{4a}{\ell}+O(\ell^{-2}),\qquad
r_3=\frac{\ell^2}{2}+O(\ell^0),
\eea 
which obey $0<r_1<r_+<r_2<r_3$. 
This gives the root structure ${\vert + \bullet - \bullet\hspace{2pt} + \rangle}$.

\item $0\le E<1$. There are only two real roots:
\bea
r_1=\frac{Qa^2}{2\ell^2}+O(\ell^{-3}),\qquad
r_2=2-\frac{4Ea}{\ell}+O(\ell^{-2}),
\eea 
which obey $0<r_1<r_+<r_2$. This gives the root structure ${\vert + \bullet\hspace{2pt} - \rangle}$.
\end{enumerate}

\subsubsection{Large $Q$ limit}

In large $Q$ limit, all the roots are real except for the bound orbits $E<1$ where there are only two real roots.
\begin{enumerate}
	\item $E=1,  0\le a<1$. The roots are 
		\bea
	r_1&=&1-\sqrt{1-a^2}+\frac{(4a\ell-8)(1-\sqrt{1-a^2})-a^2(\ell^2-4)}{2\sqrt{1-a^2}Q}+O(Q^{-2}),\nn\\
	r_2&=&1+\sqrt{1-a^2}+\frac{(8-4a\ell)(1+\sqrt{1-a^2})+a^2(\ell^2-4)}{2\sqrt{1-a^2}Q}+O(Q^{-2}),\nn\\
	r_3&=& \frac{Q}{2}+O(Q^0),
	\eea 
	which obey $0<r_1<r_+<r_2<r_3<\infty$ for $\ell \neq \ell_+(E)$. This gives the root structure $\vert + \bullet\hspace{2pt} -\bullet\hspace{2pt} + \rangle $ for $\ell \neq \ell_+(E)$. Note the solution is invalid for $a=1$, which is the  extreme Kerr black hole.  It has another scaling for extreme Kerr.

	\item $E>1,\,  0\le a < 1$, the four roots are
	\bea
	r_1&=&-\frac{\sqrt{Q}}{\sqrt{E^2-1}}+O(Q^0),\nn\\
	r_2&=&r_- - \frac{(8r_- -4a^2) E^2 - 4 a r_- E \ell +
    a^2 \ell^2}{2\sqrt{1-a^2}Q}+O(Q^{-2}),\nn\\
    r_3&=&r_++ \frac{(8r_+-4a^2) E^2 - 4 a r_+ E \ell + 
    a^2 \ell^2}{2\sqrt{1-a^2}Q}+O(Q^{-2}),\nn\\
	r_4&=& \frac{\sqrt{Q}}{\sqrt{E^2-1}}+O(Q^{0}),
	\eea 
	which obey $-\infty<r_1<0<r_2<r_+ \leq r_3<r_4<\infty$. We have $r_3=r_+$ for $\ell =\ell_+(E)$ as defined in \eqref{lpE}. This gives the root structure $\vert + \bullet\hspace{2pt} - \bullet\hspace{2pt} + \rangle$ for $\ell \neq \ell_+(E)$ and ${\vert \hspace{-7pt} \bullet\hspace{2pt} - \bullet\hspace{2pt} + \rangle}$ for $\ell =\ell_+(E)$.

	\item $0\le E<1, 0\le a<1$. There are only two real roots, 
	\bea
	r_1&=&r_- - \frac{(8r_- -4a^2) E^2 - 4 a r_- E \ell +
    a^2 \ell^2}{2\sqrt{1-a^2}Q}+O(Q^{-2}),\nn\\
	r_2&=&r_++ \frac{(8r_+-4a^2) E^2 - 4 a r_+ E \ell + 
    a^2 \ell^2}{2\sqrt{1-a^2}Q}+O(Q^{-2}),
	\eea 
	which obey $0<r_1 \leq r_+ \leq r_2<\infty$. Equality $r_2=r_+$ occurs for $\ell = \ell_+(E)$ as defined in \eqref{boundT}. This gives the root structure $\vert + \bullet\hspace{2pt} - \rangle$ for generic $\ell$ and the root structure ${\vert \hspace{-7pt} \bullet\hspace{2pt} - \rangle}$ for $\ell = \ell_+(E)$. 

\end{enumerate}
 The large $Q$ root structure is consistent with the analysis of Section \ref{sec:horizontouching} and, in particular, with Figure \ref{phasespacehorizonroots}.

\subsubsection{Large $E$ limit}
In large $E$ limit, there are only two real roots. When $0<a<1$, the real roots are
\bea
r_1&=&\frac{-3^{2/3}a^2+3^{1/3}(a^2 (-9 + \sqrt{3} \sqrt{27 + a^2}))^{2/3}}{3(a^2 (-9 + \sqrt{3} \sqrt{27 + a^2}))^{1/3}}
+O(E^{-1}),\nn\\
r_2&=&\frac{Q}{2E^2}+O(E^{-3}),
\eea
which obey $-\infty<r_1<0<r_2<1$. Instead when $a=0$, the real roots are 
\bea
r_1=-2^{1/3}(\ell^2+Q)^{1/3}E^{-2/3}+O(E^{-4/3}),\qquad r_2=0,
\eea
which obey $-\infty<r_1<r_2=0$. In both cases, this gives the root structure $\vert + \rangle$.

\subsection{Orbital classes with one root at the horizon}
\label{sec:horizontouching}

For Schwarzschild, $R(r_+)=16 E^2 \geq 0$. Motion is therefore generically ($E \neq 0$) allowed just outside the horizon.  A root touches the horizon if and only if $E=0$. In that case, $R(r)=-r(r-2)(r^2+k)$ with $k \geq 0$: the horizon root is always simple and there is no other root outside the horizon. Since for $E = 0$, $R'(r_+) < 0$, motion is disallowed just outside the horizon. We denote the root structure as ${\hspace{4pt}| \hspace{-5pt} \bullet }\hspace{2pt}- \rangle$. There is therefore no allowed motion for $E=0$. For $E \gtrsim 0$, the root structure is therefore $| + \bullet\hspace{2pt} - \rangle$.

For generic Kerr with $0 < a \leq 1$, we have at the horizon 
\bea
R(r_+) = (a \ell - 2 E r_+)^2 \geq 0. \label{Rp}
\eea 
There is a root touching the horizon if and only if 
\bea
\boxed{\ell = \frac{E}{\Omega_+} = \frac{2 E r_+}{a} }\, . \label{lpE}
\eea
Note that this condition is independent of $Q$.  Motion is generically ($\ell \neq \ell_+(E)$) allowed just outside the horizon.  The horizon root is a double root if and only if, moreover,  
\bea
Q = Q_+(E) 
\equiv \frac{r_+^2(2+r_+(E^2-1))}{r_+-2}. 
\eea 
Since $\ell_+(E)> a \sqrt{E^2-1}$, the positivity bound is $Q \geq 0$ for $\ell=\ell_+(E)$. 
We readily obtain that $Q_+(E) < 0$. We conclude that the double horizon root lies outside the phase space. After checking the sign of $\frac{\partial}{\partial Q}R'(r_+)$ we conclude that 
for $\ell=\ell_+(E)$ and $Q > Q_+(E)$  the root structure takes the form ${\hspace{4pt}| \hspace{-5pt} \bullet }- \cdots \rangle$. We do not discuss $Q < Q_+(E)$ since it is irrelevant. For further analysis, it is useful to note that the horizon root is a triple root if, moreover,
\bea
E = E_+ \equiv \sqrt{\frac{\sqrt{1-a^2}}{1+\sqrt{1-a^2}}}=\sqrt{1-r_+^{-1}}\, , \label{Eplus}
\eea
with $0 \leq E_+ < 1/\sqrt{2}$. This horizon triple root lies outside the phase space since it was already the case for the horizon double root. The horizon root cannot be a quadrupole root. We conclude that for $\ell=\ell_+(E)$ and $Q \geq 0$ there is a single root at the horizon without any further horizon touching root. 

\begin{figure}
     \centering
     \includegraphics[width=0.5\textwidth]{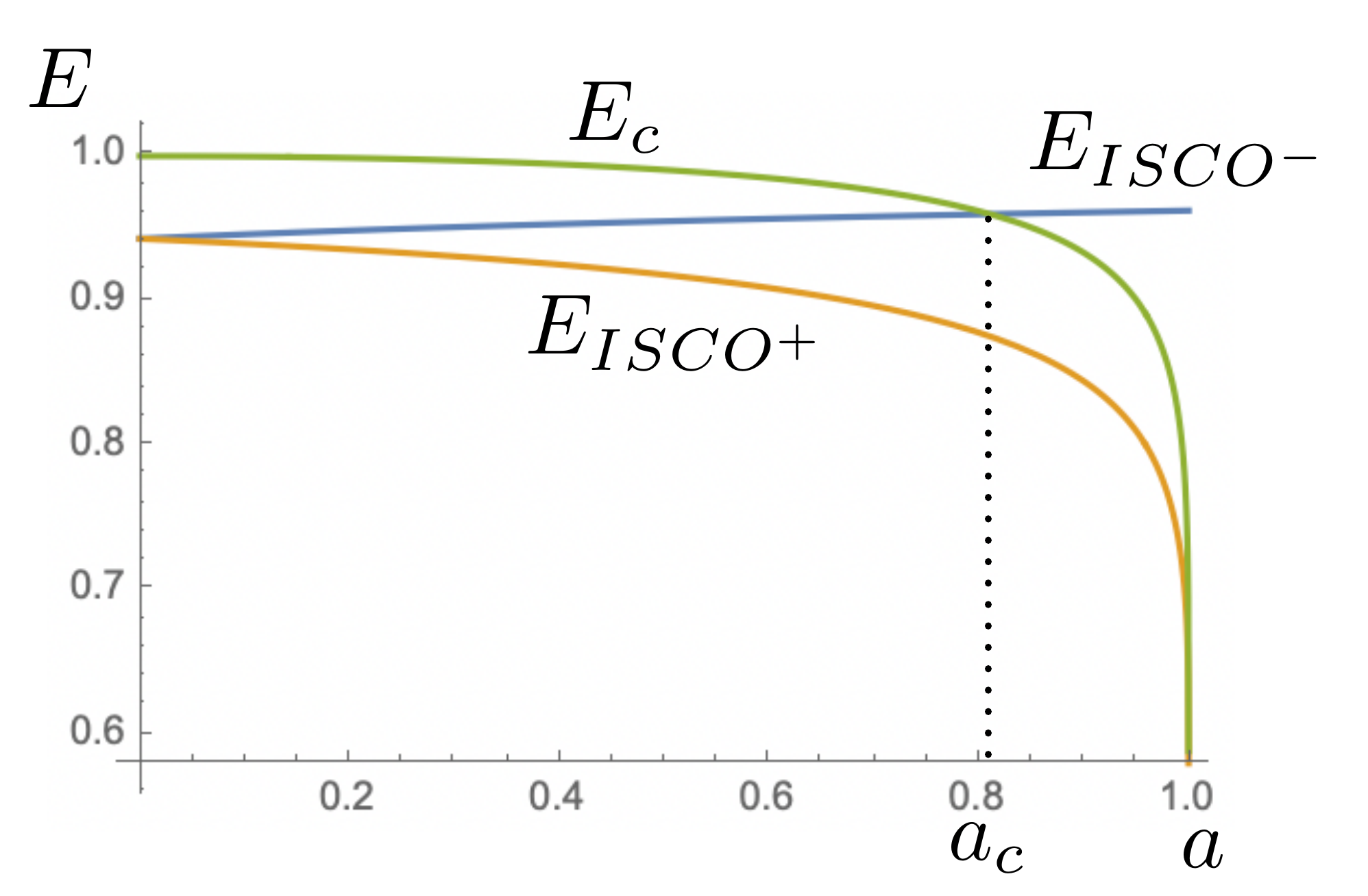}
     \caption{Energy $E_c$ (corresponding to the existence of a horizon touching root and a stable double root on the equatorial plane),  prograde ISCO energy $E_{ISCO^+}$ and retrograde ISCO energy $E_{ISCO^-}$ as a function of $a$. The critical rotation $a_c$ is defined at the intersection between $E_{ISCO^-}$ and $E_c$.}
     \label{Eminstar}
 \end{figure}

Let us now study the occurrence of double roots outside the horizon. For that purpose we impose \eqref{lpE} and consider the reduced polynomial $Y(r) \equiv R(r)/(r-r_+)$. It is cubic for $E \neq 1$ and quadratic for $E = 1$. Double roots occur for $Y(r_*)=Y'(r_*)=0$. There is a single real solution branch given by 
\begin{eqnarray}
Q &=& Q^{\text{s}}(r_*) \equiv 
\frac{r_*^2((r_+-2)r_*^2+2r_+^2r_*+(r_+-2)r_+^2)}{(r_+-2)(r_*+r_+)(r_*^2 + (r_+-3)r_* + r_+)}, \\
E &=& E^{\text{s}}(r_*) \equiv 
\frac{\sqrt{r_*}(r_*+r_+-2)}{\sqrt{(r_*+r_+)(r_*^2 + (r_+-3)r_* + r_+)}} 
\end{eqnarray}
for all $r_+ \leq r_* < \infty$. Since $R''(r_*)<0$ the double root corresponds to stable spherical orbits, here the superscript $^\text{s}$. Of course for $r_* = r_+$, one recovers the triple root at the horizon with $Q=Q^{\text{s}}(r_+) = Q_+(E_+)<0$ and $E=E^{\text{s}}(r_+)=E_+$, which lies outside the phase space. The function $E^{\text{s}}(r_*)$ is monotonously increasing along $r_*$. We call the inverse function $r_*(E)$ and $Q^{\text{s}}(E) \equiv Q^{\text{s}}(r_*(E))$.  For $r_+ < r_* < \infty$, we have $E_+ < E^{\text{s}}(r_*)< 1$. The positivity bound $Q^{\text{s}}(r_*) \geq 0$ is obeyed for $r^{\text{min}}_* \leq r_* < \infty$ where 
\begin{eqnarray}
r^{\text{min}}_* \equiv 
\frac{r_+(r_++2\sqrt{r_+-1})}{2-r_+}.
\end{eqnarray}
The function $Q^{\text{s}}(r_*)$ is monotonic between $Q^{\text{s}}(r_*^{\text{min}})=0$ and $Q^{\text{s}}(+\infty)=+\infty$. We denote the critical energy 
\begin{eqnarray}
E_c \equiv E^{\text{s}}(r^{\text{min}}_*)= 
\frac{r_++2 \sqrt{r_+-1}}{\sqrt{r_+(r_++4 \sqrt{r_+-1}+2)}}.\label{Ec}
\end{eqnarray}
This function of $a$ is plotted on Figure \ref{Eminstar}. It obeys $E_+ < E_c$ for all $0 \leq a \leq 1$.

We, therefore, obtain that for $\ell=\ell_+(E)$, $Q=Q^{\text{s}}(E)$  with $E_c \leq  E < 1$, spherical orbits exist in the phase space with root structure ${\vert \hspace{-7pt}\bullet - \bullet \hspace{-6pt}\bullet \hspace{2pt}- \rangle}$. For $\ell=\ell_+(E)$, $Q \lesssim Q^{\text{s}}(E)$, the root structure becomes ${\vert \hspace{-7pt}\bullet - \bullet + \bullet \hspace{2pt}- \rangle}$. Since there is no other double root outside the horizon and no horizon touching root, the root structure ${\vert \hspace{-7pt}\bullet - \bullet + \bullet \hspace{2pt}- \rangle}$ is valid in the entire range $0 \leq Q \leq  Q^{\text{s}}(E)$. For $\ell=\ell_+(E)$, $Q \gtrsim Q^{\text{s}}(E)$, the root structure becomes ${\vert \hspace{-7pt}\bullet  - \rangle}$. Again, since there are no further double roots and no horizon touching root, this root structure is valid for the entire range $ Q > \text{max}\{0,Q^{\text{s}}(E)\}$. For $\ell = \ell_+(E)$ and $E \geq 1$ and any $Q \geq 0$ there is a single root structure since the roots never cross in the exterior region $r_+ \leq r < \infty$ and never become double. After explicit evaluation for a particular case we find the root structure ${\vert \hspace{-7pt}\bullet - \bullet +  \rangle}$. 

\begin{figure}
     \centering
     \includegraphics[width=0.8\textwidth]{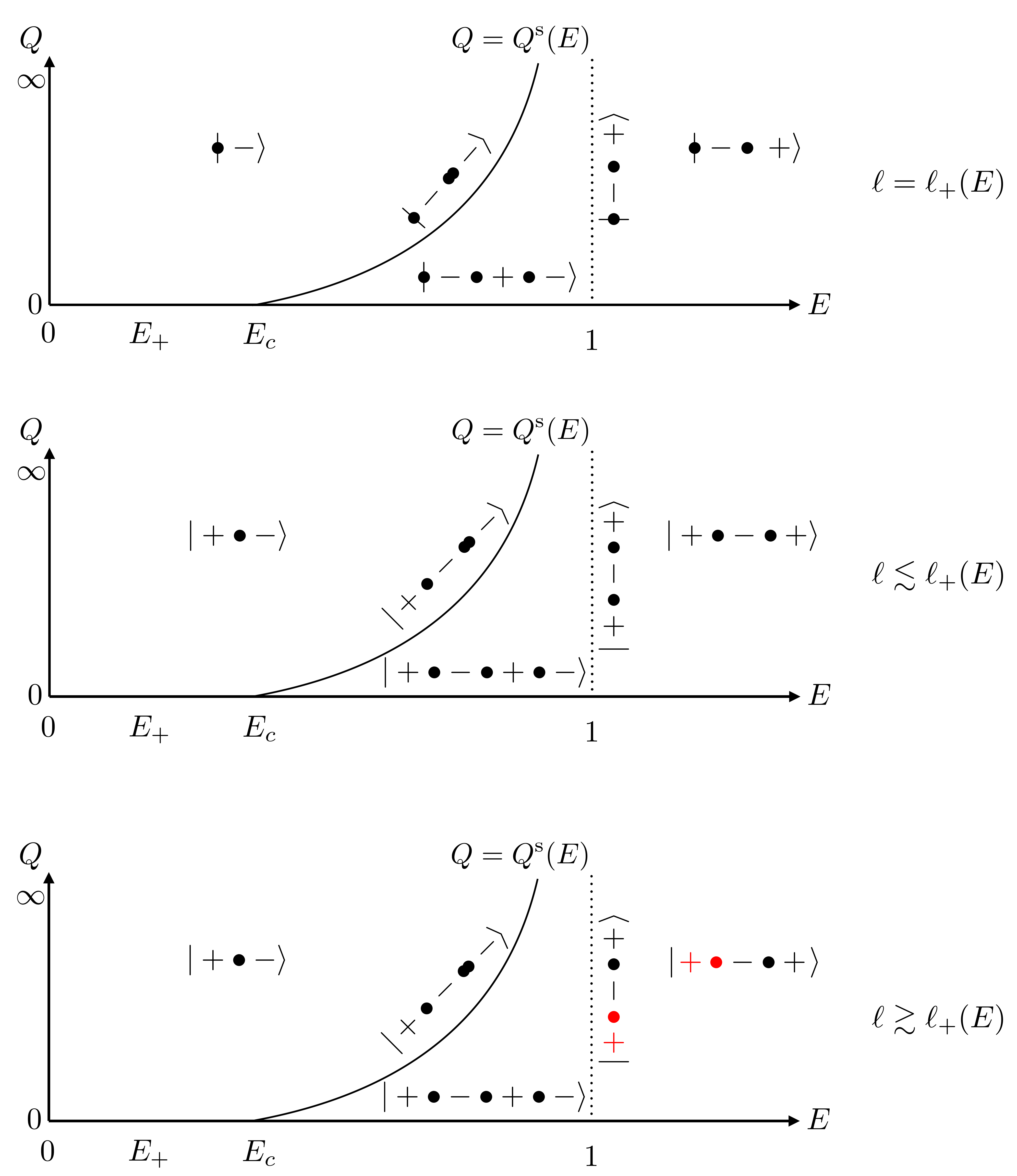}
     \caption{Phase space of root structures that admit at least one root at or close to the horizon and that obey the positivity bounds.}
     \label{phasespacehorizonroots}
 \end{figure}

For $\ell \, \simneq \,\ell_+(E)$ [$\ell$ close to but not equal to $\ell_+(E)$], the root at the horizon moves towards positive radius, which allows orbits close to the horizon. The phase space of root structures that admit at least one root at or close to the horizon is summarized on Figure \ref{phasespacehorizonroots}. 

We finally note that there is a special value
\begin{equation}
    a_c \approx 0.8109337526
\end{equation}
of $a$ such that the energy $E_c$ coincides with the energy  $ E_{ISCO^-}$ at the retrograde ISCO as defined in \eqref{EISCOm} in section \ref{app:teo}. For that special value of $a$ and the corresponding special energy $E=E_c=E_{ISCO^-} \approx 0.9598057008$, both orbit classes ${\vert \hspace{-7pt}\bullet - \bullet \hspace{-6pt}\bullet \hspace{2pt}- \rangle}$ and $\vert + \bullet \hspace{-5pt}\bullet \hspace{-5pt}\bullet \hspace{2pt}- \rangle$ are equatorial ($Q=0$) orbit classes (with a distinct angular momentum). 
\begin{table}[!tbh]    \centering
\begin{tabular}{|c|c|c|c|}\hline
 &$E_{c}$ & $E_{{ISCO}^+}$ & $E_{{ISCO}^-}$ \\\hline
$a=0$& $1$ & $\frac{2\sqrt{2}}{3}$&$\frac{2\sqrt{2}}{3}$\\\cline{1-4}
$a=1$& $\frac{1}{\sqrt{3}}$ & $\frac{1}{\sqrt{3}}$&$\frac{5}{3\sqrt{3}}$\\ \hline
\end{tabular}\caption{Energy $E_c$ and $E_{{ISCO}^\pm}$ in the Schwarzschild and extreme Kerr limit.}\label{table:energy}
\end{table}
In the Schwarzschild and extreme Kerr limit, the energy $E_{{ISCO}^{\pm}}$ and $E_c$ are shown in Table \ref{table:energy}.
When $a=0$, the two ISCO branches merge as they should for the Schwarzschild black hole.

\clearpage

\subsection{Spherical orbits}
\label{app:teo}

The timelike spherical orbits of Kerr, defined from $R(r)=R'(r)=0$,  were elegantly summarized by Teo  \cite{Teo:2020sey} based on earlier results \cite{Hughes:1999bq,Hughes:2001jr,Fayos:2007ks,Grib:2014afa,Grib:2015rea,Rana:2019bsn}. There are 4 classes of solutions $(E_i,\ell_i)$, $i=\text{a}',\text{b}',\text{c}',\text{d}'$. The first two are given by 
\bea
E_{\text{a}',\text{b}'} (Q,r)&=& \frac{r^3(r-2 )-a(a Q \mp \sqrt{\Upsilon(Q,r)})}{r^2 \sqrt{r^3(r-3)-2a (a  Q  \mp \sqrt{\Upsilon(Q,r)})}} ,\\ 
\ell_{\text{a}',\text{b}'}(Q,r) &=&- \frac{2  a r^3 +(r^2+a^2)(a Q \mp \sqrt{\Upsilon(Q,r)})}{r^2 \sqrt{r^3 (r-3)-2a(a  Q  \mp \sqrt{\Upsilon(Q,r)})}},
\eea
where 
\bea
\Upsilon(Q,r) =  r^5 - Qr^3(r-3 ) +a^2 Q^2.
\eea
The third and fourth are given by $(E_{\text{c}',\text{d}'} ,\ell_{\text{c}',\text{d}'} )=-(E_{\text{a}',\text{b}'} ,\ell_{\text{a}',\text{b}'} )$. The first and second classes continuously join. They have positive energy in the range of parameters where they exist. The range of existence is dictated by the radii of the prograde and retrograde photon orbits, respectively $r_1$ and $r_2$, 
\bea
r_{1,2} \equiv 2  \left[ 1+ \cos \left( \frac{2}{3} \arccos \left( \mp a \right) \right) \right]. 
\eea
They lie in the range $1 \leq r_1 < 3  < r_2 \leq 4 $ and are the two largest roots of \ $\Xi \equiv r^3-6 r^2 + 9 r-4a^2$. The range of existence is also determined by 
\bea
Q_1(r) = \frac{r^2}{2a^2} (r(r-3)- \sqrt{r \Xi}).
\eea
In the nonextremal case, the solution $\text{a}'$ exists for $r > r_1$: for any $0 \leq Q < \infty$ between $r_1 < r < r_2$ but for $0 \leq Q < Q_1(r)$ for $r \geq r_2 $. The solution $\text{b}'$ exists for  $r > r_2$ in the range $0 \leq Q \leq Q_1(r)$. In the extremal case $a=1$, the Boyer-Linquist radius does not resolve the near-horizon region and does not lead to a rightful parametrization at $r=1$, see \cite{Teo:2020sey} for a discussion. 

\paragraph{Prograde and retrograde orbits} The solution $\text{a}'$ is either prograde or retrograde while the solution $\text{b}'$ is retrograde. The subset of polar orbits ($\ell = 0$) within the solution $\text{a}'$ set are given by $Q = Q_0(r)$ where
\bea
Q_0(r) \equiv \frac{ r^2 (\Delta^2+4 r^2(r-1))}{(r^2+a^2)(r \Delta - (r^2-a^2))}. 
\eea
The angular momentum $\ell > 0$ for $Q < Q_0(r)$ while $\ell < 0$ for $Q > Q_0(r)$. In the range $r \geq r_2$ where $Q_1(r)$ is real, it obeys $Q_0(r) \leq Q_1(r)$. We have $Q_0(r) \geq 0$ for $r \geq r_0$ where 
\bea
r_0 \equiv 1+2 \sqrt{\frac{3-a^2}{3}}\cos \left( \frac{1}{3} \arccos \left( \frac{3(1-a^2)}{3-a^2} \sqrt{\frac{3}{3-a^2}} \right) \right). 
\eea

\paragraph{Marginally stable spherical orbits} 
The marginally stable spherical orbits $R(r)=R'(r)=R''(r)=0$ obey $Q = Q_{ms}(r)$, where 
\bea
Q_{ms}(r) \equiv - \frac{r^{5/2} [(\sqrt{\Delta}-2 \sqrt{ r})^2-4 a^2]}{4a^2 (r^{3/2}- \sqrt{r}-\sqrt{ \Delta})}.
\eea
In the range $r \geq r_2$, where $Q_1(r)$ is real, $Q_{ms}(r) \leq Q_1(r)$. The locus where $Q_{ms}(r)=Q_1(r)$ is $r=r_{ms}^*$ where 
\bea
r_{ms}^* \equiv  \left[ \frac{11}{4} + \frac{7}{2} \cos \left( \frac{1}{3} \arccos \left( \frac{143+200 a^2}{343}\right) \right) \right].
\eea

When $Q=0$, the prograde and retrograde marginally stable orbits are located at $r=r_{ms}^\pm$, where
\bea
r_{ms}^\pm &\equiv & \left\{ 3+Z_2\mp  \sqrt{ (3-Z_1)(3+Z_1 +2Z_2)} \right\},\label{rms} \\
Z_1 &\equiv & 1+(1-a^2)^{1/3}\left[(1+a)^{1/3}+(1-a)^{1/3} \right], \qquad Z_2 \equiv \sqrt{3a^2+(Z_1)^2}. \nonumber
\eea
The class $\text{a}'$ of marginally stable orbits lie in the range $r_{ms}^+ \leq r \leq r_{ms}^*$. Stability occurs for $0 \leq Q < Q_{ms}(r)$, and unstability for $Q_{ms}(r) < Q < \infty$. The class $\text{b}'$ of marginally stable orbits smoothly joins in the next range $r_{ms}^* \leq r \leq r_{ms}^-$. Unstability occurs for $0 \leq Q < Q_{ms}(r)$ and stability for $Q_{ms}(r) < Q \leq Q_1(r)$. The energy of the prograde and retrograde ISCO ($Q=0$) orbits are respectively given by
\begin{eqnarray}
E_{ISCO^+}&=& E_{\text{a}'}(0,r^+_{ms}),\label{EISCOp} \\
E_{ISCO^-}&=& E_{\text{b}'}(0,r^-_{ms})\label{EISCOm}.
\end{eqnarray}

\paragraph{Marginally bound spherical orbits} 

The marginally bound spherical orbits $R(r)=R'(r)=0$ and $E^2=1$ obey $Q = Q_{mb}(r)$ where 
\bea
Q_{mb}(r) \equiv - \frac{ r^2 [r(\sqrt{r}-2)^2-a^2]}{a^2 (\sqrt{r}-1)^2}.
\eea
When $Q=0$, the prograde and retrograde marginally bound circular orbits lie at $r=r_{mb}^+ \equiv (1+\sqrt{1-a})^2$ and $r=r_{mb}^- \equiv (1+\sqrt{1+a})^2$. 

In the range $r \geq r_2$, where $Q_1(r)$ is real, $Q_{mb}(r) \leq Q_1(r)$. Equality $Q_{mb}(r)=Q_1(r)$ occurs at $r=r_{mb}^*$, where $r_{mb}^*$ is the largest real root of the quintic $r^2 \Xi -\Delta^2 = 0$. Note that the largest root flips only at $a=\frac{1}{5^{5/2}\sqrt{2}}\sqrt{129^{3/2}-383}$. It obeys $r_{mb}^+ \leq r_{mb}^* \leq r_{mb}^-$. 

The class $\text{a}'$ of marginally bound orbits lie in the range $r_{mb}^+ \leq r \leq r_{mb}^*$. Bound orbits have $0 \leq Q < Q_{mb}(r)$, and unbound spherical orbits have $Q_{mb}(r) < Q < \infty$. The class $\text{b}'$ of marginally bound orbits lie in the range $r_{mb}^* \leq r \leq r_{mb}^-$. Unbound spherical orbits occur for $0 \leq Q < Q_{mb}(r)$, and bound spherical orbits occur for $Q_{mb}(r) < Q \leq Q_1(r)$.

\subsection{Marginal orbits}
\label{sec:marginal}

Marginal orbits are by definition orbits such that $E = 1$. The radial potential is 
\be
R(r)=2(r^3-\frac{1}{2}(Q+\ell^2)r^2+(Q+(\ell-a)^2)r-\frac{a^2 Q}{2}).
\ee 
The horizon is located at $r_+=1+\sqrt{1-a^2}$. Since $R(\infty)>0$, the root structure always takes the form $\vert \dots + \rangle$. The potential has the cubic discriminant $\Delta_3=-108(\frac{q_r^2}{4}+\frac{p^3_{r}}{27})$, where 
\bea
p_r&=&Q+(\ell-a)^2-\frac{1}{12}(Q+\ell^2)^2,\\
q_r&=&-\frac{1}{108}(Q+\ell^2)^3+\frac{1}{6}(Q+\ell^2)(Q+(\ell-a)^2)-\frac{1}{2}Q a^2.
\eea 

\subsubsection{Double root structure}

Double roots $r_*$ (where $R(r_*)=R'(r_*)=0$ or $\Delta_3=0$) occur for two solution branches denoted as $\text{a},\text{b}$ ($\text{a}$ is associated with the upper sign): 
\bea
Q = Q_{\text{a},\text{b}}(r_*)& \equiv &\frac{r_*^2(a^2-(\sqrt{r_*}\pm 2)^2r_*)}{a^2(\sqrt{r_*}\pm1)^2},\label{Qab1}\\
\ell = \ell_{\text{a},\text{b}}(r_*)&\equiv &\frac{a^2\pm 2r_*^{3/2}+r_*^2}{a(1\pm \sqrt{r_*})}.\label{lab1}
\eea 
These branches exist for the range $r_+ \leq r_* < \infty$. The branches intersect in the exterior region only at the horizon where $\ell_{\text{a},\text{b}}(r_+) = \frac{2r_+}{a}$, $Q_{\text{a},\text{b}} (r_+)= 2(2+r_+-\frac{4r_+}{a^2})<0$. Outside the horizon we have $Q_{\text a}(r_*) < Q_{\text b}(r_*)$. In terms of the double root $r_*$, the third real root is $\frac{a^2-(\sqrt{r_*}\pm 2)^2 r_*}{2(\sqrt{r_*}\pm 1)^2}$ for each solution branch. In both cases, this root is below $r_+$ and therefore irrelevant to the motion. The root structure is therefore given for both these branches as $\vert + \bullet \hspace{-4pt}\bullet \hspace{2pt} + \rangle$ for $r_*>r_+$ and ${\hspace{3pt}\vert \hspace{-9pt} \bullet \hspace{-5pt}\bullet \hspace{2pt}} + \rangle$ for $r_*=r_+$.

Orbits of either branch belong to the phase space only if the bound $Q_{\text{a},\text{b}}  \geq  0$ is obeyed. We have $\ell_+(r_*)>0$. Now $Q_{\text a}(r_*)\ge 0$ only in the range
$0\le r_*\le 2+a-2 \sqrt{1+a}$, which lies below the horizon. Such solution is then disallowed in the exterior region. With a notation that will match the one of Section \ref{sec:nonmarginal} we define 
\begin{equation}\begin{split}
r_c^{(1)}&\equiv 2+a+2\sqrt{1+a},\\r_c^{(2)} &\equiv 2-a+2\sqrt{1-a} 
\end{split}\label{rminmax}
\end{equation} 
with $r_+ < r_c^{(2)}  < 4 < r_c^{(1)}$ for $0<a<1$.  For the branch b, $Q_{\text b}(r_*)\ge 0$ with $r_* >r_+$ is satisfied for $ r_c^{(2)} \le r_*\le r_c^{(1)}$. The root structure ${\hspace{3pt}\vert \hspace{-9pt} \bullet \hspace{-5pt}\bullet \hspace{2pt}} + \rangle$ is therefore disallowed. We have $Q_{\text{b}}(r_c^{(1,2)} ) = 0$. One can check that $\ell_{\text b}(r_*)$ is a monotonically decreasing function in the range $ r_c^{(2)} \le r_*\le r_c^{(1)} $. It admits an inverse $r_*(\ell)$. At the endpoints $Q_{\text b}(r_c^{(2)})=Q_{\text b}(r_c^{(1)})=0$ and we have $\ell_{\text b}(r_c^{(2)})=2(1+\sqrt{1-a})$, $\ell_{\text b}(r_c^{(1)})=-2(1+\sqrt{1+a})$. The maximum occurs at $r_*=4$ where $Q_{\text b}(4)=16$ and $\ell_{\text b}(4)=-a$. 

\begin{figure}
     \centering
     \includegraphics[width=0.6\textwidth]{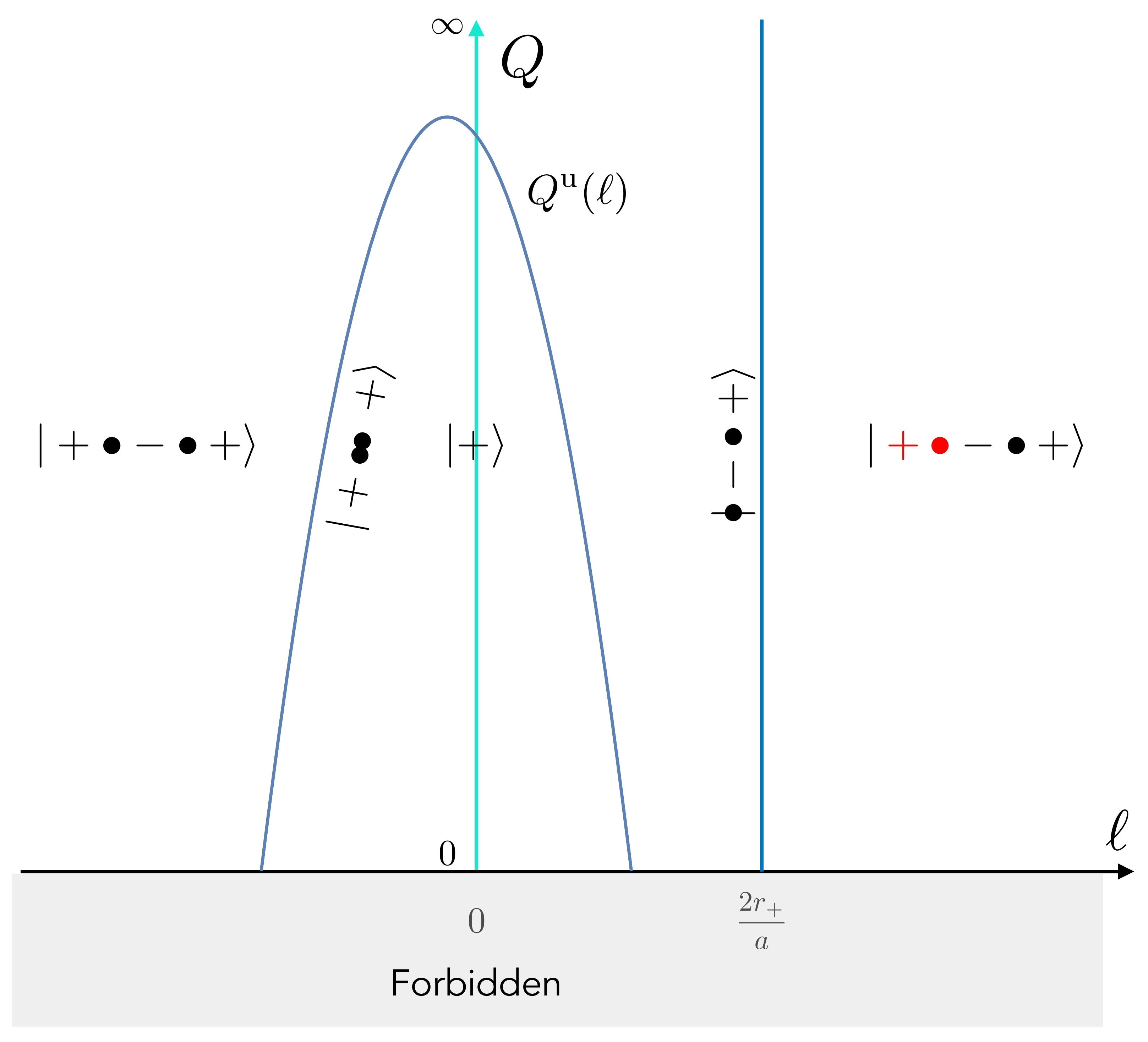}
     \caption{Root structures of marginal orbits $E=1$.}
     \label{marginalRoots}
 \end{figure}

Since $R''(r_*)>0$, spherical orbits are unstable. We define the function $Q^\text{u}(\ell) \equiv Q_{\text b}(r_*(\ell))$. For $Q=Q^\text{u}(\ell) $, the root structure is $\vert + \bullet \hspace{-4pt}\bullet \hspace{2pt} + \rangle$. The root structure for distinct values of $Q$ is given on Figure \ref{marginalRoots}. 

\subsubsection{General root structure}

Since $\Delta_3=-4a^6$ for $\ell=Q=0$, the double root becomes complex for $Q<Q^{\text{u}}(\ell) $ (i.e., the root structure is $\vert +\rangle$) while there are three distinct real roots for $Q>Q^{\text{u}}(\ell)$ (i.e., the root structure is  $\vert + \bullet - \bullet \hspace{2pt} + \rangle$ which can be read from large $Q$ expansion). 

The root structure changes when one root touches the horizon, which occurs at $\ell =\frac{2r_+}{a}$ as derived in Section \ref{sec:horizontouching}. We have $\ell_{\text b}(r_c^{(1)}) < \frac{2r_+}{a}$. Therefore, the curve $Q=Q^{\text{u}}(\ell)$ does not intersect the line $\ell = \frac{2r_+}{a}$. For $\ell > \frac{2r_+}{a}$, one also has the root structure $\vert + \bullet - \bullet \hspace{2pt} + \rangle$. Now, trapped orbits are disallowed for positive energy orbits with $\ell>\ell_+(E)$, see Section \ref{sec:boundergo}. The root structure is therefore $\vert\hspace{2pt} {\color{red} + \hspace{2pt}\bullet }- \bullet \hspace{2pt} + \rangle$. The large $\ell$ and large $Q$ limit are in agreement with the analysis of Section \ref{sec:largecharges}. We conclude that the phase space is complete for $E=1$.  The classification is depicted in Figure \ref{marginalRoots}.


\subsection{Generic nonmarginal orbits}
\label{sec:nonmarginal}

The radial potential \eqref{R} is quartic in $r$ for $E \neq 1$. The root structure for $E> 1$ takes the form $\vert \dots + \rangle$, while for $E<1$ it takes the form  $\vert \dots - \rangle$. For $\ell = \ell_+(E)$ it takes the form ${\vert \hspace{-5pt}\bullet }- \cdots \rangle$, while for $\ell \neq \ell_+(E) $ it takes the form $\vert + \cdots \rangle$. The discriminant takes the form 
\begin{equation}
\Delta_{4}= \frac{16(1-a^2)}{(E^2-1)^5}Q^5+b_4(\ell,E) Q^4+b_3(\ell,E) Q^3 + b_2(\ell,E) Q^2 + b_1(\ell,E) Q + b_0(\ell,E).
\end{equation}
It is a quintic function of $Q$ for a nonextremal black hole. 

\subsubsection{Double root structure}

The general solution of double roots at $r=r_*$  has been solved in \cite{Teo:2020sey}  by expressing $E$ and $\ell$ as a function of $r_*$ and $Q$ as we review in Section \ref{app:teo}. However, the analysis of the positivity bound on $Q$ is not straightforward for $E> 1$. Here, we will solve for the double roots by expressing $Q$ and $\ell$ as a function of $r_*$ and $E$.

\paragraph{Branches $\text{a}$ and $\text{b}$.} The two branches of the solution are ($a$ is the upper sign)
\bea
Q=Q_{\text{a},\text{b}}(E,r_*)&\equiv &\frac{r_*^2}{a^2 (r_*-1)^2} (-r_*^3+3r_*^2+(a^2-4)r_*+a^2\nn\\&&
\hspace{-1cm}+r_* (1-E^2)(r_*^3-4r_*^2+5r_* -2 a^2)\mp 2 E  \Delta(r_*) \sqrt{r_*(1+(E^2-1)r_*)}),\label{Qrs}\\
\ell = \ell_{\text{a},\text{b}}(E,r_*)&\equiv &\frac{E (r_*^2 -a^2) \pm \Delta(r_*)   \sqrt{r_* \left(\left(E^2-1\right) r_*+1\right)}}{a (r_*-1)},\label{lrs}
\eea 
where $\Delta(r) = r^2 -2 r +a^2$. These solutions formally match with \eqref{Qab1}--\eqref{lab1} for $E=1$. The solution is not valid for $r_*=1$, which coincides with the NHEK region at extremality. Since we are discussing $a<1$, the solutions are always real in the exterior region as long as $E \geq E_{\text{a} \times \text{b}}(r_*)$, where $E_{\text{a} \times \text{b}}(r_*) \equiv \sqrt{1-r_*^{-1}}$
is the energy at which the two branches meet. We have $E_+ \leq E_{\text{a} \times \text{b}}(r_*)  \leq 1$, where $E_+$ was defined in \eqref{Eplus}. Alternatively, the two branches $\text{a},\text{b}$ meet at $r_*=r_{\text{a} \times \text{b}}(E) \equiv (1-E^2)^{-1}$. For $E \geq 1$, the solutions $\text{a},\text{b}$ both exist for $r_+ \leq r_* < \infty$. For $E_+ \leq E < 1$, both solutions exist in the range $r_+ \leq r_* \leq r_{\text{a} \times \text{b}}(E)$. For $E < 1$, $Q_{\text{a},\text{b}}(E,r_{\text{a} \times \text{b}}(E)) < 0$, which violates the bound \eqref{boundQ}. Therefore, the two branches do not meet inside the phase space for $E <1$.  For $E >1$, $r_{\text{a} \times \text{b}}(E)<0$ and, therefore, the branches meet outside the phase space as well.

\paragraph{Analysis of $Q_{\text{a},\text{b}} \geq 0$.}  We have $\vert \ell_{\text{a},\text{b}}\vert > a \sqrt{E^2-1}$. Therefore, the positivity bound \eqref{boundQ} for both branches $\text{a},\text{b}$ is $Q_{\text{a},\text{b}} \geq 0$. The roots of $Q_{\text{a},\text{b}}$ might only occur at (upper sign corresponds to $\text{a}$)
\bea
E^{(1)}_{\text{a},\text{b}}(r_*)=\mp\frac{(r_*-2)\sqrt{r_*}-a}{r_*^{3/4}\sqrt{(r_*-3)\sqrt{r_*}-2a}},\label{oldE1}\\
E^{(2)}_{\text{a},\text{b}}(r_*)=\mp\frac{(r_*-2)\sqrt{r_*}+a}{r_*^{3/4}\sqrt{(r_*-3)\sqrt{r_*}+2a}}.\label{oldE2}
\eea 
We denote by $r_*^{(1)}$ the only root of $(r_*-3)\sqrt{r_*}+2a$ above the horizon, by $r_*^{(2)}$ the only root of $(r_*-3)\sqrt{r_*}-2a$ above the horizon. Explicitly, 
\bea
r_*^{(1)} &=& 2+\cos\left(\frac{2\arcsin a}{3}\right)-\sqrt{3}\sin \left(\frac{2 \arcsin a}{3}\right), \\
r_*^{(2)}&=&  2+\cos\left(\frac{2\arcsin a}{3}\right)+\sqrt{3}\sin \left(\frac{2 \arcsin a}{3}\right).
\eea
We have $r_+ < r_*^{(1)} \leq r_*^{(2)} < 4$ for $0 \leq a < 1$. Note that $E^{(1)}_{\text{a},\text{b}}$ is a real root in the range $r_*^{(2)} < r_*<\infty$, while  $E^{(2)}_{\text{a},\text{b}}$ is a real root in the range $r_*^{(1)} < r_*<\infty$. 

Let us now analyze the positivity of $Q$ in the relevant ranges of $r_*$. In the range $r_+ \le r_* \le r_*^{(1)}$, there is no real root for $Q_{\text{a},\text{b}}$, and  $Q_{\text{a},\text{b}}$ is negative, $Q_{\text{a},\text{b}}<0$. In the range $r_*^{(1)} < r_* \le r_*^{(2)}$, $E^{(1)}_{\text{a},\text{b}}$ is complex, but $E^{(2)}_{\text{a},\text{b}}$ is real. We have  $E^{(2)}_{\text{a}}\leq 0$, while $E^{(2)}_{\text{b}}\geq 0$. In the entire range $E^{(2)}_{\text{a}}<0 \leq E$, we have $Q_{\text{a}} < 0$.  For branch $\text{b}$, $Q_{\text{b}}\geq 0$ only for $E \geq E^{(2)}_{\text{b}}$. Finally, in the range $ r_*^{(2)}< r_*<\infty$, both roots $E^{(1,2)}_{\text{a,b}}$ are real for each branch. We find $E_{\text{a}}^{(1,2)}< 0$. Then, in the entire range $E_{\text{a}}^{(1,2)}< 0 \leq E$, we find $Q_{\text{a}}<0$.  Instead, we have $0  < E_{\text{b}}^{(2)} < E_{\text{b}}^{(1)}$, and we find that $Q_{\text{b}} \geq 0$ in the range $E_{\text{b}}^{(2)}\leq E \leq  E_{\text{b}}^{(1)}$. 

We conclude that the  branch $\text{a}$ lies outside the phase space and we, therefore, discard it. In the range $E_{\text{b}}^{(2)} \leq E$, branch $\text{b}$ is allowed for $r_*^{(1)} < r_* \le r_{*}^{(2)}$. In the range $E_{\text{b}}^{(2)}\leq E \leq  E_{\text{b}}^{(1)}$, branch $\text{b}$ is allowed for $r_*^{(2)} < r_* < \infty$. The final allowed range is therefore the union of the regions
\begin{equation}
    r_*^{(1)} < r_* \le r_{*}^{(2)}\quad \& \quad  E_{\text{b}}^{(2)} \leq E 
\end{equation}
and
\begin{equation}
r_*^{(2)} < r_* < \infty \quad \& \quad E_{\text{b}}^{(2)}\leq E \leq  E_{\text{b}}^{(1)}. 
\end{equation}

\paragraph{Decomposition of the region $Q_{\text{b}} \geq 0$ for $E >1$ and $E \leq 1$.} 

\begin{figure}
     \centering 
\includegraphics[width=0.5\textwidth]{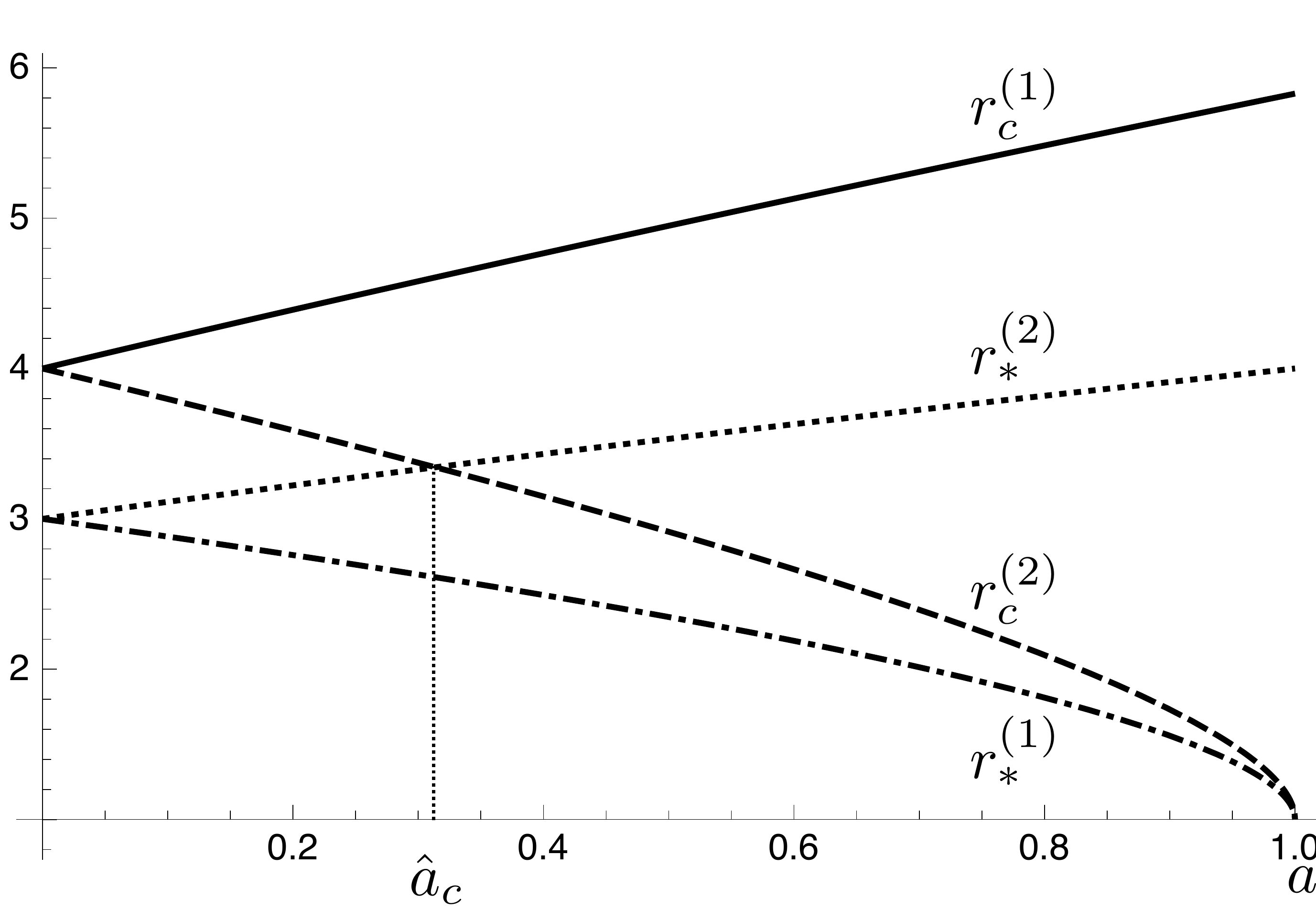}
\caption{Several critical radii of interest for the analysis of $Q \geq 0$ in the presence of double roots.}
\label{foura}
\end{figure}

\begin{figure}[htb]
     \centering 
\includegraphics[width=0.95\textwidth]{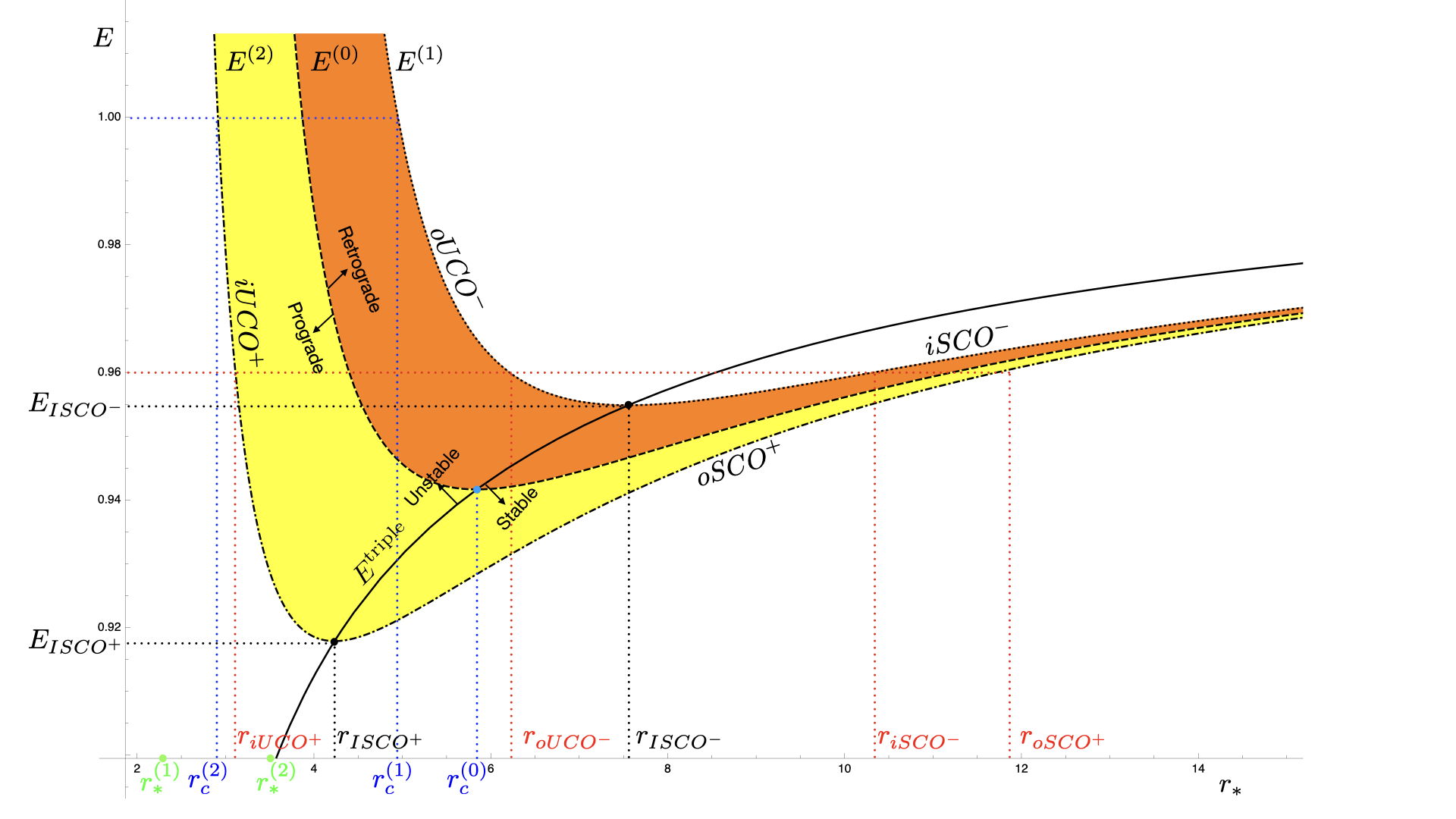}
\caption{Functions $E^{(0,1,2)}(r_*)$ and $E^{\text{triple}}(r_*)$ depicted for $a=1/2$ without loss of generality. The double roots are allowed both in the orange region (retrograde orbits: $\ell < 0$) and the yellow region (prograde orbits: $\ell > 0$). The orbits corresponding to triple roots exist (i.e., obey the positivity bound) only in the interval $r_{ISCO^+}\le r_*\le r_{ISCO^-}$ or $E_{ISCO^+}\le E\le E_{ISCO^-}$. We also depicted the four roots $r_{iUCO^+} < r_{oUCO^-} < r_{iSCO^-} < r_{oSCO^+}$ for a specific $E_{ISCO^-} < E < 1$.}
\label{Eb012}
\end{figure}

The slicing of the allowed region with fixed energy $E$ will be performed on Figure \ref{Eb012}. As a preparation, let us derive the allowed region $Q \geq 0$ for $E> 1$ and $E \leq 1$. Let us denote $r_c^{(1,2)}$ as the only root of $E_{\text{b}}^{(1,2)}-1$ above the horizon. We have
\begin{eqnarray}
r_c^{(1)}  &=& 2+a +2\sqrt{1+a} , \\ 
r_c^{(2)}  &=& 2-a +2\sqrt{1-a} .
\end{eqnarray}
This definition agrees with \eqref{rminmax}. We have $E_b^{(1)}>1$ for $r_*^{(2)}<r_*<r_c^{(1)}$, while $E_b^{(2)}>1$ for $r_*^{(1)}<r_*<r_c^{(2)}$.  The four special functions  $r_*^{(1,2)}$, $r_c^{(1,2)}$ of $a$ are plotted on Figure \ref{foura}. There is a single $a$ where there is an intersection of two such functions. We denote $\hat a_c \approx 0.313708$ the critical $a$ such that $r_*^{(2)}=r_c^{(2)}$.  For $0 \leq a < \hat a_c$, we have $r_*^{(1)} \le r_*^{(2)} < r_c^{(2)} < r_c^{(1)}$, while for $\hat a_c \leq a < 1$, we have $r_*^{(1)} < r_c^{(2)} \le r_*^{(2)} < r_c^{(1)}$.

From now on, we discard branch $\text{a}$ and drop the labels $\text{b}$ in all quantities. Instead of \eqref{oldE1}--\eqref{oldE2}, we now denote
\bea
E^{(1)}(r_*) \equiv \frac{(r_*-2)\sqrt{r_*}-a}{r_*^{3/4}\sqrt{(r_*-3)\sqrt{r_*}-2a}},\\
E^{(2)}(r_*)\equiv \frac{(r_*-2)\sqrt{r_*}+a}{r_*^{3/4}\sqrt{(r_*-3)\sqrt{r_*}+2a}}.\label{E2}
\eea
Finally, for $E < 1$ the only allowed range is  
\begin{equation}
r_c^{(2)} \leq r_* < \infty \; \& \; E^{(2)}\leq E < \text{min}(1,E^{(1)}). \label{rangeEm1}
\end{equation}
For $E\ge 1$, the range such that $Q\ge 0$ is
\begin{equation}
    r_*^{(1)}<r_*\le r_*^{(2)}\;\&\; E\ge\text{max}(1,E^{(2)}),
\end{equation}
together with
\begin{equation}
    r_*^{(2)}<r_*\le r_c^{(1)}\;\&\; \text{max}(1,E^{(2)})\le E \le E^{(1)}.
\end{equation}
For $r_*\to \infty$, both $E^{(1)}$ and $E^{(2)}$ approach $1$ from below.

\paragraph{Additional simple roots} 

After imposing \eqref{Qrs}--\eqref{lrs} (for the branch $\text{b}$), the residual potential $Y(r) \equiv R(r)/(r-r_*)^2$ is quadratic in $r$. Its discriminant is quartic in $E$ and vanishes for
\begin{equation*}
    E = \pm\sqrt{\frac{r_*(r_*-2)(r_*^2+r_*-1)+a^2 (r_*-1)(2r_*+1)+a^4 \pm 2 \sqrt{-(1-a^2)r_* \Delta(r_*)}}{(r_*^2+a^2)^2-4r_*^2}},
\end{equation*}
which are always complex numbers for $r_*>r_+$ since $(1-a^2)r_* \Delta(r_*) >0$. Since the discriminant is positive for any particular choice of parameters with $r_*>r_+$, it implies that the two residual roots are always real for $r_*>r_+$. In the large $Q$ limit, one of these roots is always below the horizon and the other one, which we will call $r_{s}$ is always above the horizon, or at the horizon in the special case $\ell = \ell_+(E)$. Since the horizon touching only occurs at $\ell = \ell_+(E)$ as studied in Section \ref{sec:horizontouching} and, in particular, does not occur at specific $Q$, we deduce that the root structure is general: only one additional root $r_s$ is above the horizon for  $\ell \neq \ell_+(E)$, and it coincides with the horizon for $\ell = \ell_+(E)$. The separatrix between the position of the roots,  $r_s < r_*$ or $r_s > r_*$, is determined by the triple root $r_s=r_*$, which will be analyzed below.

\paragraph{Triple roots (ISSO).} Triple roots occur for 
\bea
    E &=& E^{\text{triple}}(r_*) \equiv \frac{1}{2}\sqrt{\frac{\sqrt{r_*}(4r_*^4-15r_*^3+21r_*^2-(10+3a^2)r_*+3a^2)+ ( \Delta(r_*))^{3/2}}{r_*^{3/2}(r_*^3-3r_*^2+3r_*-a^2)}},\label{Eabs}\\
    \ell&=&\ell^{\text{triple}}(r_*)\equiv E^{\text{triple}}(r_*)\times \frac{a^4+6 a^2 r_*^2+ 2 \left(r_* \Delta(r_*)\right)^{3/2}-6 a^2 r_*-3 r_*^4+2 r_*^3}{a^3+a r_* \left(-4 r_*^2+9 r_*-6\right)}, \label{Labs}\\
    Q&=&Q^{\text{triple}}(r_*)\equiv \frac{-3 a^2 r_*^4-a^2 r_*^3-3 r_*^{5/2} \left( \Delta(r_*)\right)^{3/2}+r_*^6-3 r_*^5+6 r_*^4}{4 a^2 \left(a^2-r_* ((r_*-3) r_*+3)\right)}.\label{Qabs}
\eea
There are no quadrupole roots. The argument of the square root is positive in the exterior region $r_* \geq r_+$. The function $E^{\text{triple}}$ is monotonically increasing along $r_*$ and asymptotes to 1 as $r_* \rightarrow \infty$. For $E<1$, the positivity bound is $Q \geq 0$. The triple root is physically associated with the ISSO. It belongs to the allowed range \eqref{rangeEm1} as long as $E^{(2)} \leq E^{\text{triple}} \leq E^{(1)}$, which amounts to $r^{\text{min}} \leq r_* \leq r^{\text{max}}$ where $r^{\text{min}}$ is the only radius above $r_+$ such that $E^{\text{triple}} = E^{(2)}$, and $r^{\text{max}}$ is the only radius above $r_+$ such that $E^{\text{triple}} = E^{(1)}$. We have $r_c^{(2)} < r^{\text{min}}\leq r^{\text{max}}$.  Since $Q=0$ and $\ell>0$ at $E=E^{(2)}(r^{\text{min}})$, the critical radius $r^{\text{min}}$ is nothing else than the prograde ISCO radius $r_{ISCO^+}\equiv r^{\text{min}}$. Since $Q=0$ and $\ell<0$ at $E=E^{(1)}(r^{\text{max}})$, the critical radius $r^{\text{max}}$ is nothing else than the retrograde ISCO radius $r_{ISCO^-} \equiv r^{\text{max}}$. We denote $E_{ISCO^+} \equiv E^{(2)}(r_{ISCO^+})$,  $E_{ISCO^-} = E^{(1)}(r_{ISCO^-})$. The standard expressions for $r_{ISCO^\pm}$ are recalled in Section  \ref{app:teo}.

In the range $r_c^{(2)} \leq r_* < \infty$, the function $E^{\text{triple}}(r_*)$ is monotonically increasing. We denote its inverse as $r_{ISSO}(E)\equiv  r_{*}^{\text{triple}}(E)$. In the allowed range $r_{ISCO^+} < r_{ISSO}(E) < r_{ISCO^-}$ the root system is finally ${\vert + \bullet \hspace{-5pt} \bullet \hspace{-5pt} \bullet  }\hspace{2pt}- \rangle$. It describes the ISSO as well as plunging orbits. 

\paragraph{Double roots with $\ell=0$}
Branch $\text{b}$ has $\ell_{\text{b}}(r_*) =0$ for 
\begin{eqnarray}
E = E^{(0)}(r_*) \equiv \frac{\Delta(r_*)\sqrt{r_*}}{\sqrt{r_*^2+a^2}\sqrt{r_*^3-3r_*^2+a^2r_*+a^2}}. \label{El0}
\end{eqnarray}
Here $r_*^{(0)} < r_*$ where $r_*^{(0)}$ is the real root of $r_*^3-3r_*^2+a^2r_*+a^2$ above the horizon. The function $r_*^{(0)}$ of $a$ is monotonically decreasing from $r_*^{(0)} =3$ for $a=0$ to $r_*^{(0)} = 1+\sqrt{2}$ for $a=1$. The condition $1+(E^2-1)r_* \geq 0$ is precisely $r_*^{(0)} \leq  r_*$. The phase space therefore contains the branch $\ell = 0$, $E = E^{(0)}(r_*)$, $Q=Q_{\text{b}}(E^{(0)}(r_*),r_*) \equiv Q_*^{(0)}(r_*)$ for $r_*^{(0)} < r_* < \infty$ where
\begin{equation}
    Q_*^{(0)}(r_*) = \frac{r_*^2 (r_*^4+2a^2 r_*(r_*-2)+a^4)}{(r_*^2+a^2)(r_*^3-3r_*^2+a^2 r_* +a^2)}. 
\end{equation}
There is a radius $r_*=r_c^{(0)}$ such that $E^{\text{triple}}(r_*)=E^{(0)}(r_*)$. The value of $r_c^{(0)}$ could be found by searching the solution $\ell^{\text{triple}}(r_*)=0$ between $r_{\text{ISCO+}}$ and $r_{\text{ISCO-}}$. It turns out that $r_c^{(0)}$ is the unique real solution of the equation
\be
r^6-6 r^5+3 a^2 r^4+4 a^2 r^3+3 a^4 r^2-6 a^4 r+a^6=0,
\ee
which is larger than $r_+$. It is monotonically decreasing from $r_c^{(0)}(a=0)=6$ to $r_c^{(0)}(a=1)=1+\sqrt{3}+\sqrt{3+2\sqrt{3}}$.

\paragraph{Structure of double roots}

We can now deduce the root structure of the roots that contain a double root as follows. Only branch $\text{b}$ exists in the phase space.  Figure \ref{Eb012} shows the allowed region for the double and triple roots. When $E<1$, the branch $E^{(2)}$ denotes the energy of the inner unstable prograde circular orbits ($iUCO^+$), whose locations are denoted as $r_{iUCO^+}$ in the region $r_{c}^{(2)}<r_{iUCO^+}<r_{ISCO^+}$, and the energy of the outer stable prograde circular orbits ($oSCO^+$) whose locations are denoted as $r_{oSCO^+}$ in the region $r_{ISCO^+}<r_{oSCO^+}<\infty$. The branch $E^{(1)}$ denotes  the energy of the inner stable retrograde circular orbits ($iSCO^-$), whose locations are denoted as $r_{iSCO^-}$  in the region $r_{ISCO^-}<r_{iSCO^-}<\infty$, and the energy of the outer unstable retrograde circular orbits ($oUCO^-$), whose locations are denoted by $r_{oUCO^-}$ in the region $r_{c}^{(1)}<r_{oUCO^-}<r_{ISCO^-}$.  When $E\ge1$, the branch $E^{(1)}$ denotes the energy of $oUCO^-$ in the region $r_*^{(2)}<r_{oUCO^-}\le r_c^{(1)}$, while the branch $E^{(2)}$ denotes the energy of $iUCO^+$ in the region $r_*^{(1)}<r_{iUCO^+}\le r_c^{(2)}$. We discuss the structure according to the energy $E$ (which intersects as straight lines the allowed region of Figure \ref{Eb012}) as follows:

\begin{itemize}
    \item For $0 < E < E_{ISCO^+} $, there is no double root. By continuity with the large $Q$ limit, the root structure is ${\vert + \bullet \hspace{2pt}- \rangle}$ for $\ell \neq \ell_+(E)$ and ${\vert \hspace{-7pt}\bullet - \rangle}$ for $\ell = \ell_+(E)$. 
    \item For $E_{ISCO^+} \leq E \leq E_{ISCO^-}$, $Q_{\text b}(E,r_*)$ is non-negative for $r_{iUCO^+}\leq r_* \leq r_{oSCO^+}$, and it vanishes at $r_{iUCO^+}$ and $ r_{oSCO^+}$. The double root $r_*$ becomes a triple root if and only if $r_* =r_{ISSO}(E)$ with root structure ${\vert + \bullet \hspace{-5pt} \bullet \hspace{-5pt} \bullet  }\hspace{2pt}- \rangle$. 
    
    For $r_*$ in the range $r_{iUCO^+} \leq r_* < r_{ISSO}(E)$, the root structure is ${\vert + \bullet  \hspace{-3pt} \bullet  }+ \bullet \hspace{2pt}  - \rangle$. The double root $r_*$ corresponds to unstable circular orbits since $R''(r_*)>0$.  The angular momentum is monotonously decreasing for  $r_{iUCO^+} \le r_* \le r_{\text{ISSO}}(E)$. In this region, one can define a unique inverse solution $\hat r^{u}_*(\ell,E)$ of $\ell=\ell_b(E,r_*)$ at fixed $E$. The function $Q^{u}(E,\ell)$ is defined by substituting the inverse solution $\hat r^{u}_*(\ell,E)$ into $Q_b(E,r_*)$, 
    \be
    Q^{u}(E,\ell) \equiv Q_b(E,\hat r_*^{u}(\ell,E)).\label{Q1l}
    \ee
    
    For $r_*$ in the range $r_{\text{ISSO}}(E) < r_* \leq r_{oSCO^+}$ the root structure is ${\vert + \bullet - \bullet \hspace{-3pt} \bullet  }\hspace{2pt}- \rangle$. The double root $r_*$ corresponds to stable circular orbits since $R''(r_*)<0$. The angular momentum is monotonously increasing for $r_{\text{ISSO}}(E)\le r_*\le r_{oSCO^+}$. In this region, one can  define another unique inverse solution $\hat r^{s}_*(\ell,E)$ for $\ell = \ell_b(E,r_*)$ at fixed $E$. The function $Q^{s}(E,\ell)$ is defined by substituting the inverse function $\hat r^{s}_*(\ell,E)$ into $Q_b(E,r_*)$,
    \be
    Q^{s}(E,\ell) \equiv Q_b(E,\hat r_*^{s}(\ell,E)).\label{Q2l}
    \ee 
    
    The root structures containing only simple roots is straightforwardly deduced by continuity, see Figure \ref{fig:classEminus1}.

 \item  For $E_{ISCO^-}< E < 1$, $Q_\text{b}(E,r_*)$ admits four real roots $r_*$ with the order
    \be
   r_{iUCO^+}<r_{oUCO^-}<r_{iSCO^-}<r_{oSCO^+}.
    \ee
    The region $Q_b(E,r_*)\ge 0$ consists of two disconnected regions,    $r_{iUCO^+}\le r_*\le r_{oUCO^-}$ and $r_{iSCO^-}\le r_*\le r_{oSCO^+}$. The function $\ell_b(r_*)$ is monotonously decreasing in the region $r_{iUCO^+}\le r_*\le r_{oUCO^-}$, where $R''(r_*)>0$, and monotonously increasing in the region $r_{iSCO^-}\le r_*\le r_{oSCO^+}$, where $R''(r_*)<0$. Therefore, one can extend the inverse $\hat r^{u}_*(\ell,E)$ as the inverse solution of 
    \be
    \ell=\ell_b(E,r_*),\quad r_{iUCO^+}\le r_*\le r_{oUCO^-}
    \ee 
    at fixed $E$.  Similarly, one can extend the inverse $\hat r^{s}_*(\ell,E)$ as the inverse solution of 
    \be
    \ell=\ell_b(E,r_*),\quad r_{iSCO^-}\le r_*\le r_{oSCO^+}.
    \ee Then $Q^{u,s}(E,\ell)$ are still defined as \eqref{Q1l}--\eqref{Q2l}.

\item For $ E > 1$, $Q_{\text b}(E,r_*)$ is non-negative for $r_{iUCO^+}\leq r_* \leq r_{oUCO^-}$, and it vanishes at $r_{iUCO^+}$ and $ r_{oUCO^-}$. The location of the double root structure $\vert + \bullet  \hspace{-4pt} \bullet  +\rangle$ is again determined by the function $Q=Q^{u}(E,\ell)$. We find for larger $Q$ the root structure $\vert + \bullet - \bullet \hspace{2pt}  + \rangle $ and for smaller $Q$ the root structure $\vert +\rangle$. Since $E>1$, this latter root structure is now bounded $Q \geq \text{min}\{0,-(\vert \ell\vert -a \sqrt{E^2-1})^2\}$ according to \eqref{boundQ}. We denote the positive and negative values of $\ell$ such that $Q^{u}(E,\ell)=0$ as $\ell_{iUCO^+}(E)$ and $\ell_{oUCO^-}(E)$, respectively. We have $a \sqrt{E^2-1} < \vert \ell\vert < \ell_+(E)$ for both $\ell_{iUCO^+}$ and $\ell_{oUCO^-}$. The root structure for $Q<0$ is $|+\rangle$ by continuity. 

\end{itemize}
 
\begin{figure}[!htbp]
     \centering 
     \begin{subfigure}[b]{0.45\textwidth}
         \centering
         \includegraphics[width=\textwidth]{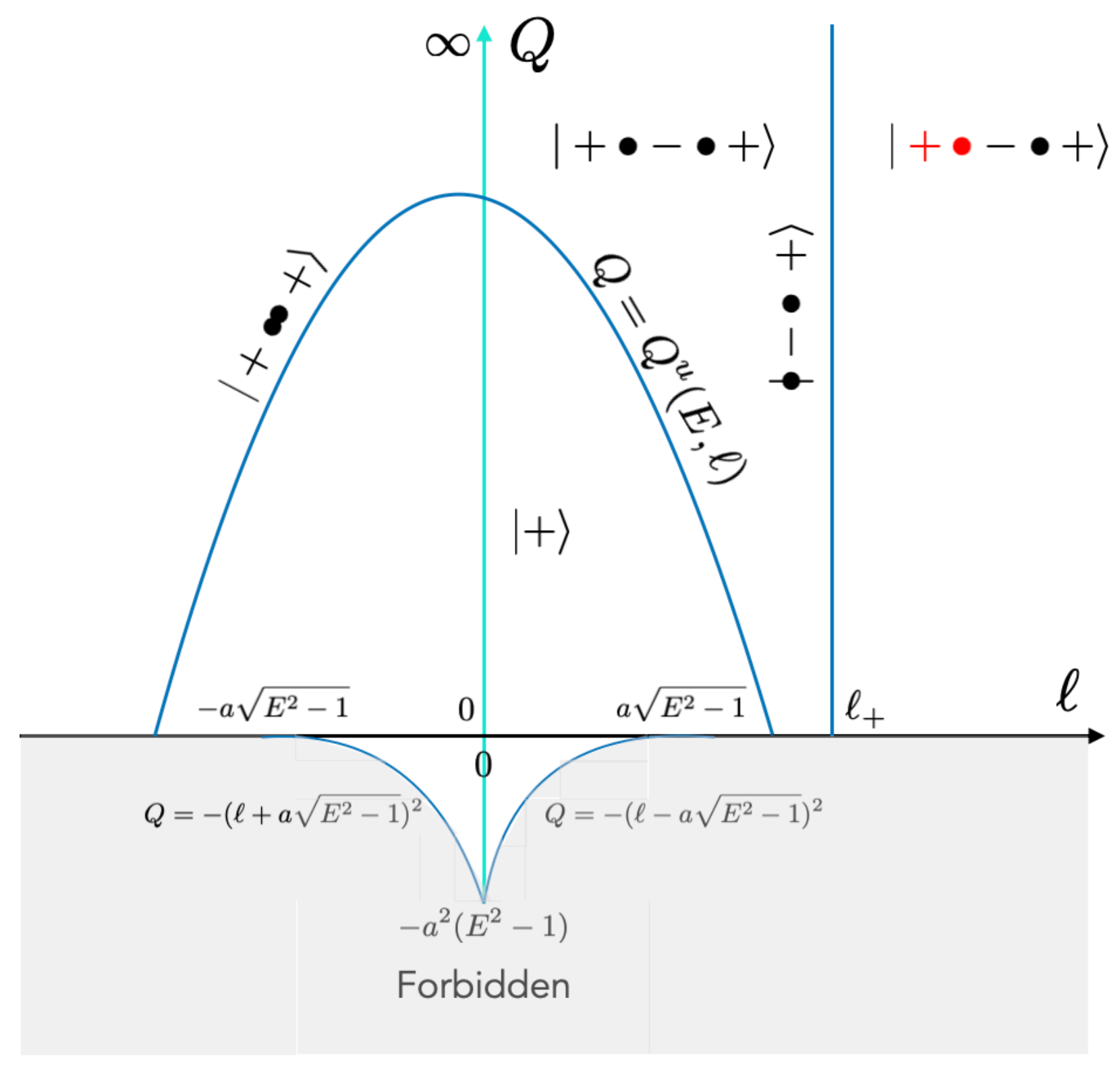}
     \end{subfigure}
     \hfill  \vspace{0.4cm}
     \begin{subfigure}[b]{0.45\textwidth}
         \centering
         \includegraphics[width=\textwidth]{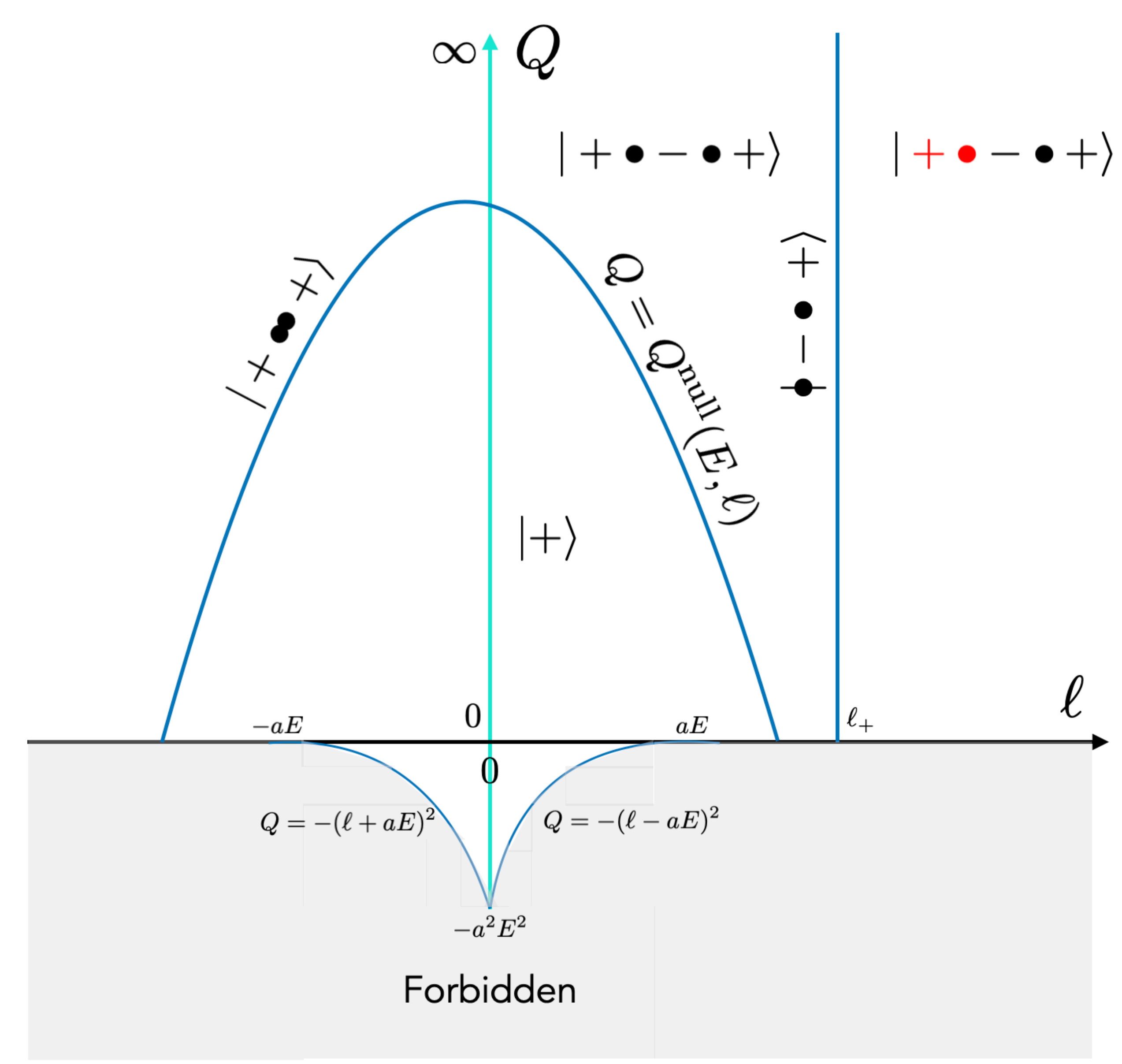}
     \end{subfigure}
     \hfill
\caption{Classification of radial motion of $E > 1$ timelike Kerr orbits (left) and $E>0$ null orbits (right).}\label{fig:Egtr1}
\end{figure}

\begin{figure}[!htbp]\vspace{-1.5cm}
     \centering

     \begin{subfigure}[b]{0.3\textwidth}
         \centering
         \includegraphics[width=\textwidth]{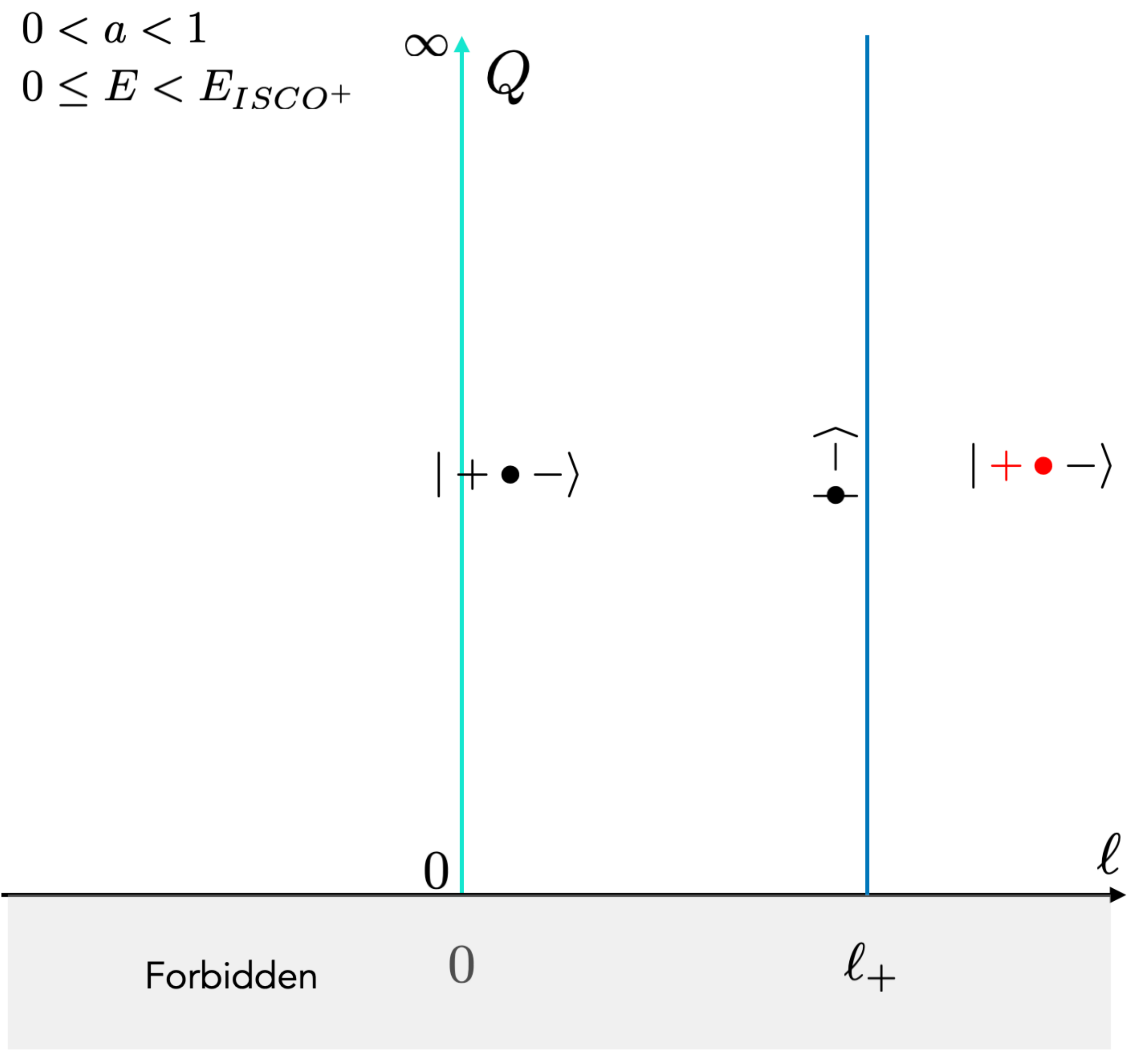}
     \end{subfigure}
     \hfill  \vspace{0.4cm}
     \begin{subfigure}[b]{0.3\textwidth}
         \centering
         \includegraphics[width=\textwidth]{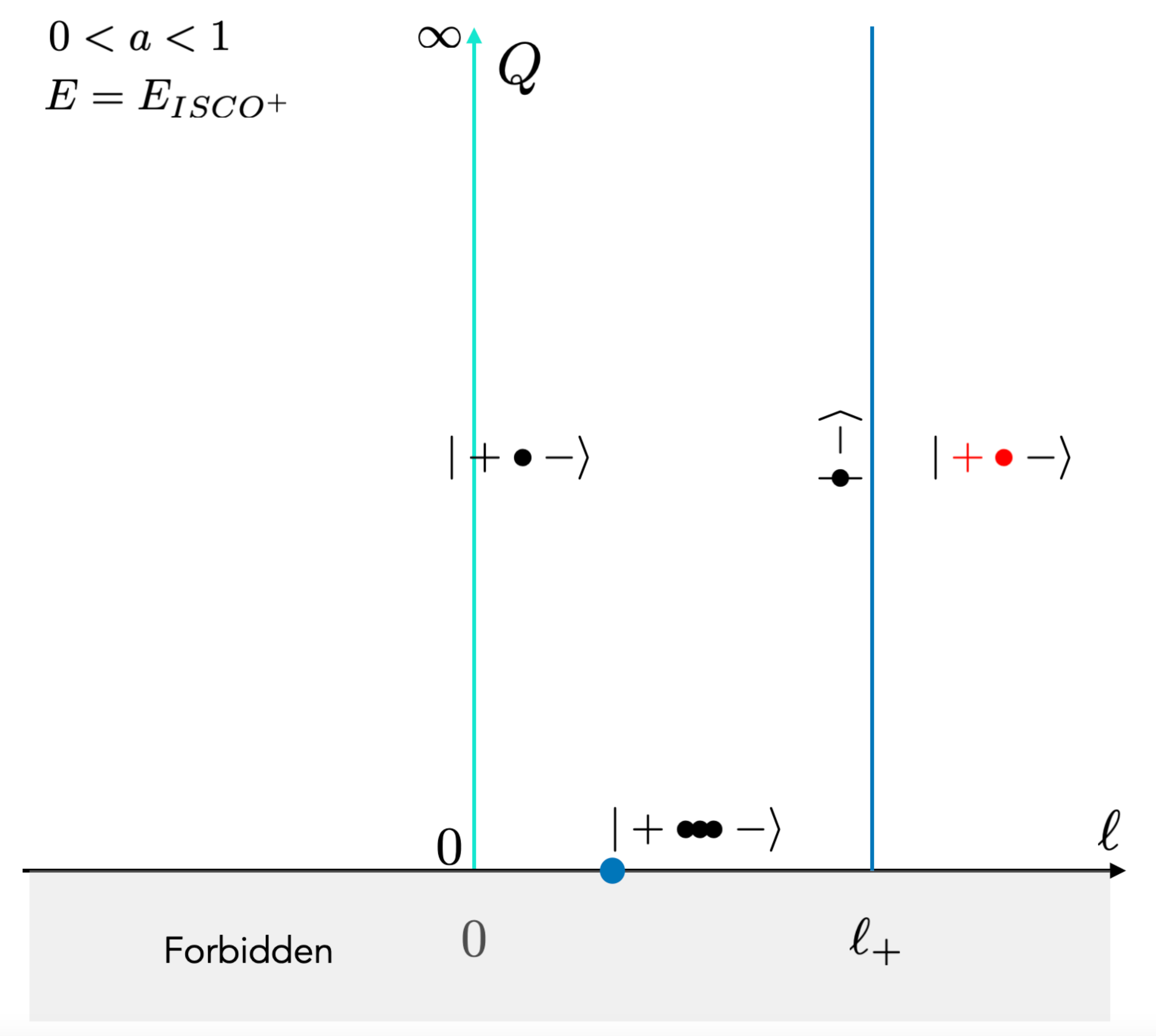}
     \end{subfigure}
     \hfill
     \begin{subfigure}[b]{0.3\textwidth}
         \centering
         \includegraphics[width=\textwidth]{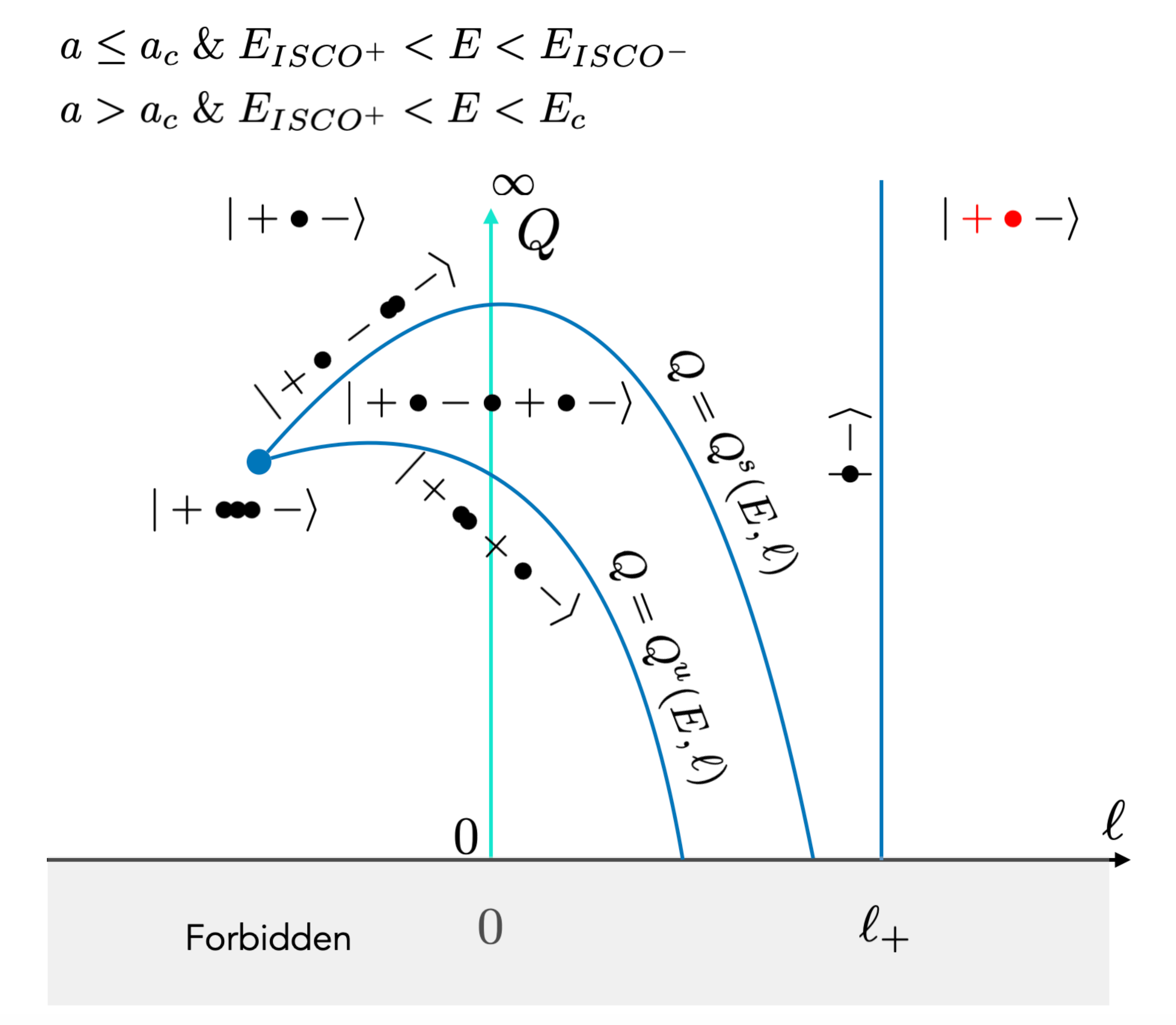}
     \end{subfigure}\break \vspace{0.4cm}
	\begin{subfigure}[b]{0.3\textwidth}
         \centering
         \includegraphics[width=\textwidth]{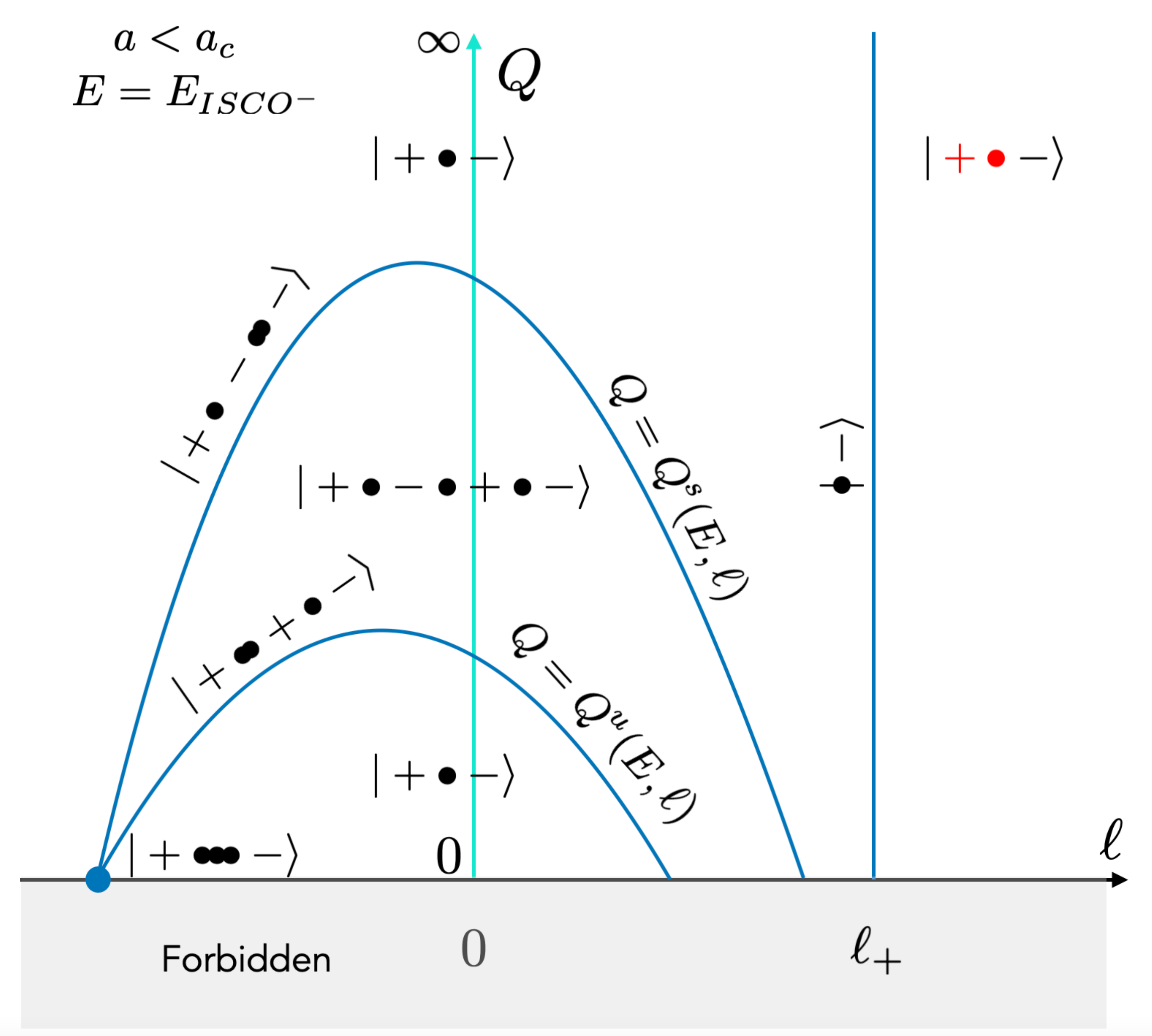}
     \end{subfigure}
     \hfill
     	\begin{subfigure}[b]{0.3\textwidth}
         \centering
         \includegraphics[width=\textwidth]{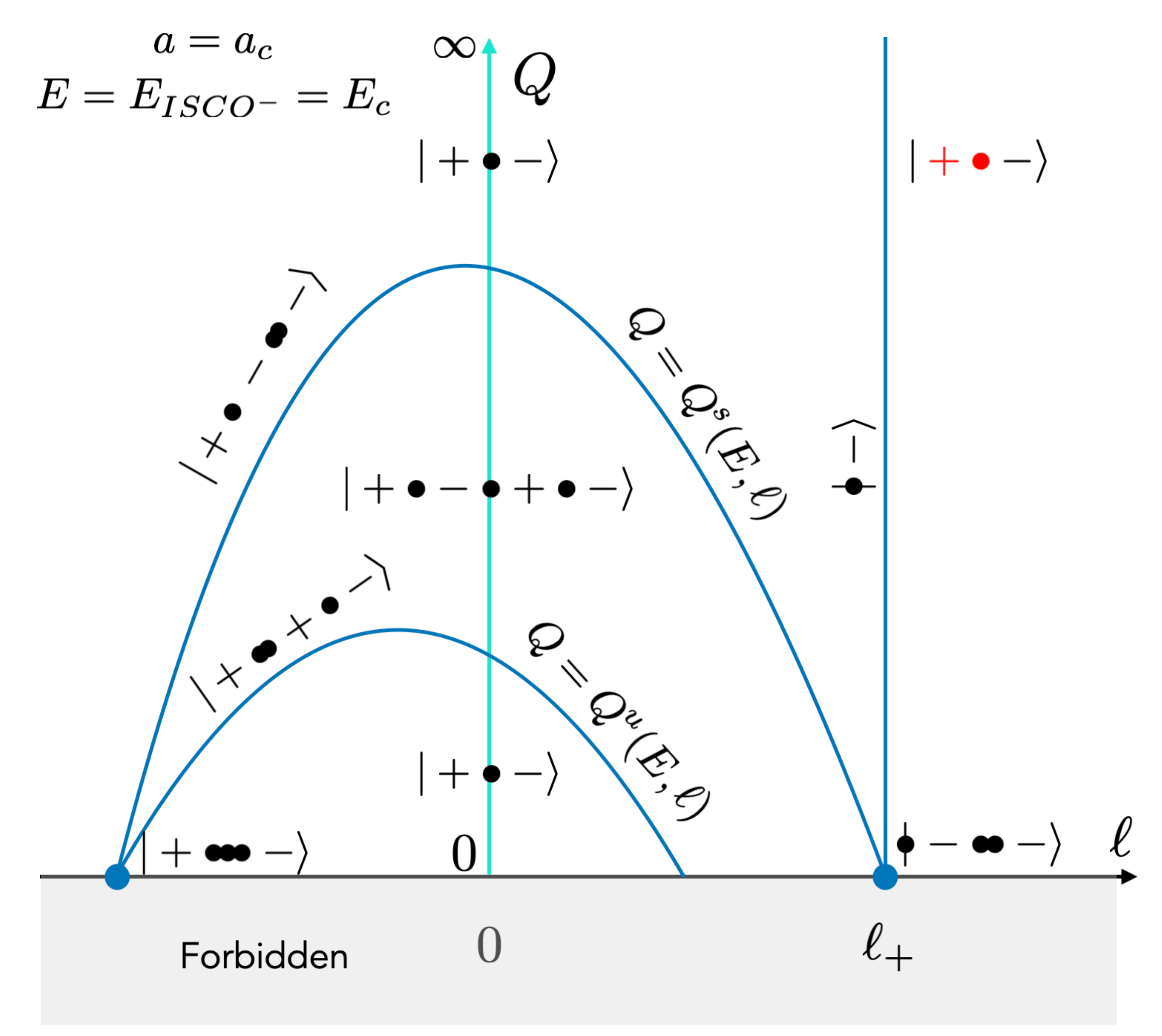}
     \end{subfigure}
     \hfill
     \begin{subfigure}[b]{0.3\textwidth}
         \centering
         \includegraphics[width=\textwidth]{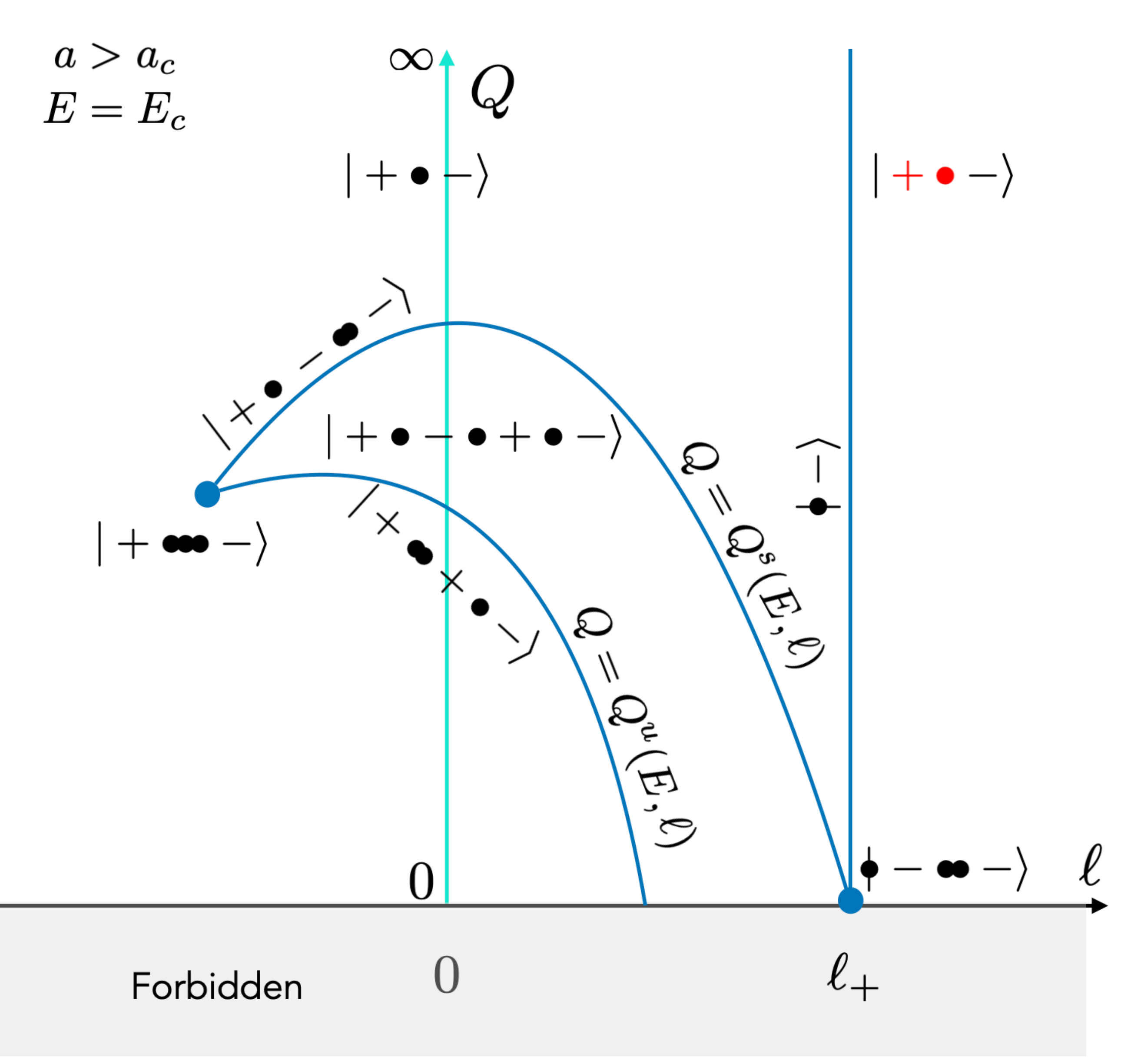}
     \end{subfigure}
     \hfill\break
      \begin{minipage}{2cm}
		\hfill\vspace{1cm}
	\end{minipage}
     \begin{subfigure}[b]{0.3\textwidth}
         \centering
         \includegraphics[width=\textwidth]{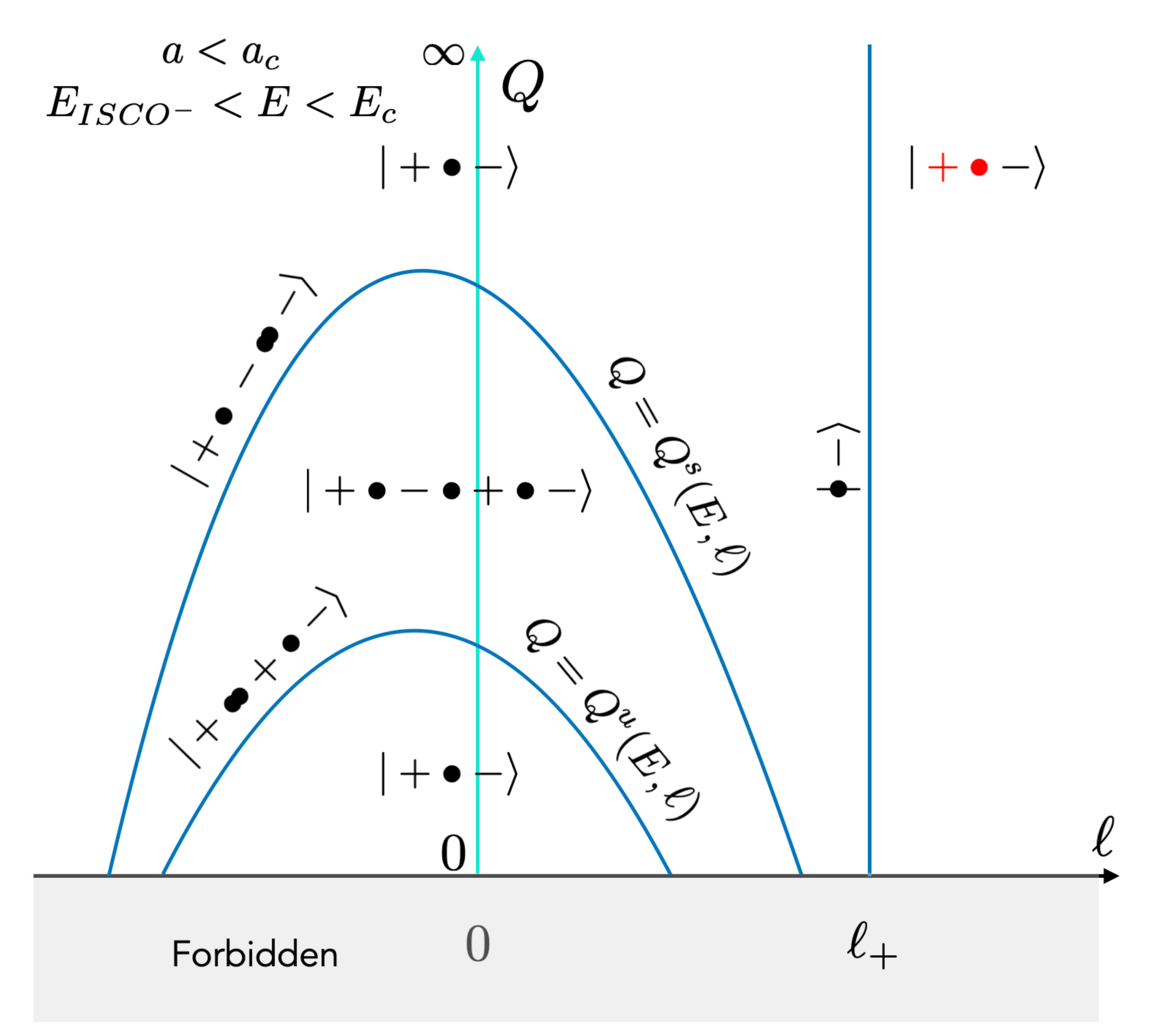}
     \end{subfigure}\hfill  \vspace{0.4cm}
     \begin{subfigure}[b]{0.3\textwidth}
         \centering
         \includegraphics[width=\textwidth]{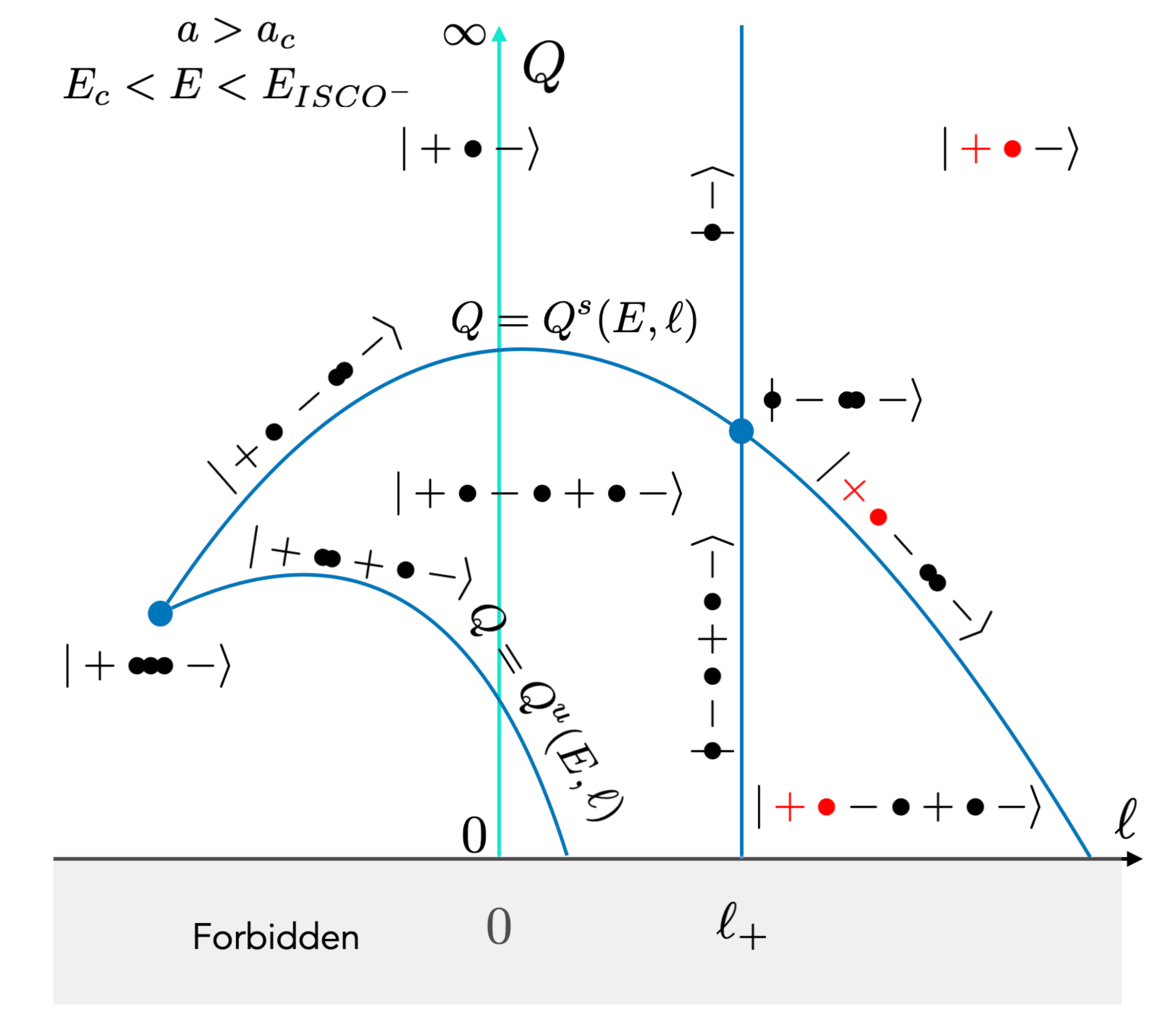}
     \end{subfigure}
       \hfill\break
       \begin{subfigure}[b]{0.3\textwidth}
         \centering
         \includegraphics[width=\textwidth]{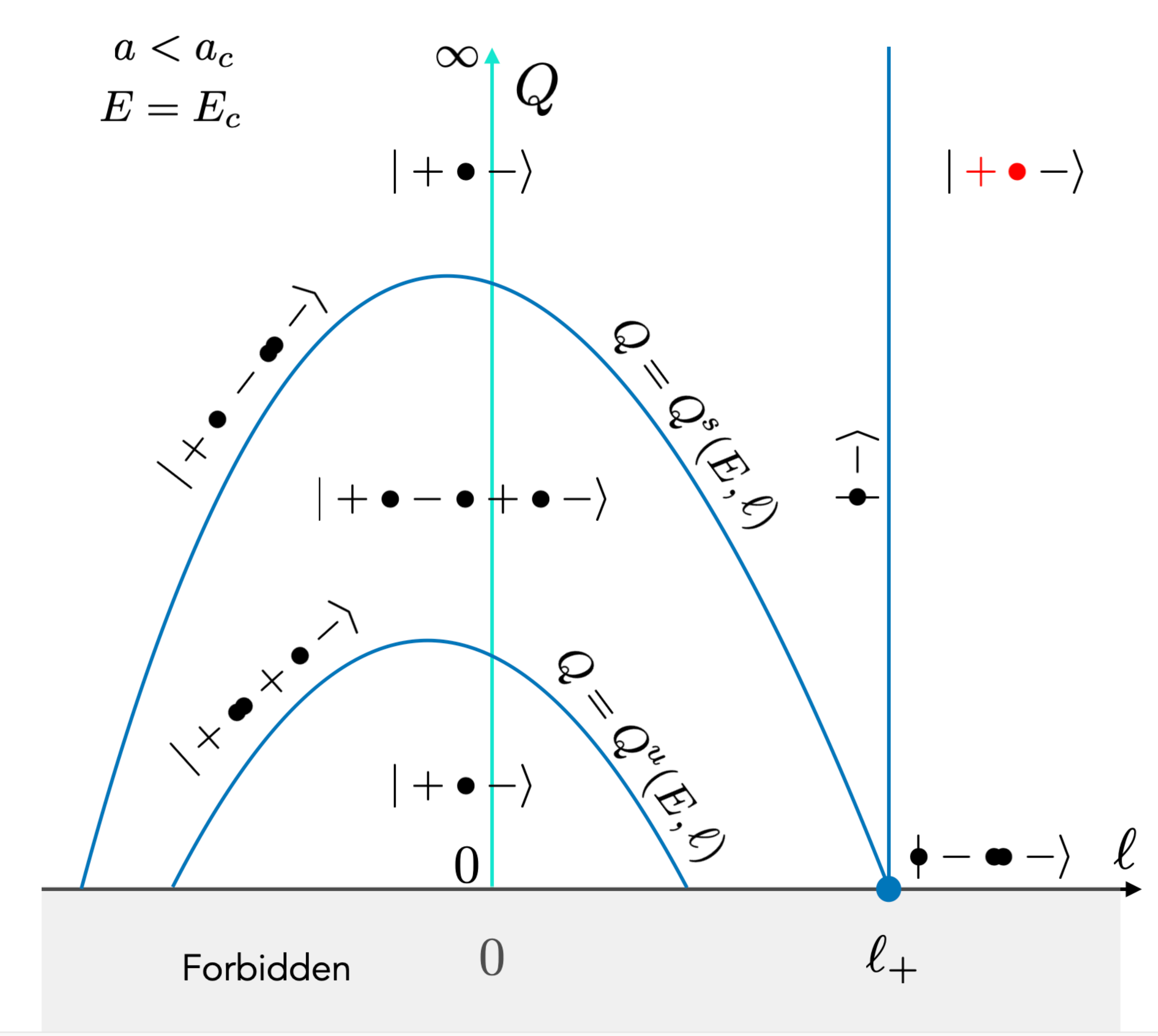}
     \end{subfigure} \hfill 
     \begin{subfigure}[b]{0.3\textwidth}
         \centering
         \includegraphics[width=\textwidth]{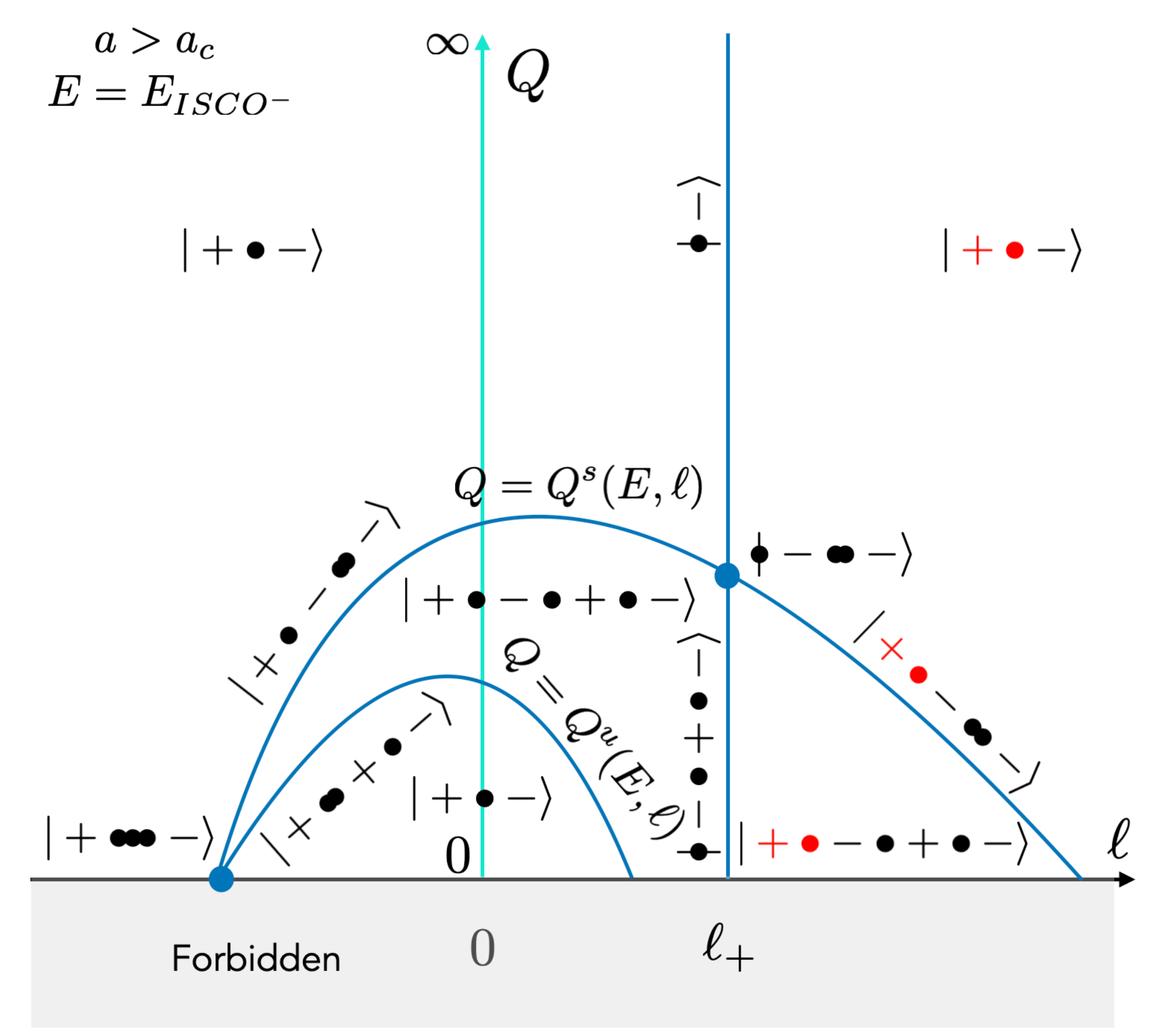}
     \end{subfigure}\hfill 
     \begin{subfigure}[b]{0.3\textwidth}
         \centering
         \includegraphics[width=\textwidth]{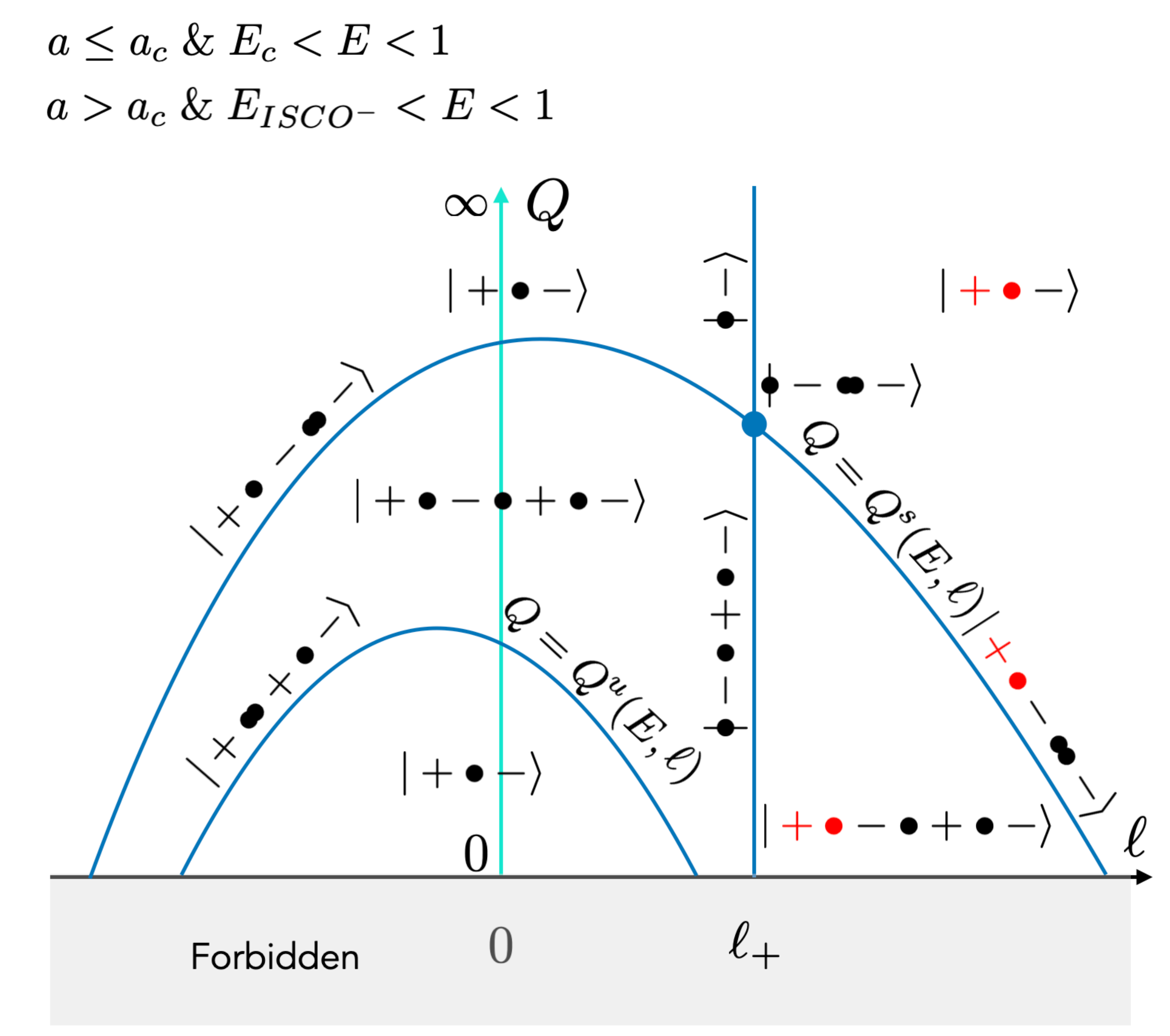}
     \end{subfigure}
        \caption{Classification of radial motion of $0 \leq E<1$ timelike Kerr orbits.}
        \label{fig:classEminus1}
\end{figure}

\begin{figure}[!htbp]
         \centering
         \includegraphics[width=\textwidth]{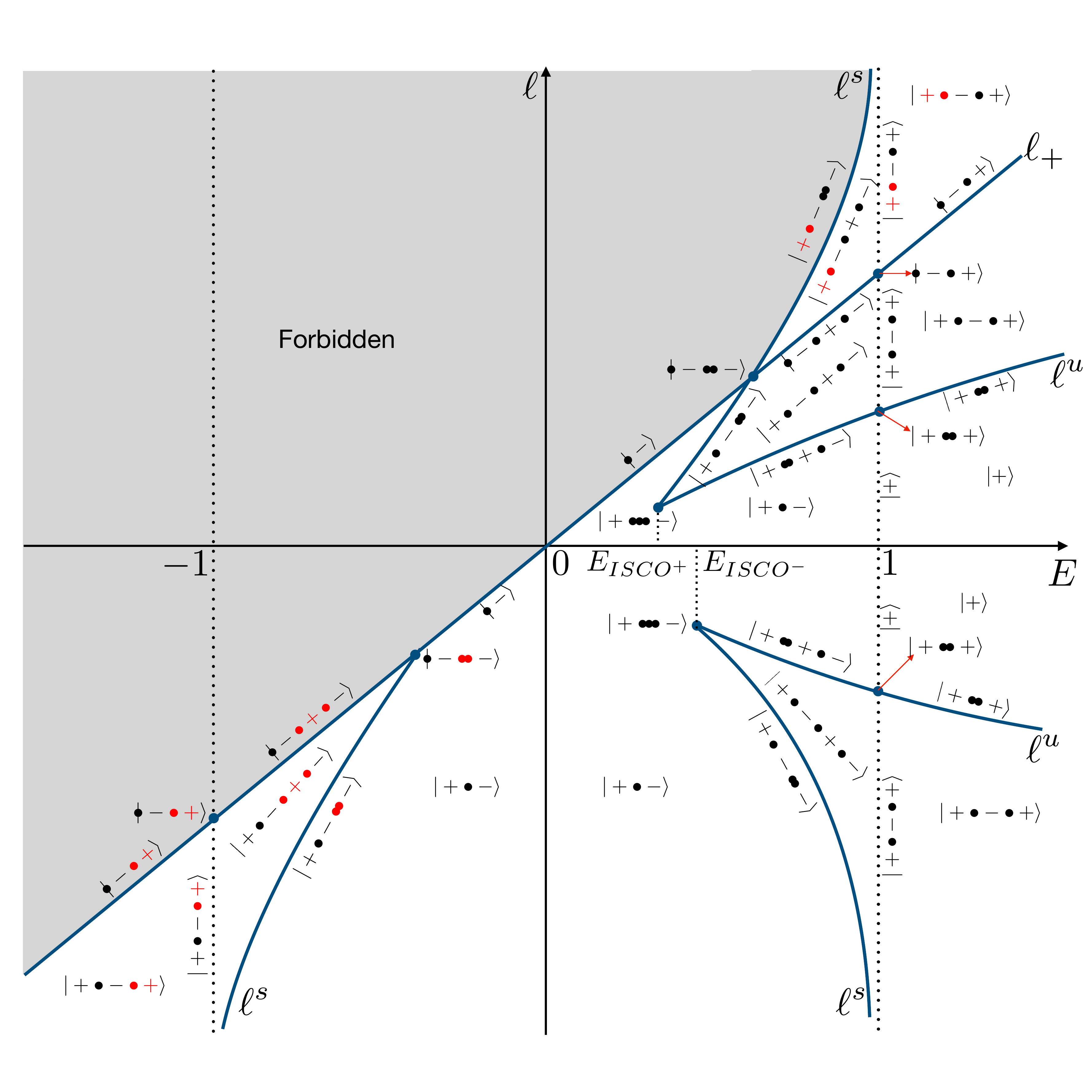}
\caption{Classification of radial motion of equatorial timelike Kerr orbits. Where $\ell^u$ and $\ell^s$ are defined in \eqref{lu} and \eqref{ls}.}\label{fig:ELQ0}
\end{figure}

The final classification of radial motion of $E<1$ timelike Kerr orbits can be found in Figure \ref{fig:classEminus1}. The classification of radial motion of $E> 1$ orbits is shown in Figure \ref{fig:Egtr1}. Moreover, we display the classification of radial motion of equatorial orbits in the $(E,\ell)$ plane in Figure \ref{fig:ELQ0}. As discussed in Section  \ref{sec:boundergo}, in the case $E>0$ and $\ell>\ell_+$,  the trapped orbits are disallowed, and the root structures have the form $\vert\hspace{2pt}{\color{red}+\bullet\cdots\rangle}$. The final list of 11 qualitatively distinct geodesic orbit classes (sometimes evaluated on specific subcases) is given in Table \ref{table:Kerrfinal} following the notations introduced for Schwarzschild in Appendix \ref{sec:schw}. The 11 distinct geodesic orbit classes already appear around the  Schwarzschild background.

\begin{table}[!tbh]    \centering
\begin{tabular}{|c|c|c|c|c|}\hline
\rule{0pt}{13pt} & \textbf{Root structure} & \textbf{Energy range} & \textbf{Radial range} & \textbf{Name} \\ \hline
Generic & $\vert + \bullet \hspace{2pt}- \rangle$ & $0 \leq E < 1$ & $r_+ \leq  r \leq r_1$ & $\mathcal T(E,\ell,Q)$ \\\cline{2-5}
 & $\vert + \bullet - \bullet + \bullet \hspace{2pt}- \rangle$ & $E_{\text{ISCO}^+} < E < 1$ & $r_+ \leq  r \leq r_1$ & $\mathcal T(E,\ell,Q)$ \\\cline{4-5}
  &  &  & $r_2 \leq  r \leq r_3$ & $\mathcal B(E,\ell,Q)$ \\\cline{2-5}
& $\vert +\rangle$ & $E \geq 1$  & $r_+ \leq  r <\infty$ & $\mathcal P(E,\ell,Q)$ \\\cline{2-5}
 & $\vert + \bullet - \bullet\hspace{2pt} + \rangle$ & $E \geq 1$  & $r_+ \leq  r \leq r_1$ & $\mathcal T(E,\ell,Q)$ \\\cline{4-5}
  &  &  & $r_2 \leq  r <\infty$ & $\mathcal D(E,\ell,Q)$ \\\cline{1-5}
 Codimen-  & $\vert\hspace{-7pt}\bullet - \rangle$ &$0 \leq E < 1$  & $\emptyset$ & $\emptyset$ \\\cline{2-5}
 sion 1 & $\vert\hspace{-7pt}\bullet -\bullet +\bullet - \rangle$ &$E_c < E < 1$  & $r_2 \leq r \leq r_3$ & $\mathcal B(E,\ell_+,Q)$ \\\cline{2-5}
 & $\vert + \bullet - \bullet\hspace{-5pt}\bullet - \rangle$  &$E_{\text{ISCO}^+} < E < 1$  & $r_+ \leq r \leq r_1$ & $\mathcal T(E,\ell,Q^s(E,\ell))$ \\\cline{4-5}
  &   &  & $r=r_2$ & $\mathcal S^s(E,\ell)$ \\\cline{2-5}
    &  $\vert +\bullet \hspace{-5pt}\bullet + \bullet -\rangle$ &$E_{\text{ISCO}^+} < E<1$  & $r_+ \leq r < r_1$ & $\mathcal W\mathcal T^u(E,\ell)$ \\\cline{4-5}
        &  & & $r=r_1$ & $\mathcal S^u(E,\ell)$ \\\cline{4-5}
           &  & & $r_1 <r \leq r_2 $ & $\mathcal H^u(E,\ell)$ \\\cline{2-5}
       & $\vert +\bullet\hspace{-5pt}\bullet +\rangle$ &$E\geq 1$ & $r_+ \leq r <r_1 $ & $\mathcal W \mathcal T^u(E,\ell)$ \\\cline{4-5}
    &  & & $r=r_1 $ & $\mathcal S^u(E,\ell)$ \\\cline{4-5}
  &  & & $r_1<r < \infty $ & $\mathcal W \mathcal D^u(E,\ell)$ \\\cline{2-5}
    & $\vert \hspace{-7pt}\bullet - \bullet +\rangle $ & $E \geq 1$& $r_2\leq r < \infty $ & $\mathcal D(E,\ell_+,Q)$ \\\cline{1-5}
  Codimen-  & $\vert \hspace{-7pt}\bullet - \bullet\hspace{-5pt}\bullet \hspace{2pt}-\rangle $ & $E_c \leq E <1$& $r=r_2$ & $\mathcal S^s(E,\ell_+,Q)$\\\cline{2-5}
   sion 2  & $\vert + \bullet\hspace{-5pt}\bullet\hspace{-5pt}\bullet \hspace{2pt}-\rangle$ & $E_{\text{ISCO}^+} \leq E \leq E_{\text{ISCO}^-}$& $r_+ \leq r < r_1$ & $\mathcal W \mathcal T_{ISSO}(E)$\\\cline{4-5}
      & & & $r=r_1$ & $\mathcal S_{ISSO}(E)$\\
     \hline
  \end{tabular}\caption{Taxonomy of the 11 qualitatively distinct classes of radial geodesic motion of timelike Kerr geodesics with $E \geq 0$. The locations of the roots are labeled in increasing order $r_1<r_2<\cdots$. The four generic geodesic classes are the following: trapped orbits $\mathcal T(E,\ell,Q)$ originating from the white hole, turning back at finite radius and plunging into the black hole; bounded orbits $\mathcal B(E,\ell,Q)$ oscillating between the turning points; plunging orbits (or outward directed orbits with initially negative radial velocity) $\mathcal P(E,\ell,Q)$ plunging into the black hole from infinity; deflecting orbits $\mathcal D(E,\ell,Q)$ coming from infinity and bouncing back to infinity. The seven nongeneric geodesic classes (such that at least one endpoint differs from a simple root, the horizon or infinity) are the following: the stable spherical orbits $\mathcal S^s(E,\ell)$, the innermost stable spherical orbits  $\mathcal S_{ISSO}(E)$, and the unstable spherical orbits $\mathcal S^u(E,\ell)$; the whirling trapped orbits $\mathcal W\mathcal T^u(E,\ell)$ either originating from the white hole and asymptotically approaching the unstable spherical orbits or originating asymptotically from the unstable spherical orbits and plunging into the black hole; the homoclinic orbits $\mathcal H^u(E,\ell)$ originating from and approaching the unstable spherical orbits after bouncing on a turning point; the whirling deflecting orbits $\mathcal W \mathcal D^u(E,\ell)$  asymptotically approaching the spherical orbits and infinity; the whirling trapped ISSO orbits $\mathcal W \mathcal T_{ISSO}(E)$ either originating from the white hole and asymptotically approaching the ISSO or originating asymptotically from the ISSO and plunging into the black hole. Note that when $\ell>\ell_+$, the trapped orbits become disallowed, which we denote as $\vert + \bullet \cdots $ $\mapsto$ $\vert {\color{red} + \bullet } \cdots $.  }\label{table:Kerrfinal}
\end{table}

\subsection{Null geodesics}
\label{nullclass}

The classification of radial motion for null geodesics can be simply obtained from the timelike classification in the limit $E \rightarrow \infty$. The corresponding phase space is provided in Figure \ref{fig:Egtr1}. In this section, we briefly review the classification of radial root structures as performed by Gralla and Lupsasca \cite{Gralla:2019ceu} (restricting our analysis to $r>r_+$) and extend it, using the constraints on motion within the ergoregion discussed in Section \ref{sec:boundergo},  to the classification of allowed radial motion. 

In this subsection we assume no mass, $\mu=0$, and positive energy $E>0$. It is convenient to define 
\be
\lambda=\frac{\ell}{E}, \qquad \eta = \frac{Q}{E^2}. 
\ee 
The radial potential is then given by 
\be
R(r)=r^4+(a^2-\eta-\lambda^2)r^2+2(a^2+\eta-2a\lambda+\lambda^2)r-a^2\eta .
\ee 
The positivity bound \eqref{boundQ} reduces to 
\be
\eta \geq \left\{ \begin{array}{ll}  0 &\text{ for } \vert \lambda\vert \geq a ;\\ 
-(\vert \lambda \vert -a)^2  &\text{ for } \vert \lambda\vert < a .\end{array} \right.
\ee 
The only double root that may obey the positivity bound is 
\bea
\eta&=&\eta_*(r_*)=-\frac{r_*^3(-4a^2+r_*(r_*-3)^2)}{a^2(r_*-1)^2},\nn\\
\lambda&=&\lambda_*(r_*)=\frac{r_*^2(r_*-3)+a^2(r_*+1)}{a(1-r_*)}.\nn
\eea 
In the radial range where $\vert\lambda_*(r_*)\vert < a$, we have $\eta_*(r_*) >0$. The bound for double roots therefore reduces to $\eta_*(r_*) \geq 0$, which amounts to 
\be
r_{ph^+}\le r_*\le r_{ph^-},
\ee 
with 
\bea
r_{ph^+}&=&2+2\cos[\frac{2}{3}\arccos a+\frac{4\pi}{3}],\nn\\
r_{ph^-}&=&2+2\cos[\frac{2}{3}\arccos a].\nn
\eea The angular momentum obeys $\lambda_{ph^-}\le\lambda_*\le  \lambda_{ph^+}$
with  $\lambda_{ph^\pm}=\lambda_*(r_{ph^\pm})$. At the horizon, $R=(2 r_+-a \lambda)^2 \geq 0$, and there is a root touching the horizon if and only if \be
\lambda=\lambda_+ \equiv \frac{2r_+}{a}.
\ee 
One has $\lambda_{ph+}<\lambda_+$ for $0<a<1$. Since $\lambda_*(r_*)$ is monotonic between $r_{ph+}\le r_*\le  r_{ph-}$, one can define an inverse function $r_*(\lambda)$, and the double root is described by 
\be
Q=Q^{\text{null}}(E,\ell)\equiv  E^2 \eta^{\text{null}}(\frac{\ell}{E}),\qquad \eta^{\text{null}}(\frac{\ell}{E}) \equiv  \eta_*(r_*(\lambda)).
\ee 
The function $\eta^{\text{null}}(\lambda)$ reaches a maximum at $\lambda = -2a $ with $\eta^{\text{null}}(-2a) = 27$ independently of $a$. The corresponding radius is $r_*=3$, which coincides with the photon sphere of the Schwarzschild solution.

Now, the bound \eqref{boundT} rules out trapped orbits within the ergoregion. This implies that the root structure $\vert +\bullet -\bullet \hspace{3pt}+\rangle$ for $\ell >\ell_+(E)=E/\Omega_+$ contains disallowed trapped orbits, ${\vert \; {\color{red} +\; \bullet }-\bullet\hspace{3pt} +\rangle}$. This condition was not analyzed in \cite{Gralla:2019ceu}. This completes the result on the classification of radial root structures by the classification of allowed radial motion. The final root structure is depicted in Figure \ref{fig:Egtr1}.

\clearpage

\section{Classification of radial geodesic motion within the ergoregion}
\label{sec:ergo}

In the previous section, we did not consider the exact location of the ergosphere in the classification of radial root systems. In the following, we will introduce the location of the ergosphere, denoted with the symbol $)$,  discuss where it appears within each root system, and present the classification of root systems and radial geodesic motion within the ergoregion, i.e., between the horizon and the ergosphere. We will start with a complete classification of such root systems and corresponding allowed radial geodesic motion on the equator, which will lead to the identification of six distinguished values of the Kerr angular momentum $a$. We will then briefly discuss the nonequatorial radial motion and finally derive the classification of radial motion within the near-horizon region of near-extremal Kerr. 

\subsection{Root structures and allowed radial motion on the equator}
\label{sec:allowedeq}

We consider the root structures with roots between the horizon and the ergosphere. There can be maximally three such roots for $E^2 \neq 1$ and maximally two such roots for $E^2=1$. The root structures have the generic form $\vert + \cdots )\cdots \rangle$ when $\ell\neq \ell_+(E)=E/\Omega_+$ or the particular form ${\vert \hspace{-7pt}\bullet - \cdots )\cdots \rangle}$ when $\ell=\ell_+(E)$. The root structure outside the ergosphere can be ignored for the classification within the ergoregion, but it is useful as a comparison with our earlier classification. 

Since 
\begin{equation}
    r_{\text{ergo}}(\theta ; a) \leq 2 < 1+\sqrt{2} \leq r_*^{(0)}(a)
\end{equation}
where $r_*^{(0)}$ is defined below \eqref{El0}, all root structures with a double root have necessarily negative angular momentum, $\ell < 0$ for $E<0$ and positive $\ell>0$ for $E>0$. By continuity, root structures with $\text{sign}(\ell)\neq \text{sign}(E)$ and with two simple roots within the ergosphere are also discarded. 

From now on, we will only discuss the equatorial case $\theta=\frac{\pi}{2}$ where, accordingly, $Q=0$ and $r_{\text{ergo}}=2$.  The potential has a root $r=0$ which can be factored out. The relevant potential becomes
\be 
R_0(r)=(E^2-1)r^3+2r^2+(a^2(E^2-1)-\ell^2)r+2(\ell-a E)^2.\label{Reqt}
\ee 
We consider only nonvanishing energy orbits $E \neq 0$ in Kerr with $0 < a \leq 1$. 

\paragraph{Inequalities} The inequalities  \eqref{eqergo1}--\eqref{eqergo2} reduce for $E<0$ to 
\begin{equation}
    (2-r)\ell \leq 2 a E < 0.\label{ineq}
\end{equation}
Negative energy orbits that reach the ergosphere $r=2$ are discarded. Moreover, all timelike orbits within the equatorial ergoregion have a negative angular momentum. The inequality \eqref{ineq} can be solved by
\be 
\ell\le \ell_{\phi}(E,r)\equiv \frac{2a E}{2-r}\label{ellEnew}
\ee  
for negative energy orbits in the ergoregion. Since $r_+\le r<2$, it implies, in particular, for all orbits with $E<0$ that 
\be 
\ell\le \frac{2a E}{2-r}\le \frac{2aE}{2-r_+}=\frac{2r_+}{a}E=\frac{E}{\Omega_+}. 
\ee 
The thermodynamic bound \eqref{boundT} is therefore obeyed for all orbits with $E<0$. The upper bound $\ell=\ell_+(E)$ corresponds to root structures with one root at the horizon.

The inequalities  \eqref{eqergo1}--\eqref{eqergo2} reduce for $E > 0$ to 
\begin{equation}
    \ell < \ell_t(E,r)\equiv \frac{r^3+r a^2+2a^2}{2a}E.\label{ellE2}
\end{equation}

For timelike geodesics, we have $u_\mu u^\mu <0$, which also implies the independent inequalities
\be 
\frac{-g_{t\phi}-\sin\theta \sqrt{\Delta}}{g_{\phi\phi}}<\frac{d\phi}{dt}<\frac{-g_{t\phi}+\sin\theta \sqrt{\Delta}}{g_{\phi\phi}}\label{upperlower}
\ee in the ergoregion. Here, $\frac{d\phi}{dt}=- (\frac{\ell}{E}g_{tt}+g_{t\phi})/(\frac{\ell}{E}g_{t\phi}+g_{\phi\phi})$. The bound \eqref{upperlower} becomes 
\be 
0<\frac{2a-r\sqrt{\Delta}}{r^3+a^2(r+2)}<\frac{d\phi}{dt}<\frac{2a+r\sqrt{\Delta}}{r^3+a^2(r+2)}.\label{upper2}
\ee  
where $\frac{d\phi}{dt}=- (\frac{\ell}{E}g_{tt}+g_{t\phi})/(\frac{\ell}{E}g_{t\phi}+g_{\phi\phi})$. We define the quantities
\bea 
\ell^{(1,2)}(E,r)=\frac{2a\mp r\sqrt{\Delta}}{2-r}E.
\eea
For $E<0$, they obey $\ell^{(2)}(E,r)<\ell_\phi(E,r)<\ell^{(1)}(E,r)$, while for $E>0$ they obey $\ell^{(1)}(E,r)<\ell_t(E,r)<\ell^{(2)}(E,r)$. We deduce from \eqref{upper2} that for $E<0$ the angular momentum should satisfy 
\begin{equation}
\ell < \ell^{(2)}(E,r)<0 \label{finalbound1}
\end{equation}
and Eq. \eqref{ellEnew} is automatically obeyed, while for $E>0$ the angular momentum should satisfy 
\begin{equation}
\ell < \ell^{(1)}(E,r)  . \label{finalbound2}
\end{equation}
and Eq. \eqref{ellE2} is automatically obeyed. 

Finally, we impose that the potential $R_0(r)$ is non-negative. This condition can be solved for $\ell$ as
\be 
\ell\ge \ell_0^{(1)}(E,r)\quad\text{or}\quad \ell\le \ell_0^{(2)}(E,r)\label{split}
\ee  within the ergoregion, where we have defined 
\be 
\ell_0^{(1,2)}(E,r)=\frac{2aE\pm \sqrt{r(2+(E^2-1)r)\Delta(r)}}{2-r}. \label{l02}
\ee 
From the definition \eqref{l02}, we can directly derive that $\ell_0^{(1)}(E,r)\ge \ell_0^{(2)}(E,r)$. Moreover, for $E>0$, $\ell_0^{(1)}(E,r)+\ell_0^{(2)}(E,r)\ge  2\ell_+(E)$, which implies for $E>0$
\be 
\ell_0^{(1)}(E,r) \ge \ell_+(E),\label{l01lpe}
\ee
while for $E < 0$, $\ell_0^{(1)}(E,r)+\ell_0^{(2)}(E,r)\le  2\ell_+(E)$ which 
\be 
\ell_0^{(2)}(E,r) \le \ell_+(E).\label{l01lpe2}
\ee

When $E<0$, we find 
\be 
\ell_0^{(2)}(E,r)<\ell^{(2)}(E,r)<\ell_0^{(1)}(E,r).
\ee When $E>0$, we find 
\be 
 \ell_0^{(2)}(E,r)<\ell^{(1)}(E,r)<\ell_0^{(1)}(E,r).
\ee 
Combining with Eqs. \eqref{finalbound1} and \eqref{finalbound2}, the bound becomes 
\be 
\ell\le \ell_0^{(2)}(E,r)\label{finalbound}
\ee for both positive and negative energy equatorial orbits within the ergoregion. This is the final inequality that supersedes all previous inequalities. In particular, for plunging orbits, the bound has to be obeyed at $r=r_+$. Since $\ell^{(2)}_0(E,r_+)= \frac{2r_+}{a}E=\ell_+(E)$, we obtain for all trapped orbits
\be 
\ell\le \ell_+(E). \label{finalboundplunging}
\ee 
From \eqref{finalbound} and \eqref{l01lpe2}, this bound is moreover obeyed for any $E<0$ orbit. 

Both prograde and retrograde orbits are therefore allowed for $E > 0$. The condition \eqref{finalbound} and \eqref{finalboundplunging} impose constraints which will be discussed below. In the limit $E=0$, the bound  \eqref{finalbound} reduces to 
\begin{equation}
    \ell \leq - \sqrt{\frac{r \Delta}{2-r}}<0, \label{eq4}
\end{equation}
and orbits are retrograde as they should. 

From \eqref{finalboundplunging}, root structures of the form $\vert + \bullet\hspace{2pt} - )$ for $\ell>\ell_+(E)$ will be denoted as $\vert {\color{red} \hspace{2pt}+\hspace{2pt} \bullet }\hspace{2pt}- )$. 

A corollary from the inequalities \eqref{l01lpe} and \eqref{split} is that for all root structures with $E>0$ and $\ell \leq \ell_+(E)$, the region $R_0(r)\ge 0$ necessarily obeys the bound \eqref{finalbound}. Therefore, all motion denoted as $+$ in root structures with $\ell \leq \ell_+(E)$ and $E>0$ are allowed. Root structures such as $\vert +\bullet -\bullet \hspace{2pt}+ )$ for $\ell>\ell_+(E)$ and $E>0$ require more care. From \eqref{finalboundplunging}, one deduces that the plunging orbits are disallowed. However, the orbits entering and escaping the ergoregion are not constrained by this inequality. We will check that such orbits obey the bound \eqref{finalbound}. We will therefore denote orbits with $E>0$ and $\ell>\ell_+(E)$ as  $\vert {\color{red} \hspace{2pt}+\hspace{2pt}\bullet }-\bullet \hspace{2pt}+ )$.

\paragraph{Special values of $a$. Roots at the horizon or at the ergosphere.}

In the following, we will define six particular values of $a$, which we will order in increasing values as 
\begin{equation}
0<a_c^{(1)} < a_c^{(2)}<a_c^{(3)}<a_c^{(4)}<a_c^{(5)}<a_c^{(6)}<1\label{speciala}
\end{equation}
where distinctive root structures will emerge.

We define the angular momentum $\ell=\ell_e(E;a)$ such that 
\be 
R_0(r_{\text{ergo}})=0.
\ee 
At the equator, the solution is unique and given by
\be 
\ell = \ell_e(E;a) \equiv \frac{(4+2a^2)E^2-a^2}{2a E}.\label{elle}
\ee 
For such angular momentum, the local root structure is given by ${\cdots ) \hspace{-7pt}\bullet \cdots }$. For $E<0$ only, the constraint $\ell_e(E;a)<0$ is obeyed only for 
\begin{equation}
    E<-\frac{a}{\sqrt{4+2a^2}}. \label{consE}
\end{equation}
A double root at the ergosphere can occur only for $a > a_c^{(1)}$, where 
\begin{equation}
a_c^{(1)}\equiv  \frac{1}{\sqrt{2}} \approx 0.707107.     
\end{equation}
The double root at the ergosphere corresponding to ${\cdots ) \hspace{-9pt}\bullet \hspace{-4pt}\bullet\cdots }$ then occurs when
\begin{eqnarray}
E &=& E^\pm_e(a) \equiv  \pm \frac{a}{2\sqrt{\sqrt{2}a-1}}, \label{Ee}\\
\ell &=& \ell_e(E^\pm_e(a);a) =  \pm \frac{2-\sqrt{2}a+\frac{a^2}{2}}{\sqrt{\sqrt{2}a-1}}. 
\end{eqnarray}
We have $\text{sign}(\ell_e(E^\pm_e(a);a))=\text{sign}(E)$ for $a> a_c^{(1)}$, in accordance with our discussion that only orbits with $\text{sign}(\ell)=\text{sign}(E)$ occur in the presence of double roots within the ergoregion. With respect to our discussion in Section \ref{sec:nonmarginal}, we have $E^\pm_e(a)=\pm E^{(2)}(2)$, where $E^{(2)}(r_*)$ is defined in \eqref{E2}.  The function $E_e(a)$ crosses $E=\pm 1$ at $a=a_c^{(2)}$, where
\begin{equation}
    a_c^{(2)}\equiv 2(\sqrt{2}-1) \approx 0.828427. 
\end{equation}
The function $\ell_e(E;a)$ crosses $\ell_+(E;a)$ defined in \eqref{lpE} only for $a > a_c^{(3)}$ at
\begin{equation}
E=E^\pm_{+e}(a)\equiv \pm \frac{a}{\sqrt{2}\sqrt{a^2-2\sqrt{1-a^2}}}    \label{Esp}
\end{equation}
where 
\begin{equation}
a_c^{(3)}  \equiv \sqrt{2(\sqrt{2}-1)}\approx 0.91018. 
\end{equation}
The root structures ${\vert \hspace{-7pt}\bullet - ) \hspace{-4pt}\bullet}$ with one root at the horizon and one root at the ergosphere therefore occur for $E$ given by \eqref{Esp} and $\ell$ given by 
\begin{equation}
\ell = \ell^\pm_{+e}(a) \equiv   \pm \frac{\sqrt{2}(1+\sqrt{1-a^2})}{\sqrt{a^2-2\sqrt{1-a^2}}}. 
\end{equation}
The energy \eqref{Esp} reaches $E^2=1$ at $a=a_c^{(5)}$, where
\begin{equation}
a_c^{(5)} \equiv 2\sqrt{\sqrt{5}-2} \approx 0.971737.  
\end{equation}
The special root structures ${\vert \hspace{-7pt}\bullet- \bullet \hspace{-6pt}\bullet \hspace{2pt}-\rangle }$ occur at the equator for $\ell=\ell_+(E;a)$ and $E=\pm E_c(a)$ where $E_c(a)$ is given in \eqref{Ec}. We have $\pm E_c(a)=E^\pm_e(a)$ at $a=a_c^{(6)}$, where
\begin{equation}
a_c^{(6)}= \frac{2}{\sqrt{3}}\sqrt{Z-24Z^{-1}}\approx 0.996175,\qquad Z \equiv 3^{1/3}(9+7\sqrt{33})^{1/3}.     
\end{equation}
For that special value of $a$, $\ell=\ell_+(E;a)$ and $E=E^\pm_e(a)$, the double root is at the ergosphere, leading to the root structure ${\vert \hspace{-7pt}\bullet- \bullet\hspace{-4pt} )\hspace{-5pt}\bullet \hspace{0pt}-\rangle }$.

Finally, a triple root at the ergosphere occurs for the particular value $a=a_c^{(4)}$, where
\begin{equation}
a_c^{(4)} \equiv \frac{2\sqrt{2}}{3} \approx 0.942809.     
\end{equation}
It is also the unique solution to the equation $r_{ISCO^+}(a)=2$. The unique triple root structure at the ergosphere ${\vert + \bullet \hspace{-2pt}) \hspace{-7pt}\bullet \hspace{-4pt}\bullet}$ therefore occurs at the two special values [$\text{sign}(E)=\text{sign}(\ell)$],
\begin{equation}
    E=\pm \sqrt{\frac{2}{3}}, \qquad \ell = \pm \frac{10}{3\sqrt{3}}, \qquad a = a_c^{(4)}. 
\end{equation}
The summary of the six distinguished values $a_c^{(i)}$ of the Kerr angular momentum is given in Table \ref{valuesa}. 


\begin{table}[!tbh]    \centering
\begin{tabular}{|c|c|c|c|c|c|}\hline
$a_c^{(1)}$ & $a_c^{(2)}$ & $a_c^{(3)}$ & $a_c^{(4)}$ & $a_c^{(5)}$ & $a_c^{(6)}$\\\hline
$\frac{1}{\sqrt{2}}$& $2(\sqrt{2}-1)$ & $\sqrt{2(\sqrt{2}-1)}$ & $\frac{2\sqrt{2}}{3}$ & $2\sqrt{\sqrt{5}-2}$ & $\frac{2}{\sqrt{3}}\sqrt{Z-24Z^{-1}}$\\\hline
0.707 & 0.828 & 0.910 & 0.943 & 0.972 & 0.996\\\hline
\end{tabular}\caption{Exact and approximate values of $a_c^{(i)}$.}\label{valuesa}
\end{table}

\paragraph{Double roots.}

So far we only defined the root structures with a double root located on the ergosphere. More generally, the double roots occur for equatorial orbits for $E=\pm E^{(2)}(r_*)$ as given by \eqref{E2} and  $\ell=\pm \ell_{\text{b}}(E^{(2)}(r_*),r_*)$ as given by \eqref{lrs} in terms of the radius of the double root $r_*$. We can rewrite these equations as 
\be 
E= E^\pm (r_*;a) \equiv \pm \frac{(r_*-2)\sqrt{r_*}+a}{\sqrt{r_*^{3/2}((r_*-3)r_*^{1/2}+2a)}},\quad \ell = \ell^\pm (r_*;a) \equiv \pm \frac{r_*^2-2a\sqrt{r_*}+a^2}{\sqrt{r_*^{3/2}((r_*-3)r_*^{1/2}+2a)}}.\label{para}
\ee 
Such double roots may lie in the ergoregion for $a \geq a_c^{(1)}$. We have 
\begin{equation}
\frac{2a E^-(r_*;a)}{2-r_*}-\ell^-(r_*;a)=-\frac{r_*^{1/4}\Delta(r_*)}{(2-r_*)\sqrt{r_*^{3/2}-3\sqrt{r_*}+2a}}<0.\label{bounddouble}
\end{equation}
Therefore, negative energy orbits with double roots are disallowed by the condition \eqref{ellE}.  This implies, in particular, that no circular orbit with negative energy is allowed in the ergoregion. By continuity, no bounded motion is allowed and only trapped orbits are allowed. The bound \eqref{finalboundplunging} therefore applies for any orbit with $E<0$. 

On the other hand, we have identically
\bea 
\ell^{(2)}_0(E^+(r_*;a),r_*)= \ell^+(r_*;a),\label{posbounddouble}
\eea 
where $\ell_0^{(2)}$ was defined in \eqref{l02}. 
Therefore, positive energy orbits with double roots are allowed by the condition 
\eqref{finalbound} (for $a \geq a_c^{(1)}$). In particular, circular orbits with positive energy are allowed in the ergoregion. However, since $r_*^{(2)}\ge 3$, circular orbits with $\ell<0$ and $E>0$ do not appear in the ergoregion.

The triple root occurs at the ISCO radius $r_*=r_{ISCO^\pm}$, which has $\text{sign}(E)=\text{sign}(\ell)=\pm 1$. In the range $r_*^{(1)}<r_*<r_{ISCO^\pm }$, one can invert $E^\pm(r_*)$ to $r_*^u(E)$ and define 
\begin{equation}
    \ell^u(E;a) \equiv \ell^\pm(r^u_*(E)).\label{lu}
\end{equation} 
As discussed in Section \ref{sec:nonmarginal}, the corresponding double root is unstable: the root structure takes the local form ${\cdots  + \bullet \hspace{-6pt}\bullet + \dots }$. Since we only consider the ergoregion in this section, we will only define $\ell^u$ when $r_*(E) \leq 2$. In the range $r_{ISCO^\pm}<r_*<\infty$, one can invert $E^\pm(r_*)$ to $r_*^s(E)$ and define 
\begin{equation}
\ell^s(E;a) \equiv \ell^\pm(r^s_*(E)). \label{ls}
\end{equation}
The corresponding double root is stable: the root structure takes the local form ${\cdots - \bullet \hspace{-6pt}\bullet - \cdots  }$. Since we only consider the ergoregion in this section, we will only define $\ell^s$ when $r^s_*(E) \leq 2$.

Let us discuss the root systems which admit both a root at the ergosphere and double roots. This occurs at the intersection of the lines $\ell = \ell_e(E;a)$ and $\ell = $ either $\ell^u(E;a)$ or $\ell^s(E;a)$. Algebraically, it amounts to find the roots $r_*$ of $\ell^+(r_*;a)=\ell_e(E^+(r_*);a)$. After analysis, there are three main cases depending upon the value of $a$. 
\begin{itemize}
    \item For $a \leq a_c^{(1)}$, there is no solution within the ergoregion. Instead, there is one real solution  $r_*=\frac{2r_+}{a^2}-1>2$ intersecting $\ell=\ell^s$, which corresponds to the root structure $\vert + \bullet \hspace{-6pt})-\bullet\hspace{-4pt}\bullet-\rangle$. 
    \item For $a_c^{(1)} < a \leq a_c^{(4)}$, there is the solution at the ergosphere $r_*=2$ intersecting $\ell=\ell^u$, corresponding to the root structure ${\vert + \bullet \hspace{-4pt})\hspace{-5pt} \bullet + \rangle }$. This structure degenerates to the triple root at the ergosphere ${\vert + \bullet \hspace{-2pt}) \hspace{-7pt}\bullet \hspace{-4pt}\bullet - \rangle}$ at $a=a_c^{(4)}$. The other real solution $r_*=\frac{2r_+}{a^2}-1$ is outside the ergosphere for $a <a_c^{(4)}$. In that case the curve $\ell=\ell_e$ intersects $\ell=\ell^s$, and the root structure is $\vert + \bullet \hspace{-6pt})-\bullet\hspace{-4pt}\bullet-\rangle$. 
    \item For $a_c^{(4)} < a \leq 1$, there is the new solution $r_*=\frac{2r_+}{a^2}-1$ intersecting $\ell=\ell^u$, but which now corresponds to the root structure $\vert +\bullet \hspace{-6pt}\bullet +\hspace{2pt})\hspace{-4pt}\bullet$, and there is still the solution $r_*=2$ intersecting now $\ell=\ell^s$, which corresponds to the root structure $\vert + \bullet - \bullet \hspace{-4pt})\hspace{-2pt} \bullet$ for $a \neq a_c^{(6)}$. The latter solution degenerates to ${\vert \hspace{-7pt}\bullet- \bullet\hspace{-4pt} )\hspace{-2pt}\bullet }$ for $a=a_c^{(6)}$. 
\end{itemize}

\paragraph{Construction of the phase diagrams.}

We defined four relevant curves to classify the root structures: 
\begin{itemize}
    \item $\ell = \ell_+(E)$. The root structure takes the local form ${\vert \hspace{-7pt}\bullet - \cdots}$. 
    \item $\ell =\ell_e(E)$. The root structure takes the local form ${\cdots ) \hspace{-4pt}\bullet }$.
     \item $\ell = \ell^u(E)$. The root structure takes the local form  ${\cdots  + \bullet \hspace{-5pt}\bullet + \dots }$.
    \item $\ell = \ell^s(E)$. The root structure takes the local form ${\cdots - \bullet \hspace{-5pt}\bullet - \cdots  }$. 
\end{itemize}
The pattern of intersection of these lines depends upon the value of the spin $a$ relatively to the special values \eqref{speciala}. Moreover, one has to impose the constraints \eqref{finalbound}--\eqref{finalboundplunging}, which qualitatively differ for $E>0$ and $E<0$ orbits. We, therefore, discuss the phase spaces for $E>0$ and $E<0$ separately.

\paragraph{Phase diagram for $E>0$.}

\begin{figure}[!htbp]\vspace{-1.5cm}
     \centering 
     \begin{subfigure}[b]{0.45\textwidth}
         \centering
         \includegraphics[width=\textwidth]{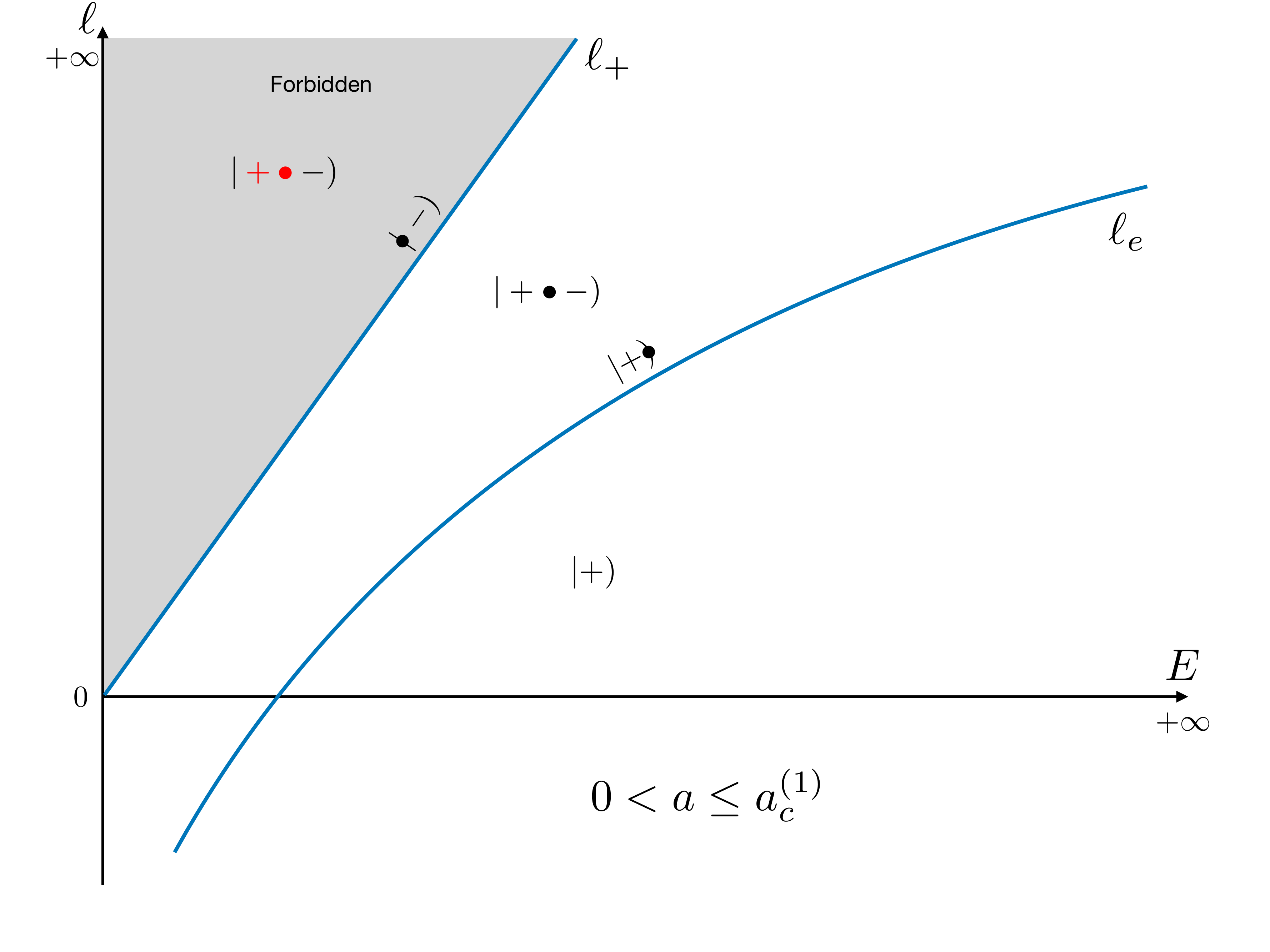}
     \end{subfigure}
     \hfill  \vspace{0.4cm}
     \begin{subfigure}[b]{0.45\textwidth}
         \centering
         \includegraphics[width=\textwidth]{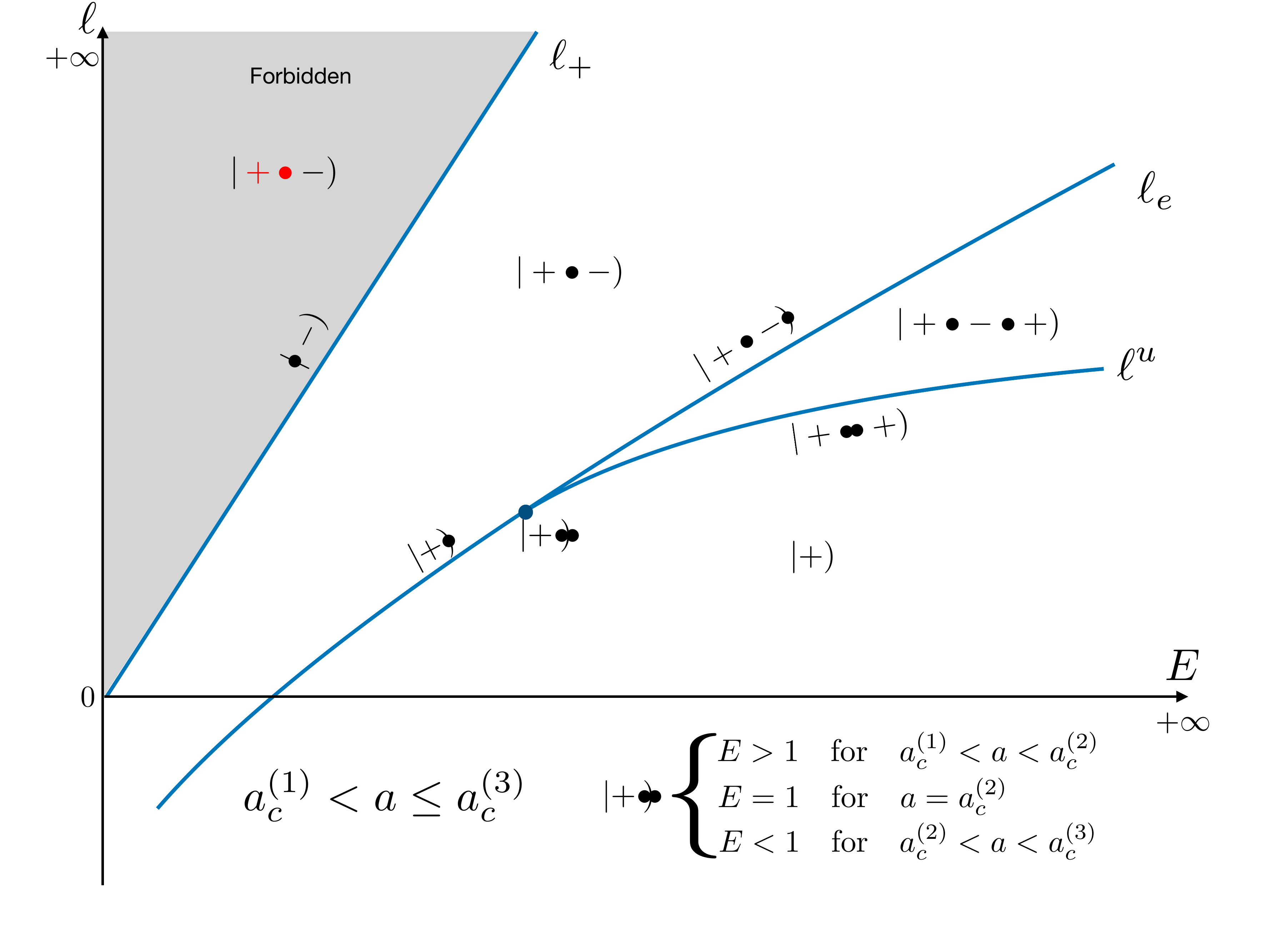}
     \end{subfigure}
     \hfill  \vspace{-0.3cm} \break
     \begin{subfigure}[b]{0.45\textwidth}
         \centering
         \includegraphics[width=\textwidth]{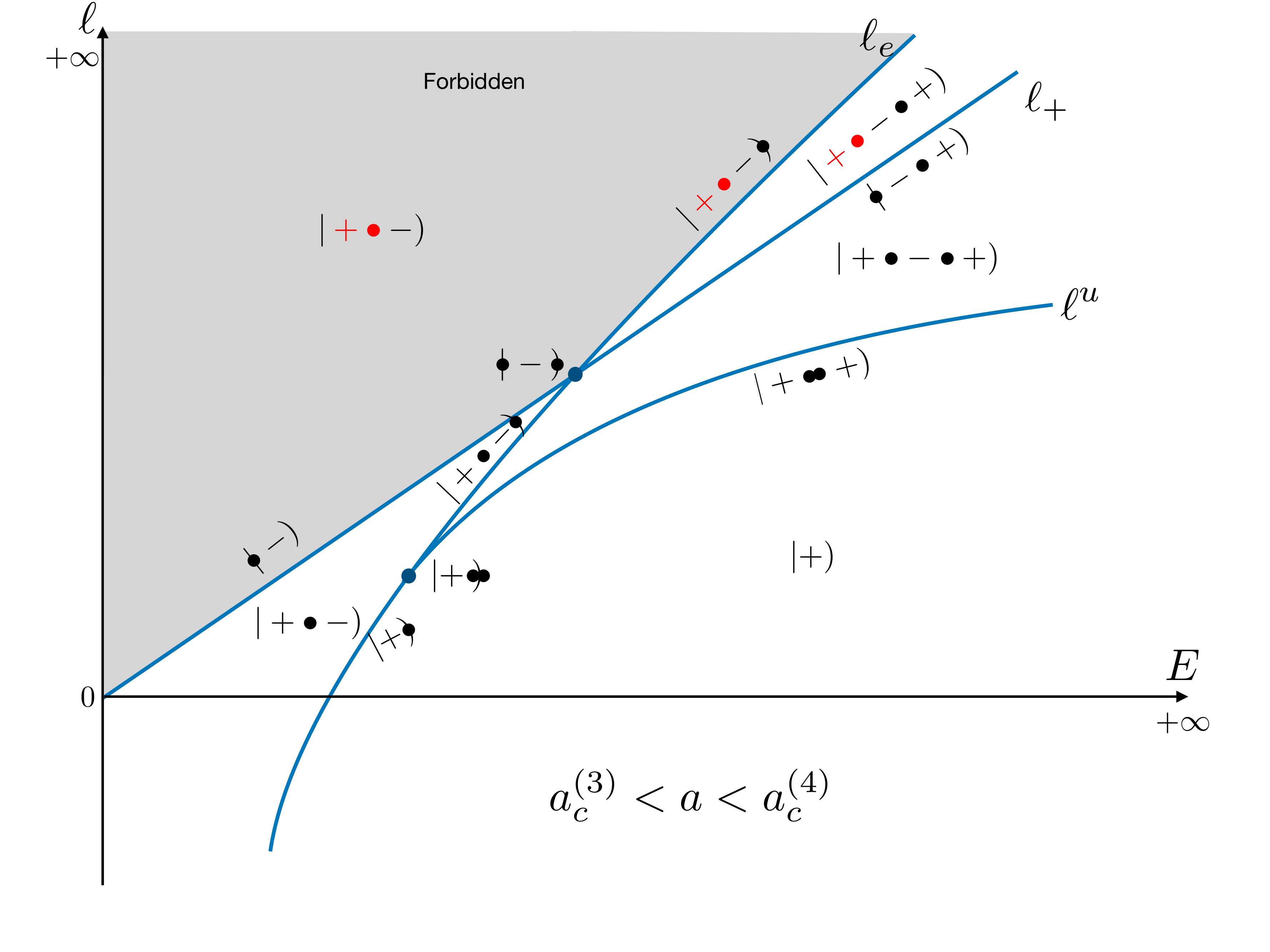}
     \end{subfigure}
     \hfill \vspace{0.4cm}
     \begin{subfigure}[b]{0.45\textwidth}
         \centering
         \includegraphics[width=\textwidth]{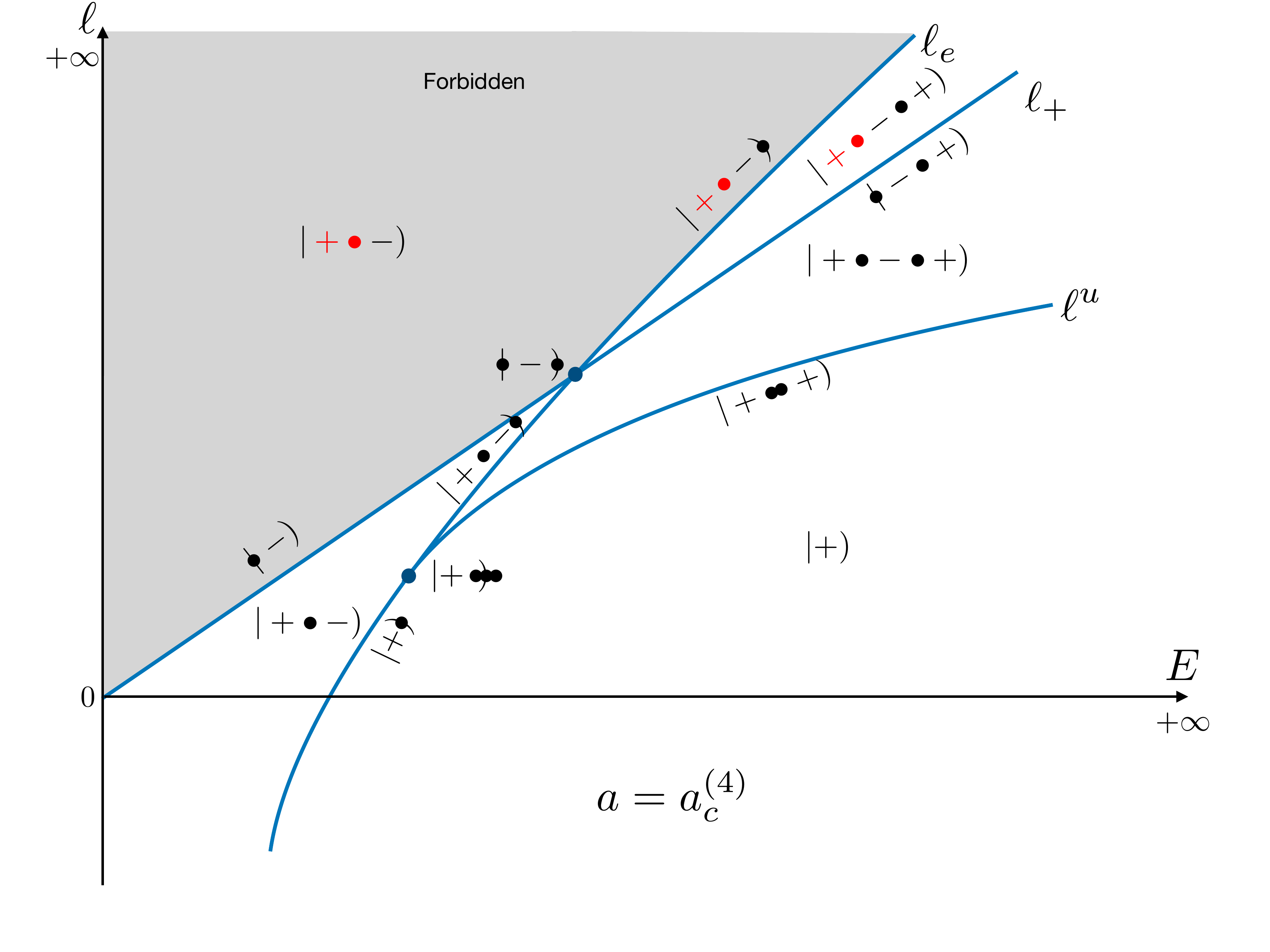}
     \end{subfigure}
     \hfill  \vspace{-0.3cm} \break
     \begin{subfigure}[b]{0.45\textwidth}
         \centering
         \includegraphics[width=\textwidth]{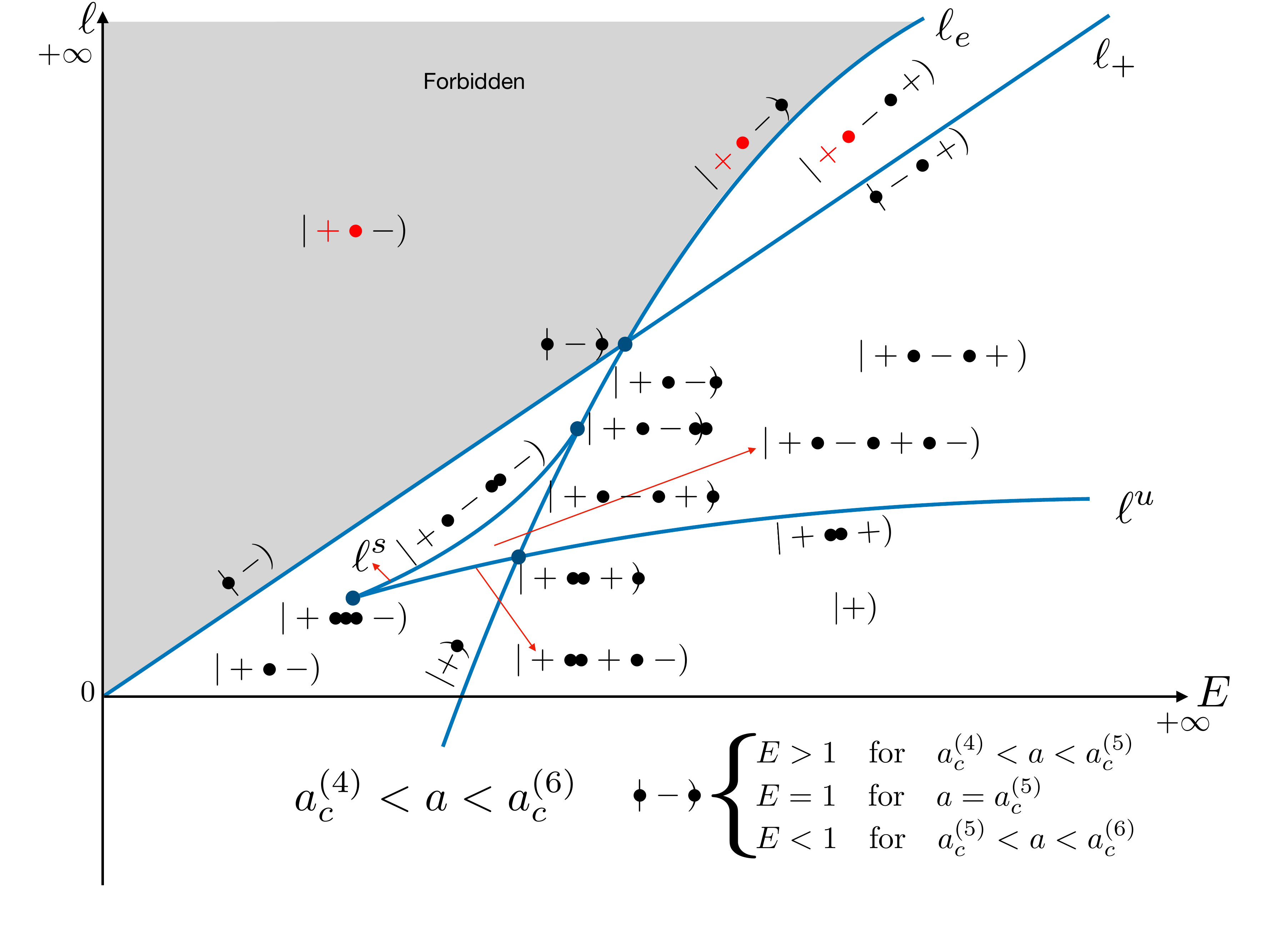}
     \end{subfigure}
     \hfill  \vspace{0.4cm}
     \begin{subfigure}[b]{0.45\textwidth}
         \centering
         \includegraphics[width=\textwidth]{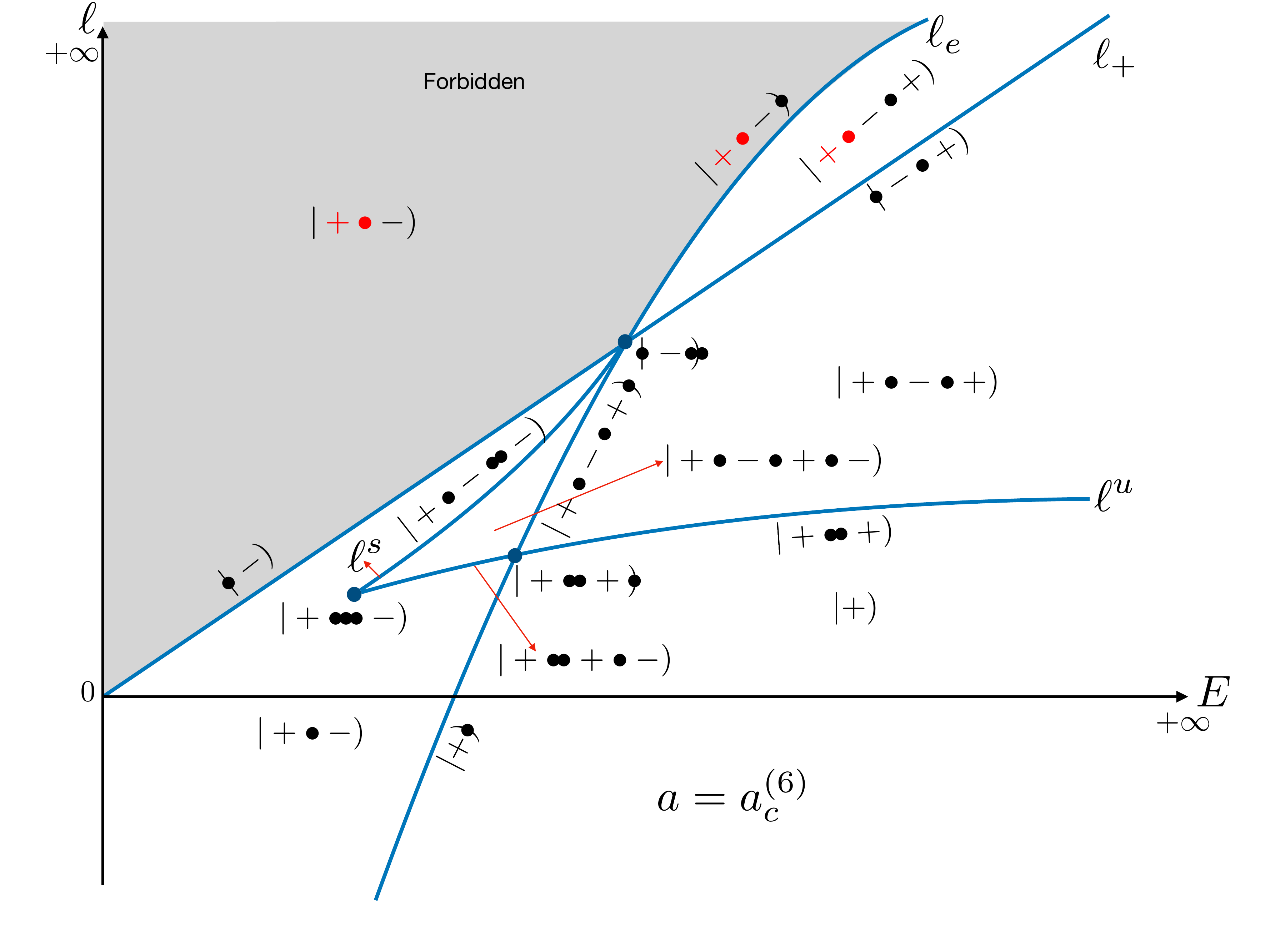}
     \end{subfigure}
     \hfill \vspace{-0.3cm}  \break
     \begin{subfigure}[b]{0.45\textwidth}
         \centering
         \includegraphics[width=\textwidth]{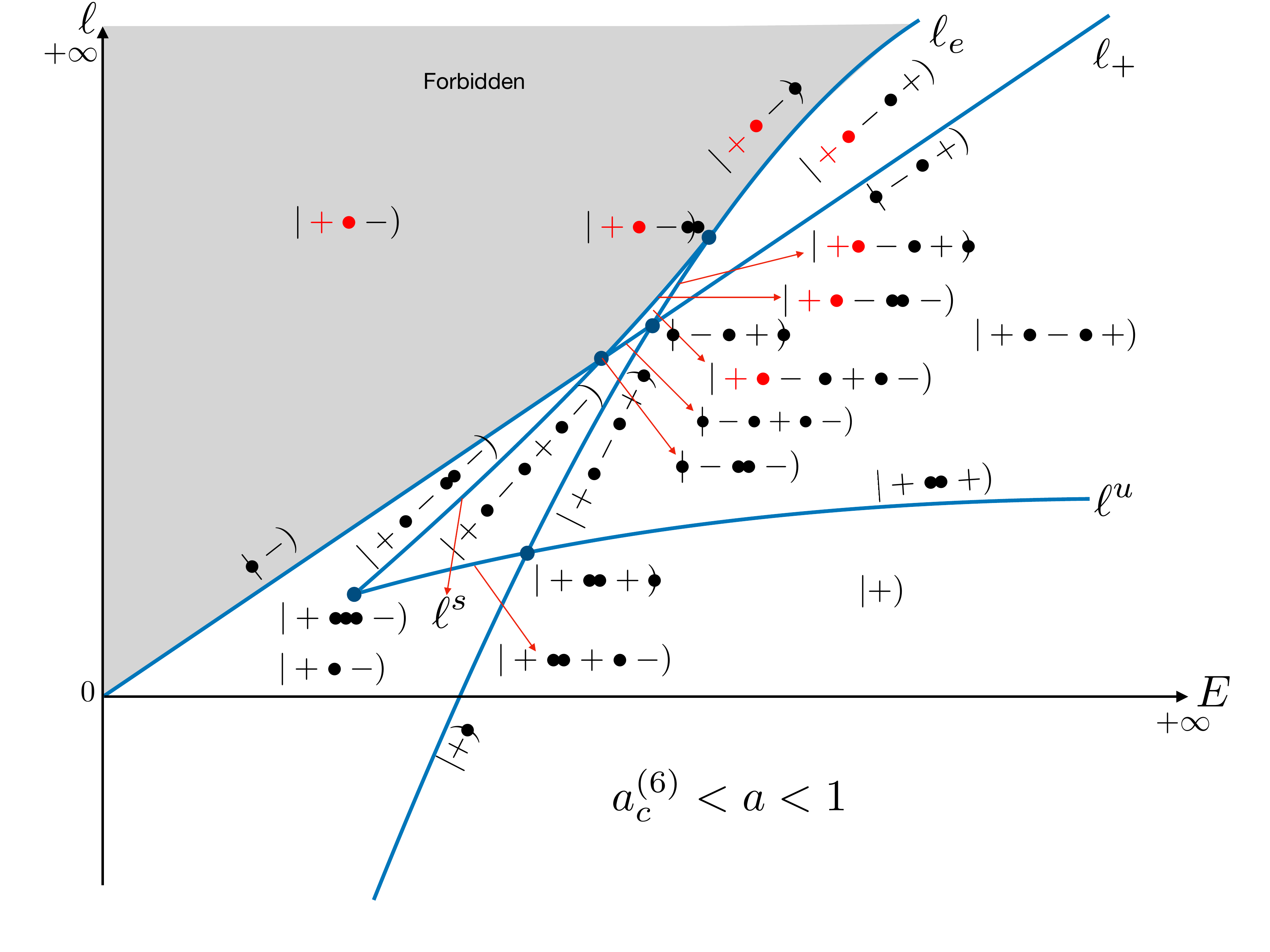}\hfill 
     \end{subfigure}
\caption{Classification of radial root structures with positive energy in the equatorial ergoregion of Kerr.}\label{fig:Ep0}
\end{figure}

The rich phase diagram for $E > 0$ is depicted in Figure \ref{fig:Ep0}. 

For $0 < a \leq a_c^{(1)}$, $\ell_+(E)$ and $\ell_e(E)$ are defined. There are no double roots within the ergoregion. Indeed, for double root systems to exist within the ergoregion, they need to cross the ergosphere upon increasing $a$, and this only occurs for $a >a_c^{(1)}$. At $\ell = \ell_e(E)$ the root structure is ${\vert + )\hspace{-4pt}\bullet}$. Since $\frac{\partial R_0}{\partial \ell}=-4a E <0$ at $\ell=\ell_e(E)$ and $r=2$, the root structure for $\ell>\ell_e(E)$ is $\vert + \bullet\hspace{2pt} - )$, while for $\ell < \ell_e(E)$ it is $\vert + )$. At the special value $\ell= \ell_+(E) > \ell_e(E)$ the latter root structure degenerates to $\vert \hspace{-7pt}\bullet -)$, and for $\ell > \ell_+(E)$, the root structure is again $\vert + \bullet\hspace{2pt} - )$ but it is not allowed by condition \eqref{finalboundplunging} and, therefore, we denote it as $\vert {\color{red} \hspace{1pt}+\hspace{2pt} \bullet}\hspace{2pt} - )$.

For $a>a_c^{(1)}$, there is a double root structure touching the ergosphere at $E=E^+_e(a)$ as defined in Eq.\eqref{Ee}. The corresponding root structure is ${\vert + \bullet \hspace{-5pt})\hspace{-2pt}\bullet }$. It continuously connects to the unstable double root branch ${\vert + \bullet\hspace{-5pt}\bullet + )}$ defined for $\ell = \ell^u(E)$. The root structure ${\vert + \bullet \hspace{-4pt})\hspace{-2pt}\bullet }$ occurs for $E>1$ as long as $a<a_c^{(2)}$, but it obeys $E<1$ for $a>a_c^{(2)}$. The root structure on the line $\ell=\ell_e(E)$ is $\vert + )\hspace{-5pt}\bullet$ for $E<E^+_e(a)$ and $\vert +\bullet \hspace{2pt}- \hspace{2pt})\hspace{-5pt}\bullet$ for $E>E^+_e(a)$. For $\ell^u(E)<\ell<\ell_e(E)$, the root structure is $\vert +\bullet -\bullet\hspace{2pt} + )$ with all motion allowed [see the corollary below Eq. \eqref{eq4}]. For $E>1$, the outer $+$ denotes deflecting orbits, while for $E<1$ it denotes bounded orbits that enter the ergoregion. 

For $a>a_c^{(3)}$, the lines $\ell_+(E)$ and $\ell_e(E)$ cross at \eqref{Esp}, which leads to the root structure ${\vert \hspace{-7pt}\bullet - \hspace{2pt}) \hspace{-4pt}\bullet}$. This root structure occurs at $E>1$ in the range $a_c^{(3)}<a<a_c^{(5)}$ but at $E<1$ in the range $a>a_c^{(5)}$. At $\ell=\ell_e(E)$ for $E>E^+_{+e}(a)$ the root structure is $\vert {\color{red}\hspace{2pt}+\hspace{2pt}\bullet }\hspace{2pt} - )\hspace{-4pt}\bullet$, while for $E^+_e(a)<E < E^+_{+e}(a)$ the root structure is $\vert +\bullet\hspace{2pt}  -\hspace{2pt} )\hspace{-4pt}\bullet$. For $\ell_+(E)<\ell<\ell_e(E)$, the root structure is $\vert {\color{red}\hspace{2pt} + \hspace{2pt}\bullet} - \bullet \hspace{2pt}+ )$. Indeed, one can numerically check on one particular value that the first root $r_1$ obeys $\ell=\ell_0^{(1)}(E,r_1)$ while the second root $r_2$ obeys $\ell=\ell_0^{(2)}(E,r_2)$. The bounds \eqref{split}--\eqref{finalbound} then imply that $r \leq r_1$ is discarded while $r \geq r_2$ is allowed.

For $a=a_c^{(4)}$, the triple root crosses the ergosphere at $E=\sqrt{2/3}$, where the root structure ${\vert + \bullet \hspace{-2pt}) \hspace{-7pt}\bullet \hspace{-4pt}\bullet}$ appears. 

For $a>a_c^{(4)}$, the triple root structure is within the ergoregion, which corresponds to ${\vert +\bullet \hspace{-5pt}\bullet \hspace{-5pt}\bullet \hspace{2pt}-)}$, and the stable branch with root structure ${\vert + \bullet - \bullet \hspace{-5pt}\bullet \hspace{2pt}-)}$ therefore appears within the ergoregion. For $a>a_c^{(4)}$, we start to have a triangular-shaped region delimited by $\ell>\ell_e(E)$, $\ell>\ell^u(E)$, and $\ell<\ell^s(E)$ with root structure $\vert + \bullet -\bullet +\bullet -)$ which contains, in particular, bounded orbits. The boundary of the triangular-shaped region contains several special root structures depicted in the figure. 

At $a=a_c^{(6)}$, the special root structure ${\vert \hspace{-7pt}\bullet- \bullet\hspace{-4pt} )\hspace{-3pt}\bullet  }$ occurs because the line $\ell =\ell^s(E)$ crosses both $\ell = \ell_e(E)$ and $\ell = \ell_+(E)$ at $E=E_e(a_c^{(6)})$. 

For $a>a_c^{(6)}$, a new triangular-shaped region occurs bounded by $\ell>\ell_+(E)$, $\ell>\ell_e(E)$, and $\ell<\ell^s(E)$. Within the triangular-shaped region, the root structure is $\vert {\color{red}\hspace{2pt} + \hspace{2pt}\bullet } - \bullet + \bullet\hspace{2pt} - )$. Indeed, the first root $r_1$ obeys $\ell=\ell_0^{(1)}(E,r_1)$ by continuity with previous cases, while the second and third roots $r_{2,3}$ obey $\ell=\ell_0^{(2)}(E,r_{2,3})$ by continuity with the double root. The conditions \eqref{split}--\eqref{finalbound} then imply that trapped orbits are disallowed while bounded orbits are allowed. The boundary of the triangular-shaped region contains  several special root structures depicted in the figure, which are continuously joined with the now square-shaped region.

Note that all $+$ orbits in root structures with $\ell \leq \ell_+$ are allowed, while for $\ell > \ell_+$  all trapped orbits in root structures are disallowed and nontrapped orbits are allowed.

\paragraph{Phase diagram for $E < 0$.}

\begin{figure}[!htbp]\vspace{1cm}
     \centering 
     \begin{subfigure}[b]{0.45\textwidth}
         \centering
         \includegraphics[width=\textwidth]{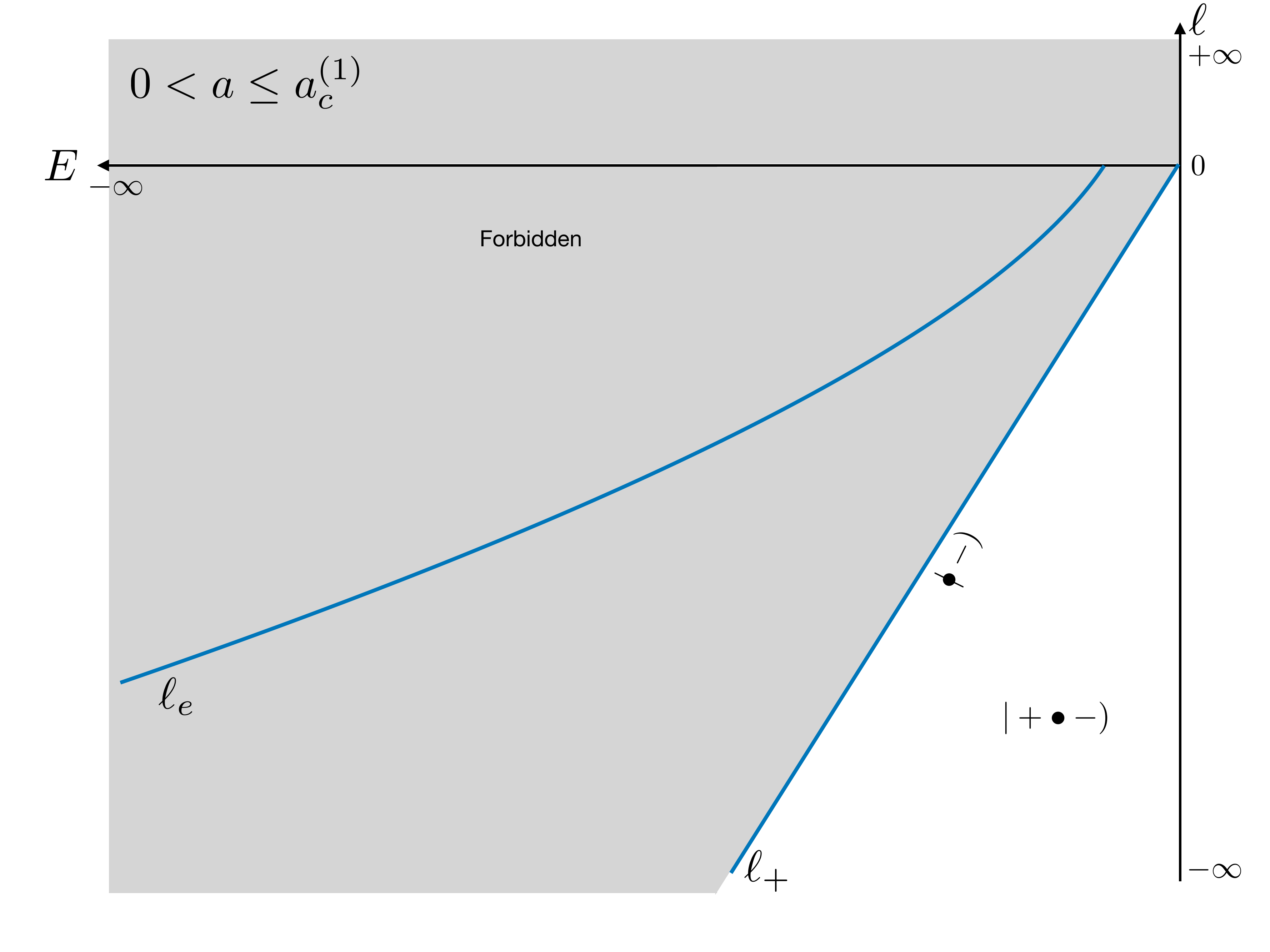}
     \end{subfigure}
     \hfill  \vspace{0.4cm}
     \begin{subfigure}[b]{0.45\textwidth}
         \centering
         \includegraphics[width=\textwidth]{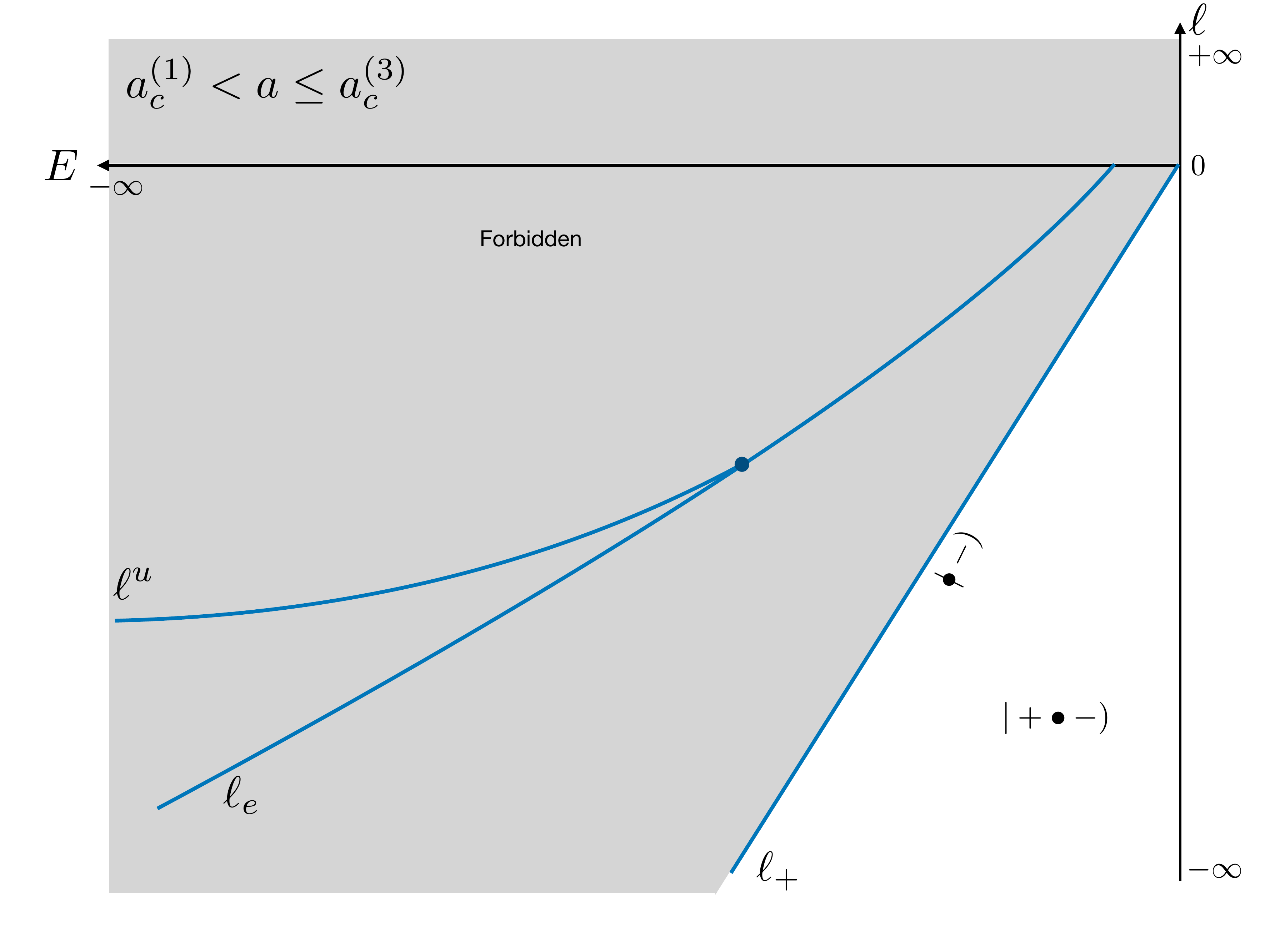}
     \end{subfigure}
     \hfill  \vspace{-0.3cm} \break
     \begin{subfigure}[b]{0.45\textwidth}
         \centering
         \includegraphics[width=\textwidth]{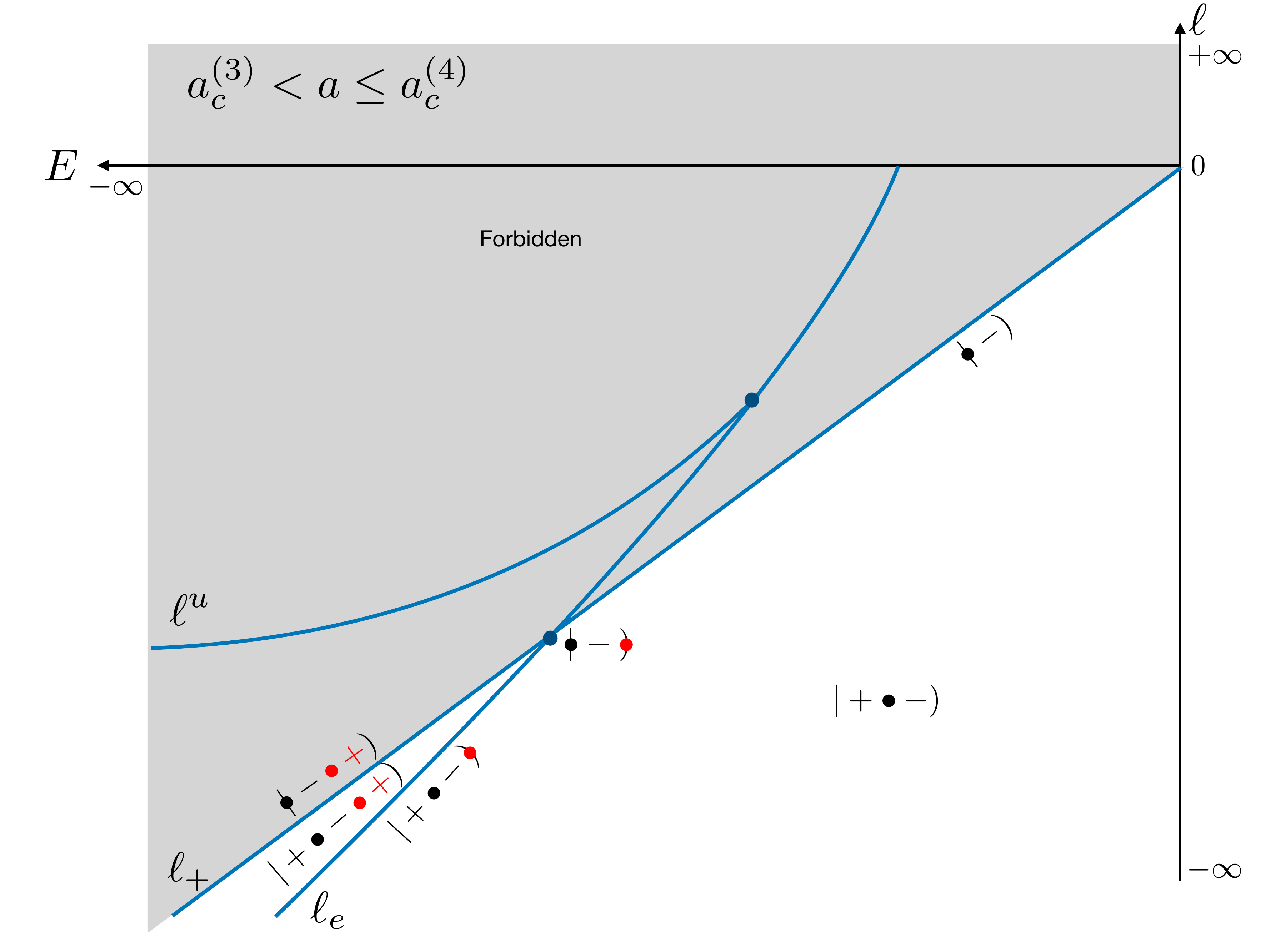}
     \end{subfigure}
     \hfill \vspace{0.4cm}
     \begin{subfigure}[b]{0.45\textwidth}
         \centering
         \includegraphics[width=\textwidth]{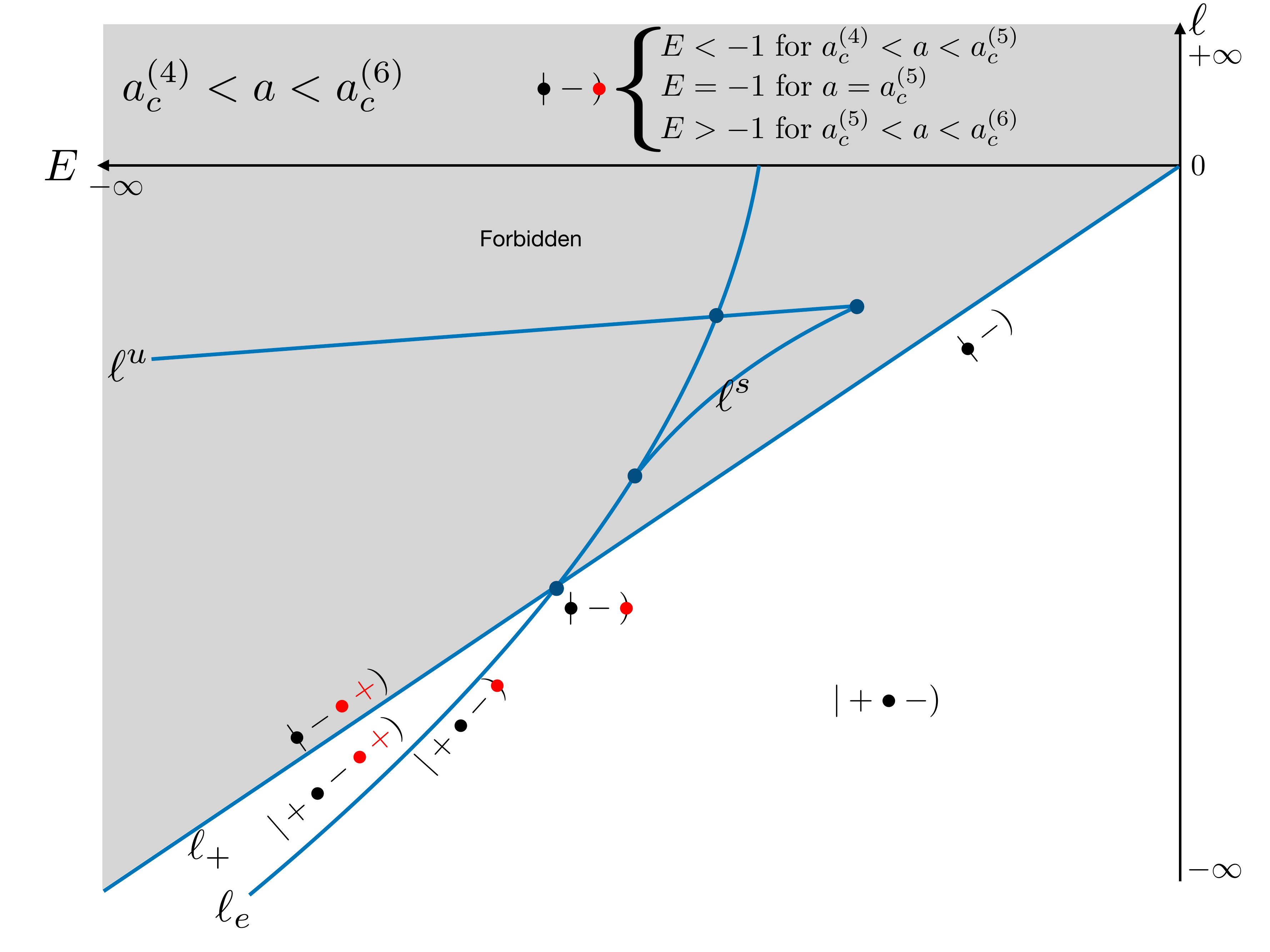}
     \end{subfigure}
     \hfill  \vspace{-0.3cm} \break
     \begin{subfigure}[b]{0.45\textwidth}
         \centering
         \includegraphics[width=\textwidth]{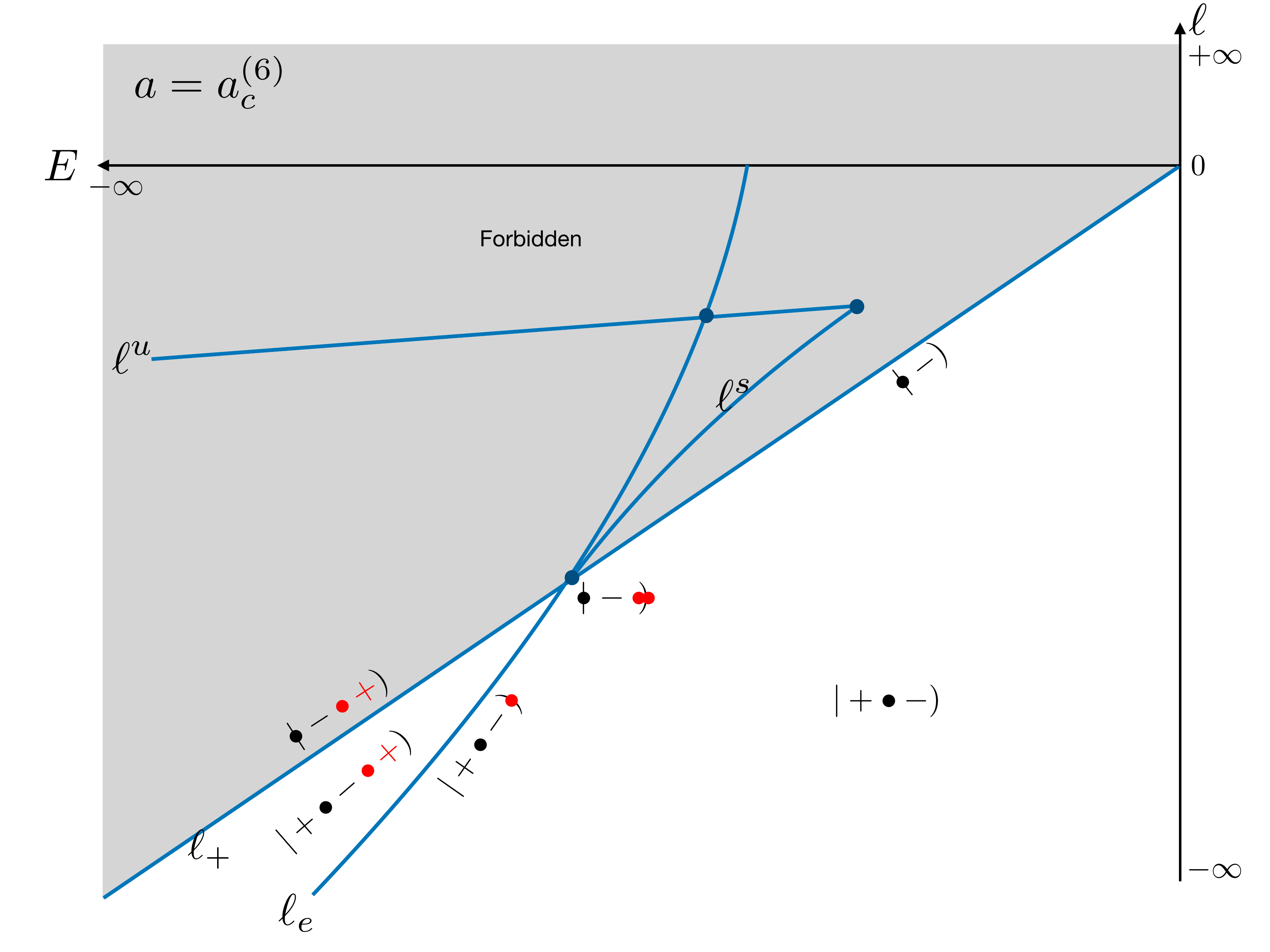}
     \end{subfigure}
     \hfill  \vspace{0.4cm}
     \begin{subfigure}[b]{0.45\textwidth}
         \centering
         \includegraphics[width=\textwidth]{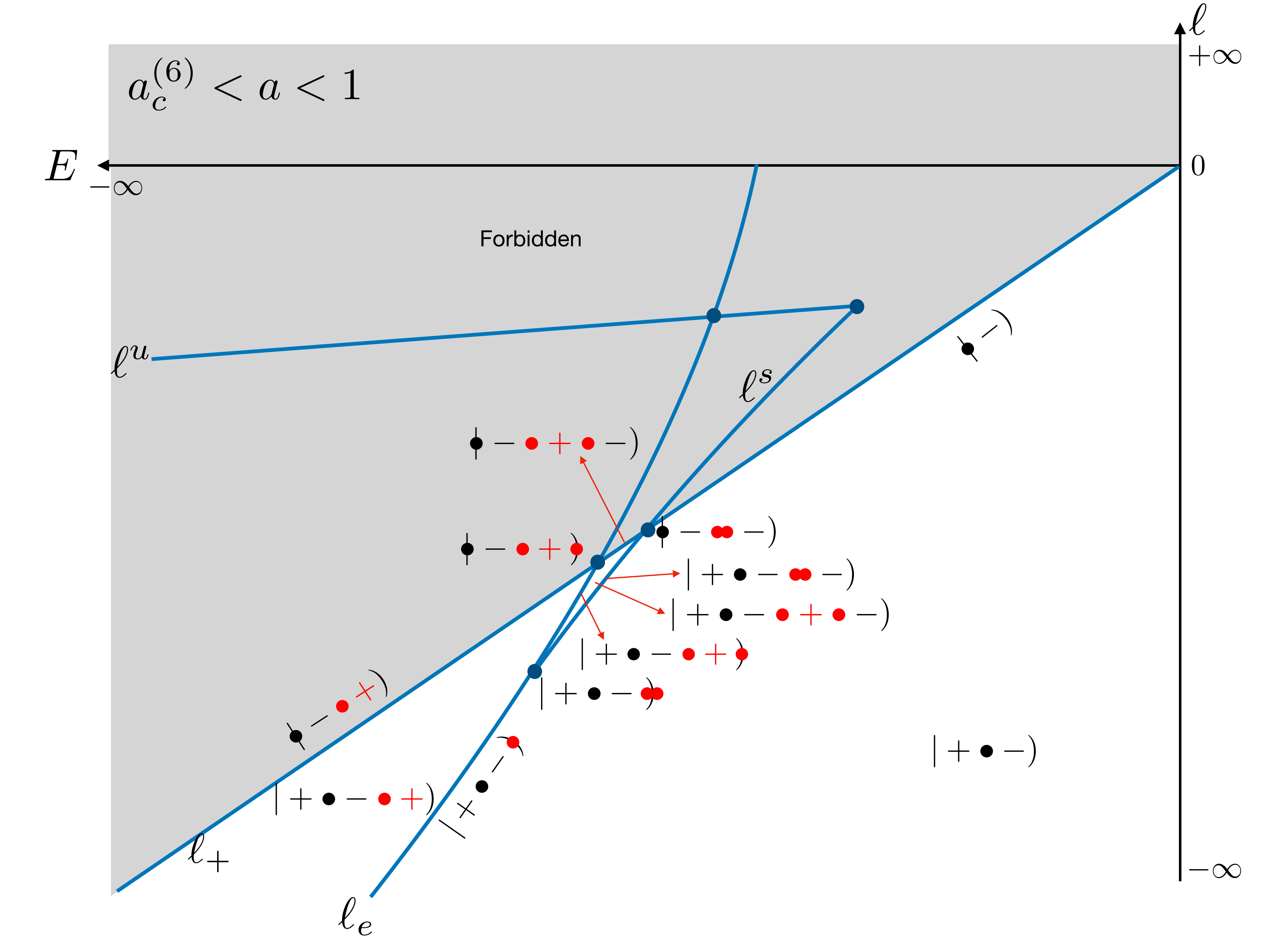}
     \end{subfigure}
     \hfill   \vspace{-0.3cm}
\caption{Classification of radial root structures with negative energy in the equatorial ergoregion of Kerr.}\label{fig:Em0}
\end{figure}

The simpler phase diagram for $E < 0$ is depicted in Figure \ref{fig:Em0}. Due to $(E,\ell) \mapsto -(E,\ell)$ symmetry, Figure \ref{fig:Em0} is related by a central flip of Figure \ref{fig:Ep0} but now with the disallowed region \eqref{finalbound} which implies the bound \eqref{finalboundplunging} for all orbits. The region $\ell>\ell_+(E)$ is therefore always discarded. Trapped orbits automatically obey the bound \eqref{finalbound}. Nontrapped orbits will always violate the bound \eqref{finalbound} as we will derive below.

For $0 < a \leq a_c^{(1)}$, $\ell_+(E)$ and $\ell_e(E)$ are defined, but $\ell_e(E)>\ell_+(E)$, and this curve is not relevant. At $\ell=\ell_+(E)$ we have the root structure ${\vert \hspace{-7pt}\bullet - )}$, while for $\ell<\ell_+(E)$ we have the root structure $\vert + \bullet\hspace{2pt} - )$. 

For $a>a_c^{(1)}$, there is a double root structure, but it is within the forbidden region, and it can be ignored. 

For $a>a_c^{(3)}$, the lines $\ell_+(E)$ and $\ell_e(E)$ cross at $E^-_{+e}(a)$ defined in Eq. \eqref{Esp}, which leads to the root structure ${\vert \hspace{-7pt}\bullet - \hspace{2pt}) \hspace{-4pt}\bullet}$. This root structure occurs at $E<-1$ in the range $a_c^{(3)}<a<a_c^{(5)}$ but at $E>-1$ in the range $a>a_c^{(5)}$.  For $\ell=\ell_+(E)$ and $E<E^-_{+e}(a)$, the root structure is ${\vert \hspace{-7pt}\bullet - {\color{red} \hspace{2pt}\bullet \hspace{2pt} +} )}$. Indeed, one can check on one particular numerical value that the largest root $r_2$ within the ergoregion obeys $\ell=\ell_0^{(1)}(E,r_2)$, and the region $r \geq r_2$ is therefore discarded from Eqs. \eqref{split} and \eqref{finalbound}. In the region $\ell_+(E) < \ell < \ell_e(E)$, the root structure is ${\vert + \bullet - {\color{red} \bullet \hspace{2pt} +} )}$. The largest root $r_2$ still obeys $\ell=\ell_0^{(1)}(E,r_2)$ by continuity. One can check on one particular numerical value that the smallest root $r_1$ instead obeys $\ell=\ell_0^{(2)}(E,r_1)$, and, therefore, the region $r \leq r_1$ is allowed. The root structure degenerates to ${\vert +\bullet \hspace{2pt}-\hspace{2pt}) {\color{red} \hspace{-5pt}\bullet}}$ at $\ell=\ell_e(E)$. 

For $a=a_c^{(4)}$, the triple root crosses the ergosphere at $E=-\sqrt{2/3}$, but it is within the forbidden region, and it can be ignored. For $a>a_c^{(4)}$, the line $\ell=\ell^s(E)$ appears within the forbidden region. 

At $a=a_c^{(6)}$, the special root structure ${\vert \hspace{-7pt}\bullet-\hspace{2pt} {\color{red}\bullet}\hspace{-4pt} )\hspace{-3pt}{\color{red}\bullet}  }$ occurs because the line $\ell =\ell^s(E)$ crosses both $\ell = \ell_e(E)$ and $\ell = \ell_+(E)$ at $E=E_e(a_c^{(6)})$. The double root $r_*$ at the ergosphere obeys $\ell=\ell_0^{(1)}(E,r_*)$ by continuity with previous cases. 

For $a>a_c^{(6)}$, a new triangular-shaped region appears bounded by $\ell<\ell_+(E)$, $\ell<\ell_e(E)$ and $\ell>\ell^s(E)$. By continuity, the second and third roots of all root structures involved obey $\ell=\ell_0^{(1)}(E,r_*)$ by continuity with the root structure ${\vert \hspace{-7pt}\bullet-\hspace{2pt} {\color{red}\bullet}\hspace{-4pt} )\hspace{-3pt}{\color{red}\bullet}  }$ and, therefore, motion is discarded from Eqs. \eqref{split} and \eqref{finalbound} except for trapped orbits. 
   
In conclusion, only trapped orbits are allowed, consistently with the analysis of \cite{1984GReGr..16...43C}.

\paragraph{Final classification of radial motion.}

The taxonomy of radial motion of positive energy Kerr geodesics in the equatorial ergoregion is listed in Tables \ref{tableT1} and \ref{tableT2}, while the taxonomy of allowed radial motion of negative energy Kerr geodesics in the equatorial ergoregion is listed in Table \ref{tableT3}. The taxonomy is consistent with the generic Kerr taxonomy in the complete exterior region as listed in Table \ref{table:Kerrfinal}.

\begin{table}[!tbh]    \centering
\begin{tabular}{|c|c|c|c|c|}\hline
\rule{0pt}{13pt} & \textbf{Root structure} & \textbf{Angular momentum} & \textbf{Radial range} & \textbf{Name} \\ \hline
Generic & $\vert+)$ & $\ell<\text{min}(\ell^u,\ell_e)$ & $r_+\le r<2$ & $\mathcal{T}(E<1,\ell,0)$\\ 
& & & & $\mathcal{P}(E\ge 1,\ell,0)$\\\cline{2-5}
 & $\vert+\bullet\hspace{2pt}-)$ & {$\ell_e<\ell<\ell_+$
 \text{and}\ ($\ell>\ell^s$
 \text{or}\ $\ell<\ell^u$) }& $r_+\le r\le r_1$ & $\mathcal{T}(E,\ell,0)$\\ \cline{2-5}
  & $\vert\hspace{2pt}{\color{red}+\hspace{2pt}\bullet}\hspace{2pt}-)$ & $\ell>\text{max}(\ell_+,\ell_e,\ell^s)$ & $\emptyset$ & $\emptyset$\\ \cline{2-5}
 & $\vert+\bullet-\bullet+)$ & $\ell^u<\ell<\text{min}(\ell_+,\ell_e)$ & $r_+\le r\le r_1$ & $\mathcal{T}(E,\ell,0)$\\ \cline{4-5}
 & & &$r_2\le r \le 2$ & $\mathcal{B}(E<1,\ell,0)$\\
 & & & & $\mathcal{D}(E\ge 1,\ell,0)$\\\cline{2-5}
  & $\vert{\color{red}\hspace{2pt}+\hspace{2pt}\bullet}-\bullet+)$ & $\ell_+<\ell<\ell_e$ & $r_2\le r\le 2$ & $\mathcal{B}(E<1,\ell,0)$\\
  & & & &$\mathcal{D}(E\ge 1,\ell,0)$ \\\cline{2-5}
 & $\vert+\bullet-\bullet+\bullet-)$ & $\text{max}(\ell^u,\ell_e)<\ell<\text{min}(\ell^s,\ell_+)$ & $r_+\le r\le r_1$ & $\mathcal{T}(E,\ell,0)$\\ \cline{4-5}
 & & & $r_2\le r \le r_3$ & $\mathcal{B}(E,\ell,0)$\\ \cline{2-5}
  & $\vert{\color{red}\hspace{2pt}+\hspace{2pt}\bullet}-\bullet+\bullet-)$ & $\text{max}(\ell_e,\ell_+)<\ell<\ell^s$ & $r_2\le r\le r_3$ & $\mathcal{B}(E,\ell,0)$\\ \hline
Codi-& $\vert+)\hspace{-5pt}\bullet$ & $\ell=\ell_e<\ell^u$ & $r_+\le r \leq 2$ & $\mathcal{T}(E,\ell_e,0)$\\ \cline{2-5}
mension & $\vert+\bullet\hspace{2pt}-\hspace{2pt})\hspace{-5pt}\bullet$ & $\text{max}(\ell^s,\ell^u)<\ell=\ell_e<\ell_+$ & $r_+\le r\le r_1$ & $\mathcal{T}(E,\ell_e,0)$\\ \cline{2-5}
   1 &$\vert{\color{red}\hspace{2pt}+\hspace{2pt}\bullet}\hspace{2pt}-\hspace{2pt})\hspace{-5pt}\bullet$ & $\ell=\ell_e>\text{max}(\ell_+,\ell^s)$ & $\emptyset$ & $\emptyset$\\ \cline{2-5}
    & $\vert+\bullet-\bullet\hspace{2pt} +\hspace{2pt})\hspace{-5pt}\bullet$ & $\ell^u<\ell=\ell_e<\text{min}(\ell^s,\ell_+)$ & $r_+\le r\le r_1$ & $\mathcal{T}(E,\ell_e,0)$\\\cline{4-5}
    & & &$r_2\le r\le 2$ & $\mathcal{B}(E,\ell_e, 0)$\\\cline{2-5}
     & $\vert{\color{red}\hspace{2pt}+\hspace{2pt}\bullet}-\bullet\hspace{2pt} +\hspace{2pt})\hspace{-5pt}\bullet$ & $\ell_+<\ell=\ell_e<\ell^s$ & $r_2\le r\le 2$ & $\mathcal{B}(E,\ell_e, 0)$\\\cline{2-5}
    & $\vert\hspace{-7pt}\bullet-)$ & $\ell=\ell_+<\text{min}(\ell_{e},\ell_{s})$ & $\emptyset$ & $\emptyset$\\ \cline{2-5}
    & $\vert\hspace{-7pt}\bullet-\bullet+\bullet-)$ & $\ell^{s}<\ell=\ell_+<\ell_{e}$ & $r_2\le r\le r_3$ & $\mathcal{B}(E,\ell_+,0)$\\\cline{2-5}
    &$\vert\hspace{-7pt}\bullet-\bullet +)$ & $\ell=\ell_+>\ell_{e}$ & $r_2<r\le 2$ & $\mathcal{B}(E<1,\ell_+, 0)$\\ 
    & & & & $\mathcal{D}(E\ge 1,\ell_+)$\\\cline{2-5}
   & $\vert+\bullet\hspace{-4pt}\bullet+)$ & $\ell=\ell^u>\ell_e$ &$r_+\le r< r_1$ & $\mathcal{WT}^u(E,\ell^u)$\\ \cline{4-5}
 & & & $r=r_1$ & $\mathcal{C}^u(E,\ell^u)$\\ \cline{4-5}
 & & & $r_1<r\le 2$ & $\mathcal{H}^u(E<1,\ell^u)$\\
 & & & & $\mathcal{WD}^u(E\ge 1,\ell^u)$\\\cline{2-5}
 & $\vert+\bullet\hspace{-4pt}\bullet+\bullet-)$ & $\ell_{ISCO^+}<\ell=\ell^u<\ell_e$ &$r_+\le r<r_1$ & $\mathcal{WT}^u(E,\ell^u)$\\\cline{4-5}
 & & & $r=r_1$ & $\mathcal{C}^u(E,\ell^u)$\\\cline{4-5}
 & & & $r_1<r\le r_2$ & $\mathcal{H}^u(E,\ell^u)$\\\cline{2-5}
 &$\vert+\bullet-\bullet\hspace{-4pt}\bullet-)$ & $\ell^u<\ell=\ell^s<\text{min}(\ell_+,\ell_e(E_e^+))$ & $r_+\le r\le r_1$ & $\mathcal{T}(E,\ell^s,0)$\\\cline{4-5}
 & & & $r=r_2$ & $\mathcal{C}^s(E,\ell^s)$\\\cline{2-5}
  &$\vert{\color{red}\hspace{2pt}+\hspace{2pt}\bullet}-\bullet\hspace{-4pt}\bullet-)$ & $\ell_+<\ell=\ell^s<\ell_e$ & $r= r_2$ & $\mathcal{C}^s(E,\ell^s)$\\\hline
 \end{tabular}\caption{Taxonomy of radial motion of positive energy Kerr geodesics in the equatorial ergoregion: codimension 0 and 1 root structures. The range of $a$ such that all quantities are real can be deduced from Figure \ref{fig:Ep0}. As a matter of convention, whether bounds are complex or there is no corresponding curve in Figure \ref{fig:Ep0}, they do not lead to constraints.}\label{tableT1}
\end{table}

\begin{table}[!tbh]    \centering
\begin{tabular}{|c|c|c|c|c|}\hline
\rule{0pt}{13pt} & \textbf{Root structure} & \textbf{Angular momentum} & \textbf{Radial range} & \textbf{Name} \\ \hline
 Codimen-& $\vert\hspace{-7pt}\bullet-\hspace{2pt})\hspace{-5pt}\bullet$ & $\ell=\ell_{e}=\ell_+>\ell^s$ & $\emptyset$ & $\emptyset$\\\cline{2-5}
 sion 2 & $\vert\hspace{-7pt}\bullet-\bullet+\hspace{2pt})\hspace{-5pt}\bullet$ & $\ell=\ell_{e}=\ell_+<\ell^s$ & $r_2\le r\le 2$ & $\mathcal{B}(E,\ell_+,0)$\\\cline{2-5}
  & $\vert+\hspace{2pt})\hspace{-9pt}\bullet\hspace{-4pt}\bullet$ & $\ell=\ell_e=\ell^u$ ($a<a_c^{(4)}$)& $r_+\le r<2$ & $\mathcal{WT}^u(E,\ell_e)$\\\cline{4-5}
  & & & $r=2$ & $\mathcal{C}^u(E,\ell_e)$ \\\cline{2-5}
  & $\vert +\bullet\hspace{-4pt}\bullet+\hspace{2pt})\hspace{-5pt}\bullet$ & $\ell=\ell_e=\ell^u$ ($a>a_c^{(4)}$)& $r_+\le r<r_1$ & $\mathcal{WT}^u(E,\ell_{e})$\\\cline{4-5}
  & & & $r=r_1$ & $\mathcal{C}^u(E,\ell_{e})$\\\cline{4-5}
 & & & $r_1<r<2$ & $\mathcal{H}^u(E,\ell_{e})$\\\cline{2-5}
 & $\vert+\bullet\hspace{2pt}-\hspace{4pt})\hspace{-9pt}\bullet\hspace{-4pt}\bullet$ & $\ell=\ell_e=\ell^s<\ell_+$ & $r_+\le r\le r_1$ & $\mathcal{T}(E,\ell_e,0)$\\\cline{4-5}
 & & & $r=2$ & $\mathcal{C}^s(E,\ell_e)$\\\cline{2-5}
  & $\vert{\color{red}\hspace{2pt}+\hspace{2pt}\bullet}\hspace{2pt}-\hspace{4pt})\hspace{-9pt}\bullet\hspace{-4pt}\bullet$ & $\ell=\ell_e=\ell^s>\ell_+$& $ r=2$ & $\mathcal{C}^s(E,\ell_e)$\\\cline{2-5}
 & $\vert\hspace{-7pt}\bullet-\bullet\hspace{-4pt}\bullet\hspace{2pt}-)$ & $\ell=\ell_+=\ell^s$ & $r= r_2$ & $\mathcal{C}^s(E,\ell_+)$\\\cline{2-5}
 & $\vert+\bullet\hspace{-4pt}\bullet\hspace{-4pt}\bullet\hspace{2pt}-)$ & $\ell=\ell_{ISCO^+}$ & $r_+\le r<r_{ISCO^+}$ &$\mathcal{WT}_{ISCO}(E)$\\\cline{4-5}
 & & & $r=r_{ISCO^+}$ & $\mathcal{C}_{ISCO}(E)$ \\\hline
 Codimen- & $\vert\hspace{-7pt}\bullet-\hspace{3pt})\hspace{-9pt}\bullet\hspace{-4pt}\bullet$ & $\ell=\ell_e=\ell^s=\ell_+$ & $r=2$ & $\mathcal{C}^s(E,\ell_+)$\\\cline{2-5}
 sion 3 & $\vert+\hspace{7pt})\hspace{-12pt}\bullet\hspace{-4pt}\bullet\hspace{-2pt}\bullet$ & $\ell=\ell_e=\ell_{ISCO^+}$ &$r_+\le r<2$ &$\mathcal{WT}_{ISCO}(E)$\\\cline{4-5}
 & & & $r=2$ & $\mathcal{C}_{ISCO}(E)$\\\hline
 \end{tabular}\caption{Taxonomy of radial motion of positive energy Kerr geodesics in the equatorial ergoregion: codimension 2 and 3 root structures. The range of $a$ such that all quantities are real can be deduced from Figure \ref{fig:Ep0}. As a matter of convention, whether bounds are complex or there is no corresponding curve in Figure \ref{fig:Ep0}, they do not lead to constraints.}\label{tableT2}
\end{table}

\begin{table}[!tbh]    \centering
\begin{tabular}{|c|c|c|c|c|}\hline
\rule{0pt}{13pt} & \textbf{Root structure} & \textbf{Angular momentum} & \textbf{Radial range} & \textbf{Name} \\ \hline
Generic & $\vert+\bullet\hspace{2pt}-)$ & $\ell<\text{min}(\ell_e,\ell_+,\ell^s)$ & $r_+\le r\le r_1$ & $\mathcal{T}(E,\ell,0)$\\\cline{2-3}
& $\vert+\bullet-{\color{red}\bullet\hspace{2pt}+})$ &$\ell_e<\ell<\ell_+$ &  & \\\cline{2-3}
& $\vert+\bullet-{\color{red} \bullet+\bullet}\hspace{2pt}-)$ & $\ell^s<\ell<\text{min}(\ell_+,\ell_e)$ & &  \\\cline{1-3}
Codimen-& $\vert+\bullet\hspace{2pt}-)\hspace{-4pt}{\color{red}\bullet}$ & $\ell=\ell_e<\text{min}(\ell_+,\ell^s)$&  & \\\cline{2-3}
sion 1 & $\vert+\bullet-{\color{red}\bullet\hspace{-2pt}\bullet}\hspace{2pt}-)$ & $\ell=\ell^s$, $\ell_e<\ell<\ell_+$ & &\\\cline{2-3}
& $\vert+\bullet-{\color{red}\bullet\hspace{2pt}+\hspace{2pt}})\hspace{-4pt}{\color{red}\bullet}$ & $\ell=\ell_e$, $\ell^s<\ell<\ell_+$ & &\\\cline{1-3}
Codimen-&$\vert+\bullet\hspace{2pt}-\hspace{2pt})\hspace{-6pt}{\color{red}\bullet\hspace{-2pt}\bullet}$ & $\ell=\ell_e=\ell^s<\ell_+$ & &\\
sion 2 & & & &\\\hline
 \end{tabular}\caption{Taxonomy of allowed radial motion of negative energy Kerr geodesics in the equatorial ergoregion. Root structure with only disallowed motion are not listed. As a matter of convention, whether bounds are complex, they do not lead to constraints.}\label{tableT3}
\end{table}

\clearpage

\subsection{Nonequatorial orbits within the ergoregion}
\label{sec:details}

Let us first discuss $Q<0$ orbits. From the analysis of Section \ref{sec2}, all such orbits have $E^2>1$. From Figure \ref{fig:Egtr1} the root structure of such orbits is $\vert + \rangle$. Since the ergosphere needs to be crossed, negative energies are discarded, and the root structure taking into account the ergoregion is $\vert + ) + \rangle$. There is therefore a single root structure within the ergoregion for $Q<0$ namely $\vert + )$ valid for $E>1$. The bound on $Q$ \eqref{boundQ} needs to be obeyed. The polar motion is $\text{Vortical}(E,Q)$ in the denomination of \cite{Compere:2020eat} (see also \cite{Kapec:2019hro}). 

In the following we will only discuss $Q \geq 0$ orbits. The polar motion of all orbits $Q>0$ is librating around the equator. These are the $\text{Pendular}(E,Q)$ in the terminology of \cite{Compere:2020eat}, see their Figures 1 and 3 (see also \cite{Kapec:2019hro}). Nonequatorial $Q=0$ orbits also exist for $E^2>1$ and are asymptotically approaching the equator at early and late proper times. These are the $\text{Equator-attractive}(E)$ in the terminology of \cite{Compere:2020eat}. In both cases, the classification of radial motion will necessarily match the equatorial case from continuity or as a limiting behavior from $Q=0$. The phase diagram displayed in Figures \ref{fig:Ep0} and  \ref{fig:Em0} therefore directly extend to $Q>0$ orbits. 

More precisely, the potential $R(r)$ is quadratic in $\ell$. The coefficient of $\ell^2$ is $(2-r)r$, which is positive strictly inside the ergoregion. Since $R(r) \geq 0$, the angular momentum then obeys
\begin{equation}
    \ell \leq \ell_0^{(2)}(E,Q,r) \qquad \text{or} \qquad  \ell \geq \ell_0^{(1)}(E,Q,r)\label{split2}
\end{equation}
where 
\begin{equation}
    \ell_0^{(1,2)}(E,Q,r)\equiv \frac{2 a E \pm \sqrt{\Delta(r)} \sqrt{(2-r)(r+\frac{Q}{r})+E^2r^2}}{(2-r)} \label{l012}
\end{equation}
are manifestly real for $Q \geq 0$, obey $\ell_0^{(2)}(E,Q,r) \leq \ell_0^{(1)}(E,Q,r)$ inside the ergoregion, and  $\ell^{(1)}(E,Q,r_+)=\ell^{(2)}(E,Q,r_+)= \frac{2r_+}{a}E=\ell_+(E) \leq 0$ at the horizon. Note that there is another solution $\ell_0^{(1)}=\ell_0^{(2)}$ for $Q=Q(E,r)<0$, but it is irrelevant since $Q < 0$ orbits were already discussed and are now disregarded. For $Q=0$ we demonstrated that the bound \eqref{finalbound} is always valid. By continuity or as a limiting case from $Q=0$, this implies that allowed motion for any $Q>0$ also obeys 
\begin{equation}
\ell\le\ell_0^{(2)}(E,Q,r). \label{bound5}
\end{equation} 
In particular, for $E=0$, the bound $\ell\le\ell_0^{(2)}(Q,E,r)$ reduces to 
\begin{equation}
    \ell \leq - \sqrt{\frac{(r^2+Q) \Delta}{r(2-r)}}<0. 
\end{equation}
Therefore, the orbits for $E=0$ are allowed only for negative angular momentum $\ell<0$. Since $\ell_0^{(2)}(E,Q,r_+)=\ell_+(E) $, all trapped orbits obey $\ell \leq \ell_+(E)$. This rules out the trapped orbits for $\ell> \ell_+(E)$. Since $\ell^{(2)}(E,Q,r) \leq \ell_+(E)<0$ for $E<0$ orbits within the ergoregion, the bound $\ell \leq \ell_+(E)$ is obeyed for all $E<0$ orbits.

Remember that the potential $R(r)$ is invariant under the symmetry $(E,\ell)\mapsto (-E,-\ell)$. Since $\ell_0^{(1)}(E,Q,r)=-\ell_0^{(2)}(-E,Q,r)$ or, equivalently, $\ell_0^{(2)}(E,Q,r)=-\ell_0^{(1)}(-E,Q,r)$, a given allowed orbit labeled by $(E,\ell,Q,r)$ will be disallowed for $(-E,-\ell,Q,r)$ and vice versa. For any value of $(r,Q)$ there is therefore a single pair $(E,\ell)$ corresponding to allowed motion. This explains why opposite roots are respectively allowed in the root structures depicted in the diagrams $E>0$ and $E<0$. In particular, since spherical orbits are allowed for $E>0$ they are disallowed for $E<0$. In addition, since $r_*^{(0)}$ as defined after Eq. \eqref{El0} is larger than the radius of the ergosphere, spherical orbits within the ergoregion have necessarily positive angular momentum, $\ell>0$.

\subsection{(near-)NHEK orbits}

The (near-)NHEK limit \cite{Bardeen:1999px,Amsel:2009ev,Dias:2009ex,Bredberg:2009pv} consists in a near-extremal limit $a \mapsto 1$ combined with a near-horizon limit $r \mapsto 1$ and a corotating limit. In the (near-)NHEK limit, all orbits lie entirely within the ergoregion and, therefore, the (near-)NHEK orbits are a subset of the orbits studied earlier in this section. In the NHEK limit the finite energy and radius are 
\begin{equation}
R=(r-1)\lambda^{-2/3}, \qquad \hat E=(2E-\ell)\lambda^{-2/3}, \label{defRE}
\end{equation}
where $\lambda\equiv \sqrt{1-a^2}$.
In the near-NHEK limit the finite energy and radius are 
\begin{equation}
 \hat r=\kappa(r-r_+)/\lambda, \qquad \hat e=\kappa(2E-\ell)/\lambda. \label{defre}
\end{equation}
Therefore, the (near-)NHEK region can be identified as an infinitesimally narrow band around the line $\ell = \ell_+(E) = 2E$ in the last Figures \ref{fig:Ep0} and \ref{fig:Em0} for $a_c^{(6)}<a<1$. Note that negative (near)-NHEK energy orbits $\hat E<0$ or $\hat e<0$ correspond to orbits with $\ell>\ell_+(E)$, which have necessarily $E>0$.
The ISSO angular momentum at extremality is given by 
\begin{equation}
\ell_* \equiv \frac{E_{ISSO}(1)}{\Omega_+(1)}= \frac{2}{\sqrt{3}}\sqrt{1+Q}.     
\end{equation}
As deduced in Proposition 2 of \cite{Compere:2020eat}, the classification of radial motion can be obtained from the classification of equatorial motion since all dependence on $Q$ is through the dependence in $\ell_*$. We will therefore specialize to equatorial motion $Q=0$, $\theta=\pi/2$ in the following without loss of generality. We will reproduce the classification of \cite{Compere:2020eat} for the NHEK case. For the near-NHEK case, we will reproduce the classification of \cite{Compere:2020eat} up to a correction in the range of deflecting orbits, which is in fact larger than previously stated.

\subsubsection{NHEK orbits}

In the NHEK region the radial potential on the equatorial plane is 
\be
v_N(R)=\hat E^2+2\hat E\ell R+\frac{R^2}{4}(3\ell^2-4)
\ee
where $R$ and $\hat E$ were defined in \eqref{defRE}. The limit of $\ell\le \ell_0^{(2)}(E,r)$ becomes
\be 
\ell\le \ell +(\hat{E}+\ell R-\frac{1}{2}\sqrt{4+\ell^2}R)\lambda^{2/3}+o(\lambda^{2/3}).
\ee This equation is satisfied only for  
\bea
\hat{E}+\ell R-\frac{1}{2}\sqrt{4+\ell^2}R\ge 0
\eea which is solved for 
\bea 
\ell\ge \frac{2 \left(\sqrt{\hat{E}^2 +3 R^2}-2 \hat{E} \right)}{3 R}.\label{NHEKbound}
\eea 
 This condition implies $v_N(R)\ge 0$ and $\frac{dT}{d\lambda}>0$ and rules out, in particular, past-oriented motion. As a consequence of our derivation for the general equatorial case in Section \ref{sec:allowedeq}, the bound \eqref{NHEKbound} is the strongest bound imposed on radial motion from the existence of time and azimuthal motion.

The inequality $\ell \leq \ell_+(E)$ is equivalent in the NHEK limit to 
\be
\hat E \geq 0. 
\ee
From the discussion of Section \ref{sec:allowedeq} we infer that when $\hat E>0$, all orbits plunging into the black hole are allowed; when $\hat E=0$, all orbits plunging into the black hole are disallowed for $\ell<0$; when $\hat E<0$, all  orbits plunging into the black hole are disallowed.

The two simple roots of $v_N(R)$ are
\be
R_{1,2}=\frac{-4\hat E \ell\pm 2|\hat E|(\sqrt{\ell^2+4}))}{3\ell^2-4}. 
\ee
When $\ell=\pm\ell_*=\pm\frac{2}{\sqrt{3}}$, the radial potential is
\be
v_N(R)=\hat E^2\pm\frac{4\hat ER}{\sqrt{3}}
\ee
and the simple root is $R_1=\mp\frac{\sqrt{3}\hat E}{4}$.\\
For $\hat E>0$, 
\begin{itemize}
    \item When $\ell>\ell_*$, $R_2<R_1<0$, then $v_N(R)>0$ for $R\ge 0$. The root structure is $\vert+\vert_N$. 
    \item When $\ell<-\ell_*$, $0<R_2<R_1$, then $v_N(R)>0$ for $0<R<R_2$ and $R>R_1$. However, when $R>R_1$, the orbits disobey the bound \eqref{NHEKbound}. The root structure is $\vert+\bullet-{\color{red}\bullet\hspace{2pt}+\hspace{2pt}}\vert_N$. 
    \item When $\ell=-\ell_*$, $R_1>0$, then $v_N(R)>0$ for $0\le R<R_1$. The root structure is $\vert+\bullet-\vert_N$.
    \item When $\ell=\ell_*$, $R_1<0$, then $v_N(R)>0$ for $R\ge0$. The root structure is $\vert+\vert_N$.
    \item When $|\ell|<\ell_*$, $R_1<0<R_2$, then $v_N(R)>0$ for $0<R<R_2$. The root structure is $\vert+\bullet-\vert_N$.
\end{itemize}

For $\hat E<0$,
\begin{itemize}
    \item When $\ell>\ell_*$, $R_1>R_2>0$, then $v_N(R)>0$ for $0<R<R_2$ and $R>R_1$. Only when $R>R_1$, the orbits obey the bound \eqref{NHEKbound}. The root structure is $\vert\hspace{2pt}{\color{red}+\hspace{2pt}\bullet}-\bullet+\vert_N$. 
    \item When $\ell<-\ell_*$, $R_2<R_1<0$, then $v_N(R)>0$ for $R\ge 0$. The root structure is $\vert\hspace{2pt}{\color{red}+}\hspace{2pt}\vert_N$. 
    \item When $\ell=-\ell_*$, $R_1<0$, then $v_N(R)>0$ for $R>0$. The root structure is $\vert\hspace{2pt}{\color{red}+}\hspace{2pt}\vert_N$.
    \item When $\ell=\ell_*$, $R_1>0$, then $v_N(R)>0$ for $0\le R<R_1$. The root structure is $\vert\hspace{2pt}{\color{red}+\hspace{2pt}\bullet}-\vert_N$.
    \item When $|\ell|<\ell_*$, $R_1<0<R_2$, then $v_N(R)>0$ for $0<R<R_2$. The root structure is $\vert\hspace{2pt}{\color{red}+\hspace{2pt}\bullet}-\vert_N$.
\end{itemize}

For $\hat E=0$, the potential is 
\be
v_N(R)=\frac{R^2}{4}(3\ell^2-4)
\ee
It is easy to see the following:
\begin{itemize}
    \item When $|\ell|>\ell_*$, $v_N(R)>0$ for $R>0$. When $\ell>\ell_*$ the root structure is $\;{\vert\hspace{-9pt}\bullet\hspace{-5pt}\bullet+\vert_N}$. When $\ell<-\ell_*$ the root structure is $\;\vert\hspace{-7pt}{\color{red}\bullet\hspace{-4pt}\bullet+\hspace{2pt}}\vert_N$.
    \item When $|\ell|=\ell_*$, $v_N(R)=0$ for any $R$. We denote this special root structure as $\vert\hspace{2pt}0\hspace{2pt}\vert_N$ since the potential is always $0$. When $\ell=\ell_*$ the root structure is $\vert\hspace{2pt}0\hspace{2pt}\vert_N$. When $\ell=-\ell_*$ the root structure is $\vert\hspace{2pt}{\color{red}0}\hspace{2pt}\vert_N$. 
    \item When $|\ell|<\ell_*$, $v_N(R)<0$ for $R>0$. The root structure is $\;\vert\hspace{-7pt}{\color{red}\bullet\hspace{-2pt}\bullet}-\vert_N$.
\end{itemize}
We display the root structure in Figure \ref{fig:nNHEK0}. This classification exactly matches with the classification obtained in \cite{Compere:2020eat} (see their Figure 5).

\subsubsection{near-NHEK orbits}

In the near-NHEK region, the radial potential on the equatorial plane is 
\be
v_n(\hat r)=\frac{1}{4}\hat r(\hat r+2\kappa)(3\ell^2-4)+2\hat e\ell \hat r+(\hat e+\kappa\ell)^2
\ee
where $\hat r$, $\hat e$ were defined in \eqref{defre}. 

The limit of $\ell\le\ell_0^{(2)}$ becomes
\be
\ell\le\ell+\frac{\hat e+\ell \hat r-\frac{1}{2}\sqrt{\hat r(\hat r+2\kappa)(4+\ell^2)}}{\kappa}\lambda+\mathcal{O}(\lambda^2)
\ee
This is satisfied only for 
\be
\ell\ge\frac{2(-2\hat e(\hat r+\kappa)+\sqrt{\hat r(\hat r+2\kappa)(\hat e^2+\mathcal{A})})}{\mathcal{A}}\label{boundnNHEK}
\ee
where 
\be
\mathcal{A}=3\hat r^2+6\hat r\kappa+4\kappa^2
\ee
The condition \eqref{boundnNHEK} implies $v_n(\hat r)\ge 0$ and $\frac{d\hat t}{d\tau}>0$. As a consequence of our derivation for the general equatorial case in Section \ref{sec:allowedeq}, the bound \eqref{boundnNHEK} is the strongest bound imposed on radial motion from the existence of time and azimuthal motion.

The inequality $\ell \leq \ell_+(E)$ is equivalent in the near-NHEK limit to 
\be
\ell + \frac{\hat e}{\kappa} \geq 0. 
\ee
From the discussion of Section \ref{sec:allowedeq} we infer that when $\ell + \frac{\hat e}{\kappa}>0$, all orbits plunging into the black hole are allowed; when $\ell + \frac{\hat e}{\kappa}=0$, all orbits plunging into the black hole are disallowed for $\ell<0$; when $\ell + \frac{\hat e}{\kappa}<0$, all orbits plunging into the black hole are disallowed.

Solving $v_n(\hat r)=0$, the two simple roots are
\be
\hat r_{1,2}=\frac{-4\hat e\ell+\kappa(4-3\ell^2)\pm\sqrt{(4+\ell^2)(4\hat e^2+\kappa^2(4-3\ell^2))}}{3\ell^2-4}. 
\ee
Here $\hat r_{1,2}$ are real when $-\ell_n\le\ell\le\ell_n$, where $\ell_n=\frac{2\sqrt{\hat e^2+\kappa^2}}{\sqrt{3}\kappa}\geq \ell_*$. Note that when $\hat r_2=0$, $\ell=-\frac{\hat e}{\kappa}$.

When $\ell=\pm\ell_n$, there is a double root at
\be
\hat r_*=\mp\kappa(\frac{\kappa\ell_n}{\hat e}\pm 1). 
\ee
It is positive (and therefore relevant) only for $\ell=\ell_n$ and $\hat e<0$ or for $\ell=-\ell_n$ and $\hat e>0$. When $\ell=\pm\ell_*$, the radial potential is 
\be
v_n(\hat r)=(\hat e\pm\frac{2\kappa}{\sqrt{3}})^2\pm\frac{4\hat e\hat r}{\sqrt{3}}
\ee
and the simple root is $\hat r_1=\mp\frac{(2\sqrt{3}\kappa  \pm  3e)^2}{12\sqrt{3}e}$. 

We conclude that when $\hat e>0$,
\begin{itemize}
    \item When $|\ell|>\ell_n$, $v_n(\hat r)>0$ for $\hat r>0$, the root structure is $\vert+\vert_n$.
    \item When $\ell=\ell_n$, $\hat r_*<0$, then $v_n(\hat r)>0$ for $\hat r>0$, the root structure is $\vert+\vert_n$.
    \item When $\ell_*<\ell<\ell_n$, $\hat r_2<\hat r_1<0$, then $v_n(\hat r)>0$ for $\hat r>0$, the root structure is $\vert+\vert_n$.
    \item When $\ell=\ell_*$, $\hat r_1<0$, then $v_n(\hat r)>0$ for $\hat r>0$, the root structure is $\vert+\vert_n$.
    \item When $-\ell_*<\ell<\ell_*$, $\hat r_1<0<\hat r_2$, then $v_n(\hat r)>0$ for $0<\hat r<\hat r_2$, the root structure is $\vert+\bullet-\vert_n$. 
    \item When $\ell=-\ell_*$, $\hat r_1>0$, then $v_n(\hat r)>0$ for $0<\hat r<\hat r_1$, the root structure is $\vert+\bullet-\vert_n$.
    \item When $-\ell_n<\ell<-\ell_*$, $0<\hat r_2<\hat r_1$, then $v_n(\hat r)>0$ for $0<\hat r<\hat r_2$ and $\hat r>\hat r_1$, the root structure is $\vert+\bullet-\bullet+\vert_n$.
    \item When $\ell=-\ell_n$, $\hat r_*>0$, then $v_n(\hat r)\ge 0$ for $\hat r>0$, the root structure is $\vert+\bullet\hspace{-2pt}\bullet+\vert_n$.
\end{itemize}
When $\hat e<0$,
\begin{itemize}
    \item When $|\ell|>\ell_n$, $v_n(\hat r)>0$ for $\hat r>0$, the root structure is $\vert+\vert_n$.
    \item When $\ell=\ell_n$, $\hat r_*>0$, then $v_n(\hat r)\ge 0$ for $\hat r>0$, the root structure is $\vert+\bullet\hspace{-2pt}\bullet+\vert_n$.
    \item When $\ell_*<\ell<\ell_n$, $0<\hat r_2<\hat r_1$, then $v_n(\hat r)>0$ for $0<\hat r<\hat r_2$ and $\hat r>\hat r_1$, the root structure is $\vert+\bullet-\bullet+\vert_n$.
    \item When $\ell=\ell_*$, $\hat r_1>0$, then $v_n(\hat r)>0$ for $0<\hat r<\hat r_1$, the root structure is $\vert+\bullet-\vert_n$.
    \item When $-\ell_*<\ell<\ell_*$, $\hat r_1<0<\hat r_2$, then $v_n(\hat r)>0$ for $0<\hat r<\hat r_2$, the root structure is $\vert+\bullet-\vert_n$.
    \item When $\ell=-\ell_*$, $\hat r_1<0$, then $v_n(\hat r)>0$ for $\hat r>0$, the root structure is $\vert+\vert_n$.
    \item When $-\ell_n<\ell<-\ell_*$, $\hat r_2<\hat r_1<0$, then $v_n(\hat r)>0$ for $\hat r>0$, the root structure is $\vert+\vert_n$.
\end{itemize}

When $\hat e=0$, 
\begin{itemize}
\item When $\ell=\ell_n=\ell_*$, there is no root. The root structure is $\vert+\vert_n$.
    \item When $|\ell|\ge \ell_*$, $v_n(\hat r)>0$ for $\hat r>0$, the root structure is $\vert+\vert_n$.
    \item When $-\ell_*<\ell<\ell_*$, $\hat r_1<0<\hat r_2$, then $v_n(\hat r)>0$ for $0<\hat r<\hat r_2$, the root structure is $\vert+\bullet-\vert_n$.
\end{itemize}

\begin{figure}[!htbp]\vspace{0.5cm}
     \centering 
     \begin{subfigure}[b]{0.45\textwidth}
         \centering
         \includegraphics[width=\textwidth]{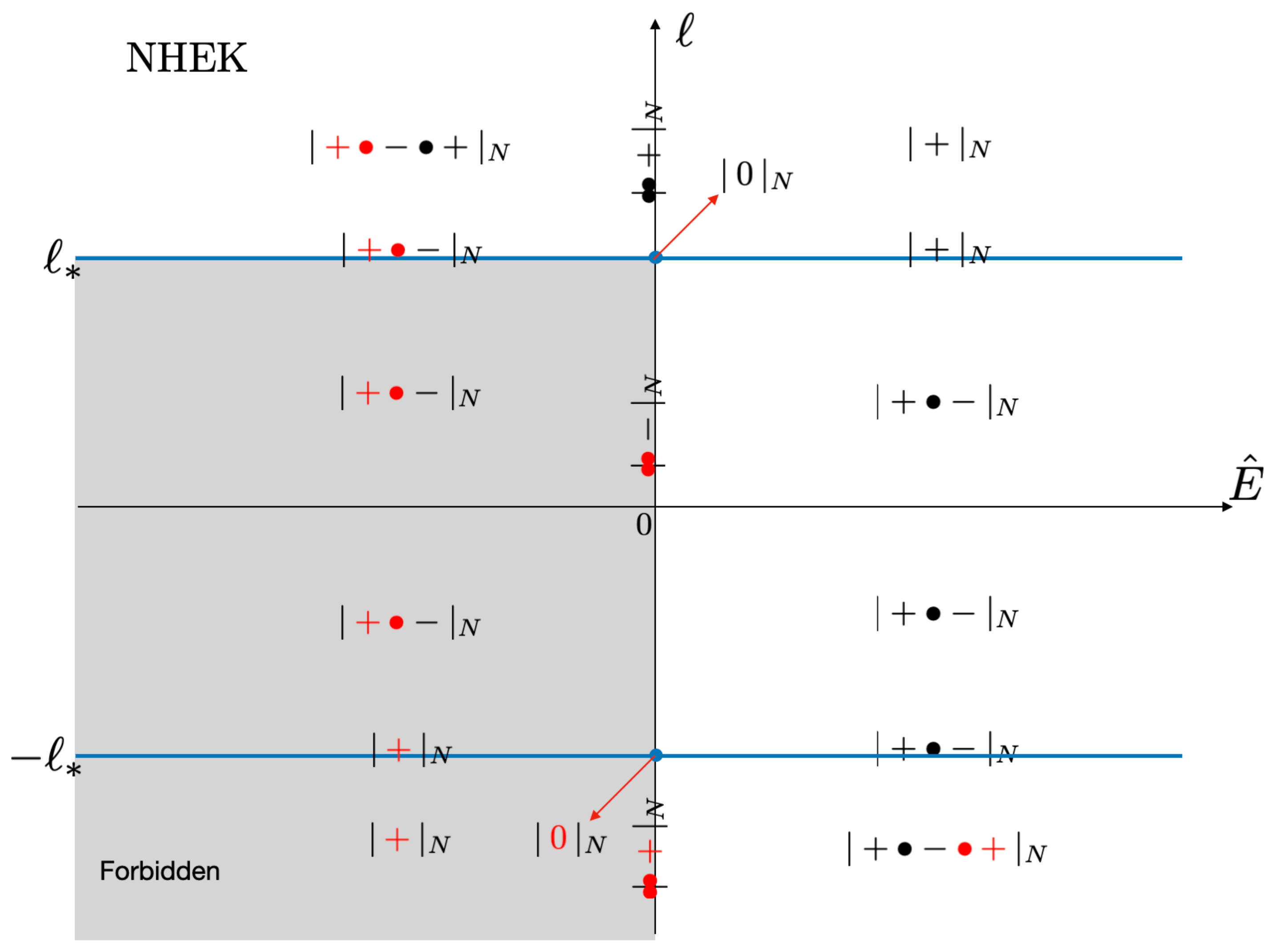}
     \end{subfigure}
     \hfill  \vspace{0.4cm}
     \begin{subfigure}[b]{0.45\textwidth}
         \centering
         \includegraphics[width=\textwidth]{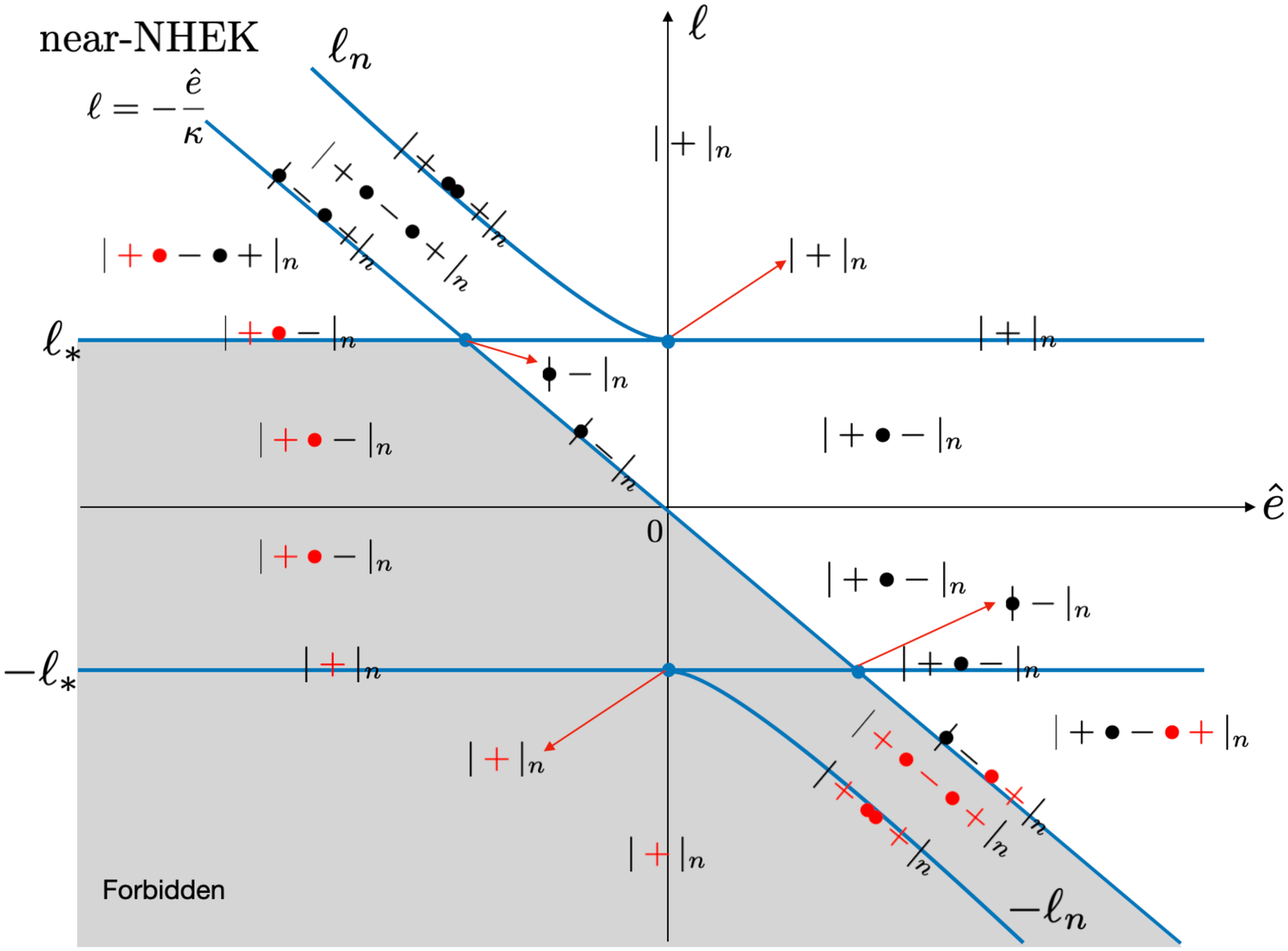}
     \end{subfigure}
     \hfill   \vspace{-0.3cm}
\caption{Classification of radial root structures in NHEK and near-NHEK.}\label{fig:nNHEK0}
\end{figure}

Considering the bound \eqref{boundnNHEK}, we display the root structure including the red color code for disallowed orbits in Figure \ref{fig:nNHEK0}. Comparing with the results of \cite{Compere:2020eat}, we find one forgotten range for deflecting orbits\footnote{The reason for this forgotten range is that the classification of equatorial orbits in \cite{Compere:2017hsi} (consequently used in \cite{Compere:2020eat}) assumed that the parametrization of such deflecting orbits was given in generality as in Eq. (2.17) of \cite{Hadar:2016vmk}, which is only valid in the region $\ell>-\hat e/\kappa$ while a parametrization of larger range exists covering the region $\ell_* < \ell \leq -\hat e/\kappa$ as well as we now showed.}. In fact, deflecting orbits are allowed in the range $\hat e<-\kappa \ell_*$, $\ell_*<\ell\le-\frac{\hat e}{\kappa}$.\footnote{Here is the following correction in the notation of \cite{Compere:2020eat}. In Table 5 of \cite{Compere:2020eat} the range of the class $\text{Deflecting}(e,\ell)$ should be $-\infty < e < -\kappa \sqrt{-\mathcal C}<0$, not $-\kappa \ell < e < -\kappa \sqrt{-\mathcal C}<0$. In Table 7, the upper left red triangle $\ell>\ell_*$, $e<-\kappa \ell$ should not be disallowed but instead is allowed with the class $\text{Deflecting}(e,\ell)$.}

\clearpage

\section{Separatrix between generic radial geodesic classes}
\label{sec3}

The separatrix is defined as the codimension 1 region in phase space such that the root structure contains a double root. Since negative energy orbits admit no double root, we assume $E>0$ in this section. In the following, we will show that the separatrix can be entirely described in terms of a single quartic that appeared previously in related contexts in \cite{OShaughnessy:2002tbu,Glampedakis:2002ya,Glampedakis:2005hs,Levin:2008yp},
\bea\label{quartic}
p^2(p-2e-6)^2-2a^2(1+e)p((3-e)p+14+2e^2)+a^4(3-e)^2(1+e)^2=0.
\eea 
The interpretation of $p$ and $e$ will differ depending upon the region of the phase space considered. As we will discuss, the separatrix is the union of three distinct regions respectively obtained when (1) the pericenter of bound motion becomes a double root (in the region $E<1$), (2) the eccentricity of bound motion becomes zero (in the region $E<1$), (3) the turning point of unbound motion becomes a double root (in the region $E \geq 1$). Only in region (1), $p$ is the standard semilatus rectus and $e$ the eccentricity of the  bound geodesics existing in that region.  
 
\subsection{Bounded orbits: Pericenter becoming a double root}

Bounded orbits occur when the root structure contains the structure $\bullet + \bullet$, where the bullets indicate the turning points, namely, the pericenter $r_p$ and apocenter $r_a$ that radially bound the orbit. Given our classification of bounded orbits, we can now simply read off Figure \ref{fig:classEminus1} to deduce in the phase space spanned by the parameters $(E,\ell,Q)$ which are the root structures that bound the root structures that contain the sequence $\bullet + \bullet$ corresponding to bounded orbits. Bounded orbits around a Kerr black hole only occur in the three-dimensional region bounded as
\begin{equation}
  \text{max}(0,  Q^{u}(E,\ell)) \leq Q \leq Q^{s}(E,\ell)
\end{equation}
which is defined for $E_{ISCO^+} \leq E < 1$. The phase space boundary of bound motion, which is the part of the separatrix for $E<1$, is therefore the union of the locations $Q=\text{max}(0,  Q^{u}(E,\ell))$ and $Q=Q^{s}(E,\ell)$, which were 
implicitly defined in \eqref{Q1l}--\eqref{Q2l}. 

We will discuss in this section the lower separatrix $Q=\text{max}(0,  Q^{u}(E,\ell))$, while the upper separatrix will be discussed in Section \ref{uppersep}. The radial phase angle $\psi$, eccentricity $e$, and semilatus rectum $p$ are defined by parametrizing bounded orbits as quasi-Keplerian orbits, 
\be
r=\frac{p}{1+e\cos\psi} , \qquad r_p\le r\le r_a,
\ee 
with $0 \leq e<1$. The pericenter and apocenter radii are defined as
\bea
r_p\equiv \frac{p}{1+e},\quad r_a \equiv \frac{p}{1-e}.
\eea
The condition $r_p > r_+$ translates into the range of $p$, \begin{equation}
(1+e)(1+\sqrt{1-a^2}) < p < \infty  .  \label{rangep}
\end{equation}
The radial potential vanishes exactly at the turning points $r_a$ and $r_p$, 
\bea
R(r_p)=R(r_a)=0.\label{sp1}
\eea
In order to write the lower separatrix $Q=\text{max}(0,  Q^{u}(E,\ell))$ in simplest terms, we will use the parameters $(e,p,Q)$. The bound $0 \leq Q < \infty$ is then trivially enforced. (We can think of the inclination $\cos\iota = \ell/\sqrt{\ell^2+Q^2}$ as an auxiliary parameter.) At the location $Q=Q^{u}(E,\ell)$ in phase space, the root structure $\vert + \bullet - \bullet + \bullet - \rangle$ becomes $\vert + \bullet \hspace{-5pt}\bullet + \bullet - \rangle$. The pericenter therefore becomes a double root,  
\bea
R'(r_p)=0.\label{sp2}
\eea 
The three equations \eqref{sp1} and \eqref{sp2} lead to a unique solution for $(Q,E,\ell)$ in terms of the parameters $(p,e)$. Indeed, the equations $R(r_p)=R'(r_p)=0$ are equivalent to 
\bea
Q=Q_b(E,r_*=r_p),\quad \ell=\ell_b(E,r_*=r_p)\label{Qlhere}
\eea 
as shown in Section \ref{sec:nonmarginal}, see Eq. \eqref{Qrs}. Upon substitution of $Q$ and $\ell$ in $R(r_a)=0$, and after some manipulations involving taking a square, we find a quadratic equation for $E^2$, $16 A (E^2)^2 - 8B E^2 + C^2 = 0$ where 
\bea
A(e,p) &\!\!\equiv \!\!&p^3 \Big\{ -a^2 (1 - e)^2 (1 + e)^3 + 
      p \Big( (1+e)^2(3-4e+2e^2)-(1+e)(3-2e+e^2)p+p^2 \Big) \Big\} ; \nonumber\\
B(e,p) &\!\!\equiv \!\!&  p^2 a^2(1-e)^2(1+e)^3\Big( (1+e)(3-2e)-3p \Big) +p^3 \Big( -2(1-e)(1+e)^3(5-5e+2e^2) \nonumber \\
&& +(3-e)(1+e)^2(7-7e+2e^2)p -(1+e)(15-10e+3e^2)p^2+ 4p^3 \Big);\label{AB}\\
C(e,p) &\!\!\equiv \!\!& -a^2 (1 - e)^2 (1 + e)^3 + 
   p \Big(2(3-e)(1-e)(1+e)^2 - (3-e)^2(1+e)p + 4p^2 \Big).  \nonumber
\eea
The discriminant
\bea
\Delta(e,p)\equiv B^2(e,p)-A(e,p) C^2(e,p) &=& (1-e)^4(1+e)^5p^3\Big( a^2(1+e)^2+p(-2-2e+p)\Big)^2  \nonumber \\
&& \times \Big( a^2(1-e)^2+p(-2+2e+p)\Big)\label{Delta}
\eea
is always positive in the range \eqref{rangep}. Only the solution 
\begin{equation}
E= E^u(e,p)\equiv  \frac{1}{2}\sqrt{\frac{B(e,p)+ \sqrt{\Delta(e,p) }}{A(e,p)} }
\end{equation} 
is physical because the other solution of the quadratic equation does not obey the equation $R(r_a)=0$. It only appeared from taking a square to obtain the quadratic equation. The solution is therefore unique. Note that for Schwarzschild, $a=0$, we have correctly $E^u(0,6)=\frac{2\sqrt{2}}{3}$. Upon substituting $E$ in \eqref{Qlhere}, we also obtain explicitly 
\begin{eqnarray}
Q &=& Q^{u}(e,p) \equiv Q_b(E^u(e,p),\frac{p}{1+e}), \label{Q1}\\
\ell &=& \ell^{u}(e,p) \equiv \ell_b(E^u(e,p),\frac{p}{1+e}). \label{l1}
\end{eqnarray}

The bound $Q^{u}(e,p)\ge0$ is obeyed if and only if $p$ is restricted to a finite interval, 
\bea
p_2(e;a)\le p\le p_1(e;a). \label{separatrixp}
\eea 
The upper and lower bounds are obtained at $Q^{u}(e,p)=0$, which corresponds to equatorial orbits. Now, the roots of $Q_{\text{b}}(E,r_*)$ only occur at $E^{(1)}_{\text{b}}$ or $E^{(2)}_{\text{b}}$ given in \eqref{oldE1}--\eqref{oldE2}. We deduce that the functions $p_i(e;a)$, $i=1,2$ are the only solutions outside the horizon of 
\bea
E^u(e,p_i(e)) = E_{\text{b}}^{(i)}(\frac{p_i(e)}{1+e}), \qquad i=1,2,
\eea 
with the dependence in $a$ understood. (One can easily disentangle the cases $i=1$ from $i=2$ by studying a special case, see below.) These two expressions involve nested square roots. After taking twice the square in an appropriate fashion, one can reduce these equalities in terms of two polynomials in $p$. After factoring out polynomials with unphysical roots, one is left in both cases $i=1,2$ with a single fourth-order polynomial, which is exactly given by Eq. \eqref{quartic}. 

There are exactly two roots outside the horizon, which are precisely $p_1(e)$ and $p_2(e)$. This fourth-order polynomial in $p$ precisely agrees with Eq. (22) of \cite{Levin:2008yp} which was obtained earlier in \cite{OShaughnessy:2002tbu,Glampedakis:2002ya,Glampedakis:2005hs}. The reason why the same fourth-order polynomial is found is simply that $E_{\text{b}}^{(1)}$ and $E_{\text{b}}^{(2)}$ are related by a flip of $a$, while $E^u(e,p)$ and the polynomial \eqref{quartic} are invariant under $a \mapsto -a$.  The two functions $p_1(e)$ and $p_2(e)$, with $0 \leq e < 1$, are plotted on Figure \ref{perelation}. This completely specifies this branch of the separatrix in its simplest form. Special cases of these  functions are the following: 
\begin{figure}
    \centering
    \includegraphics[width=0.8\textwidth]{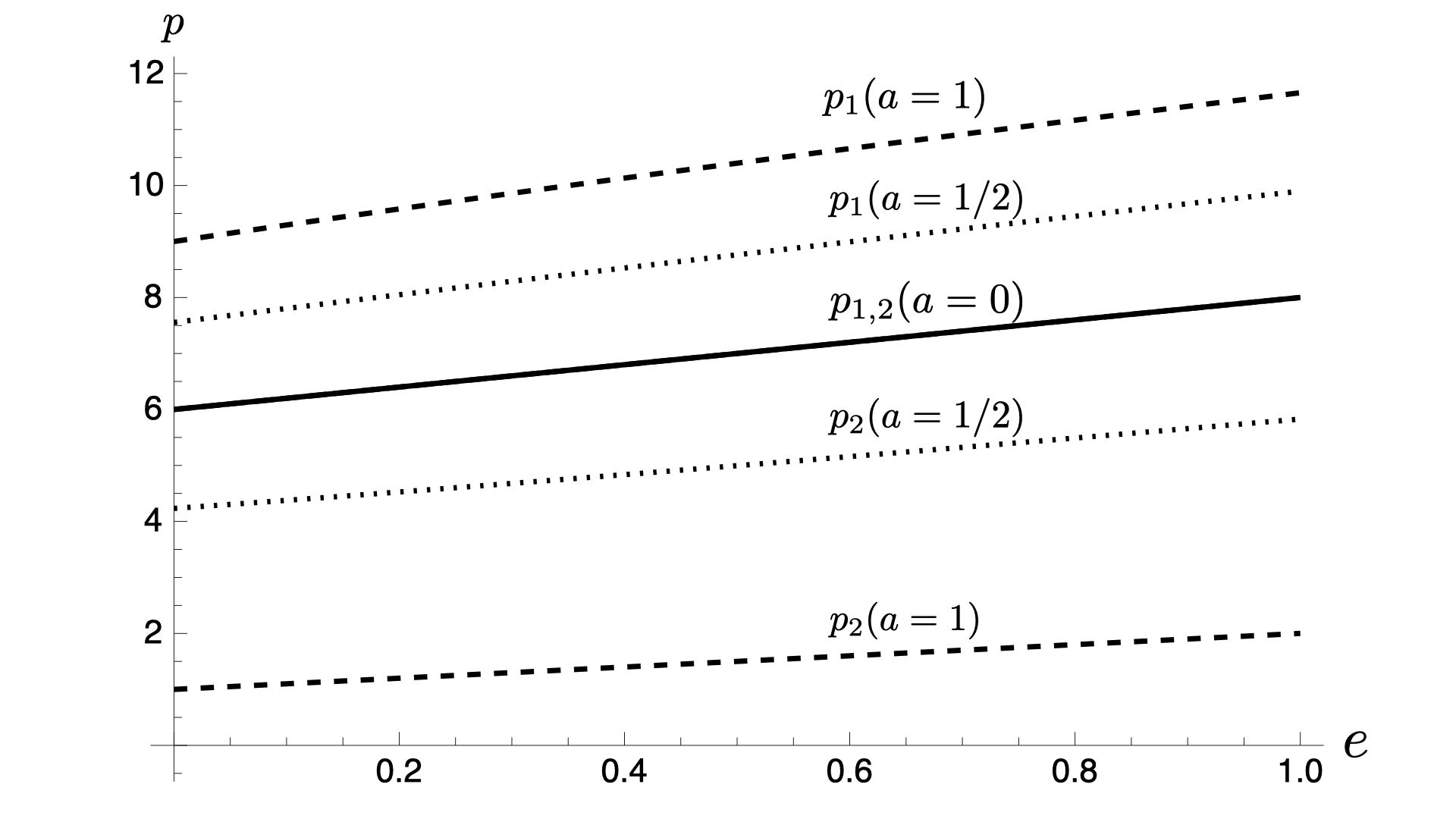}
    \caption{Part of the separatrix with root structure $\vert + \bullet \hspace{-3pt}\bullet + \bullet - \rangle$ occurring for $E_{ISCO^+} \leq E<1$. The corresponding bounded orbits are infinitely  whirling around the limiting spherical orbit. The parameters $(e,p)$ obey the bound $Q \ge 0$  in the region $p_2(e ;a)\le p\le p_1(e;a)$ for $0 \leq e<1$. The allowed region expands with increasing spin $a$.}
    \label{perelation}
\end{figure} 

\begin{enumerate}
\item For Schwarzschild, $a=0$, $p_1(e;0)=p_2(e;0)=6+2e$ and the finite region \eqref{separatrixp} degenerates into a line. In this case, 
    \bea
    Q^{u}(e,p_{1,2};a=0)&=&\frac{4 (e+3)^2}{(3-e) (e+1)},  \\
 \ell^{u}(e,p_{1,2};a=0)&= &0,\\
 E^u(e,p_{1,2};a=0)&=&\sqrt{\frac{8}{9-e^2}} . 
    \eea 
Note that for the Schwarzschild black hole, the geodesics only depend upon the combination $k=Q+\ell^2$. Upon performing a $SO(3)$ rotation, one can reach the equatorial plane with $Q=0$ and the angular momentum becomes
    \bea
    \ell^{u}(e,p;a=0)=\frac{2(3+e)}{\sqrt{(3-e)(1+e)}}.
    \eea 
    \item At the edges of the domain \eqref{separatrixp}: for circular orbits without eccentricity, $e=0$, we find the ISCO, 
    \bea
    p_1(e=0,a)&=&r_{ISCO^-}(a),\\
    p_2(e=0,a)&=&r_{ISCO^+}(a).
    \eea 
In this case, we recover the standard values 
    \bea
    Q^{u}(e=0,p_2;a)&=& 0,  \\
 \ell^{u}(e=0,p_2;a)&= &  \frac{p_2^2-2a\sqrt{p_2}+a^2}{\sqrt{2ap_2^{3/2}+(p_2-3)p_2^2}}, \\
 E^u(e=0,p_2;a)&=&\frac{\sqrt{p_2}(p_2-2)+a}{\sqrt{2ap_2^{3/2}+(p_2-3)p_2^2}} ,
    \eea 
    and
    \bea
    Q^{u}(e=0,p_1;a)&=& 0,  \\
 \ell^{u}(e=0,p_1;a)&= &  \frac{-p_1^2-2a\sqrt{p_1}-a^2}{\sqrt{-2ap_1^{3/2}+(p_1-3)p_1^2}}, \\
 E^u(e=0,p_1;a)&=& \frac{\sqrt{p_1}(p_1-2)-a}{\sqrt{-2ap_1^{3/2}+(p_1-3)p_1^2}} .
    \eea  
Inside the domain \eqref{separatrixp}: for spherical orbits without eccentricity, $e=0$, we have the generic expressions \eqref{Q1}--\eqref{l1}, while 
     \begin{equation}
       E^{u}(e=0,p;a)=\frac{1}{2}\sqrt{\frac{\sqrt{p(a^2+p(p-2))^3}+p(p-1)(p(10+p(4p-11))-3a^2)}{p^2(p(p(p-3)+3)-a^2)}}.
     \end{equation}
    
 \item In the parabolic limit $e\to 1$, we have at the edges of the domain
    \begin{equation}\label{p1p2}\begin{split}
    p_1(e=1 ; a)&=2 \left(a+2 \sqrt{1+a}+2\right)=2r_c^{(1)},\\
    p_2(e=1 ;a)&=2\left(-a+2 \sqrt{1-a}+2\right)=2r_c^{(2)},
    \end{split}
    \end{equation}
where $r_c^{(i)}$ were defined in \eqref{rminmax}.  In this case, 
    \bea
   Q^{u}(e=1,p_ 2;a)&=&0,  \\
 \ell^{u}(e=1,p_ 2;a)&= & 2(1+\sqrt{1-a}), \\
 E^{u}(e=1,p_2;a)&=&1 ,
    \eea 
and 
  \bea
     Q^{u}(e=1,p_1;a)&=&0,  \\
 \ell^{u}(e=1,p_1;a)&= &-2(1+\sqrt{1+a}) ,\\
    E^{u}(e=1,p_1;a)&=&1.
    \eea 
In the domain, we have the expressions
    \bea
     \!\!  Q^{u}(e=1,p;a)&\!\!=\!\!&\frac{p^2 (8 a^2 - 16 p + 4 a^2 p + 6 p^2 -  p^3 + 
   2 \sqrt{2} \sqrt{p} (4 a^2 + (-4 + p) p))}{8 a^2 (-2 + p)^2},  \\
\!\! \ell^{u}(e=1,p;a)&\!\!=\!\! &\frac{2p^2-8a^2-\sqrt{2p}(4a^2+p(p-4))}{4a(p-2)}, \\
\!\! E^{u}(e=1,p;a)&\!\!=\!\!& 1 .
    \eea

    \item In the extremal case, $a=1$, we have exactly
        \bea
    p_1(e ;a= 1)&=& 5+e+4\sqrt{1+e},\\
    p_2(e ;a=1)&=&1+e.
    \eea 
    For $p=p_1(e;a=1)$, we find
    \bea
     Q^{u}(e,p_1;a=1)&=& 0 \\
 \ell^{u}(e,p_1;a=1)&= &-\frac{2 (11 (1 + \sqrt{1 + e}) + e (8 + e + 3 \sqrt{1 + e}))}{\sqrt{(3 - e) (1 + e)} (3 + e + 3 \sqrt{1 + e})}, \\
 E^{u}(e,p_1;a=1)&=& \frac{3 - e + 2 \sqrt{1 + e}}{\sqrt{3 - e} (2 + \sqrt{1 + e})}. 
    \eea
    For $p=p_2(e;a=1)$, one has $r_p=1,r_a=\frac{1+e}{1-e}$, and the solution for $(Q,\ell,E)$ is not uniquely determined in terms of $e$. Solving instead \eqref{sp1}--\eqref{sp2}, one finds the two-parameter family
\bea
    Q^{u}(e,p_2;a=1)&=&\frac{(1+e)((3-e)\ell^2-4(1+e))}{4(1-e)^2},\\
  E^{u}(e,p_2;a=1)&=&\frac{1}{2}\ell,
\eea     
which is parametrized by $(e,\ell)$. Positivity of $Q$ requires $\ell \geq 2\sqrt{1+e}/\sqrt{3-e}$. Since $r_p=1$, the pericenter lies in the NHEK region. At zero eccentricity $e=0$, the apocenter also lies in the NHEK region and Carter's constant reduces to the value for the NHEK separatrix $Q=3\ell^2/4-1$ \cite{Compere:2020eat} since the entire geodesics lies in the NHEK region. At nonzero eccentricity, the orbit is partly in the NHEK region and partly in the exterior extremal Kerr region. When $ \ell < 2/\sqrt{2-e}$, we have $Q^{u}(e,p_2;a=1) < 3\ell^2/4-1$ or $\ell>\ell_*$ with $\ell_*=\frac{2}{\sqrt{3}}\sqrt{1+Q}$. Such orbits can match with the $\text{Deflecting}(E,\ell)$ NHEK orbits as denoted in \cite{Compere:2020eat}, during their motion within the NHEK region.

        \item Finally, note that the linear approximation in $e$ is around $5\%$ accurate, 
    \bea    
    p_1(e;a)&\approx&r_{ISCO^-}+(2r_c^{(1)}-r_{ISCO^-})e,\\
    p_2(e; a)&\approx& r_{ISCO^+}+(2r_c^{(2)}-r_{ISCO^+})e.
    \eea 
\end{enumerate}
In summary, the lower separatrix is spanned by $(e,p)$ in the range 
\bea
0 \leq e < 1,\qquad p_2(e;a)\le p\le p_1(e;a),\label{separatrixpSUM}
\eea 
where $p_i(e;a)$ are the two solutions to the quartic \eqref{quartic} outside the horizon. The explicit manifestly real values of $(Q,\ell,E)$ are given in terms of $(e,p)$ by 
\begin{equation}\label{solQLE}
\begin{split}
Q &=Q^{\text{u}}(e,p)\equiv Q_{\text{b}}(E^{u}(e,p),\frac{p}{1+e}), \\
\ell &= \ell^{\text{u}}(e,p)\equiv  \ell_{\text{b}}(E^{u}(e,p),\frac{p}{1+e}), \\
E &=E^{\text{u}}(e,p) \equiv \frac{1}{2}\sqrt{\frac{B+ \sqrt{\Delta }}{A} },
\end{split}
\end{equation} 
where $Q_{\text{b}}$, $\ell_{\text{b}}$ are defined in \eqref{Qrs}--\eqref{lrs}, and $A,B,\Delta$ are defined in \eqref{AB}--\eqref{Delta}. Special cases are shown above. This provides a more explicit  parametrization of this branch of the separatrix than in terms of the twelveth-order polynomial defined in \cite{Stein:2019buj}.

\subsection{Bounded orbits: Zero eccentricity limit}
\label{uppersep}

Let us now discuss the separatrix $Q=Q^{s}(E,\ell)$. At this location in phase space, the root structure $\vert + \bullet - \bullet + \bullet \hspace{2pt} - \rangle$ turns into $\vert + \bullet - \bullet \hspace{-5pt} \bullet - \rangle$. The double root gives spherical orbits. The pericenter and apocenter coincide, the eccentricity goes to 0, $e=0$, $r_a=r_p=p$. The parametrization in terms of $(e,p)$ therefore degenerates and becomes inappropriate. In the parametrization $(E,p)$, the upper limit of the separatrix is simply given by  $Q=Q_{\text{b}}(E,p)$, $\ell= \ell_{\text{b}}(E,p)$, which are defined in \eqref{Qrs}-\eqref{lrs}. 

The root system $\vert + \bullet - \bullet \hspace{-5pt} \bullet - \rangle$ admits a simple root and the larger double root that labels the radial location of the spherical orbits. We can therefore parametrize the three roots as
\be
r_{\text{double}}=\frac{p}{1+e},\quad r_{\text{simple}}=\frac{p}{1-e},\quad -1 < e \leq 0,\label{upper1}
\ee 
where $(p,e)$ are new parameters. The new parameter $e$ is now exactly minus the relative distance between the simple and the double root, $e = - (r_\text{double}-r_\text{simple})/(r_\text{double}+r_\text{simple})$. The roots degenerates to a triple root at $e=0$. The potential should satisfy 
\bea
R(r_{\text{double}})=R'(r_{\text{double}})=0,\quad R(r_{\text{simple}})=0.
\eea 
The explicit solution of these equations is exactly \eqref{solQLE} as before with the new interpretation of $(p,e)$ and superscripts $\mbox{}^\text{u}$ to $\mbox{}^\text{s}$ since the spherical orbits are now stable. Explicitly, 
\begin{equation}\label{solQLEs}
\begin{split}
Q &=Q^{\text{s}}(e,p)\equiv Q_{\text{b}}(E^{s}(e,p),\frac{p}{1+e}), \\
\ell &= \ell^{\text{s}}(e,p)\equiv  \ell_{\text{b}}(E^{s}(e,p),\frac{p}{1+e}), \\
E &=E^{\text{s}}(e,p) \equiv \frac{1}{2}\sqrt{\frac{B+ \sqrt{\Delta }}{A} },
\end{split}
\end{equation}
for $-1 < e  \leq 0$. We have that $\Delta >0$ and $E$ is real. The separatrix obeying $Q \geq 0$ is therefore given by the range \bea
-1 < e \leq  0,\qquad p_2(e;a)\le p\le p_1(e;a),\label{separatrixpNEG}
\eea 
where $p_i(e;a)$ are the two real solutions to the quartic \eqref{quartic}. The same quartic therefore controls this part of the separatrix. This is shown in Figure \ref{upper}. We note the following special values: 
\begin{figure}
    \centering
    \includegraphics[width=0.8\textwidth]{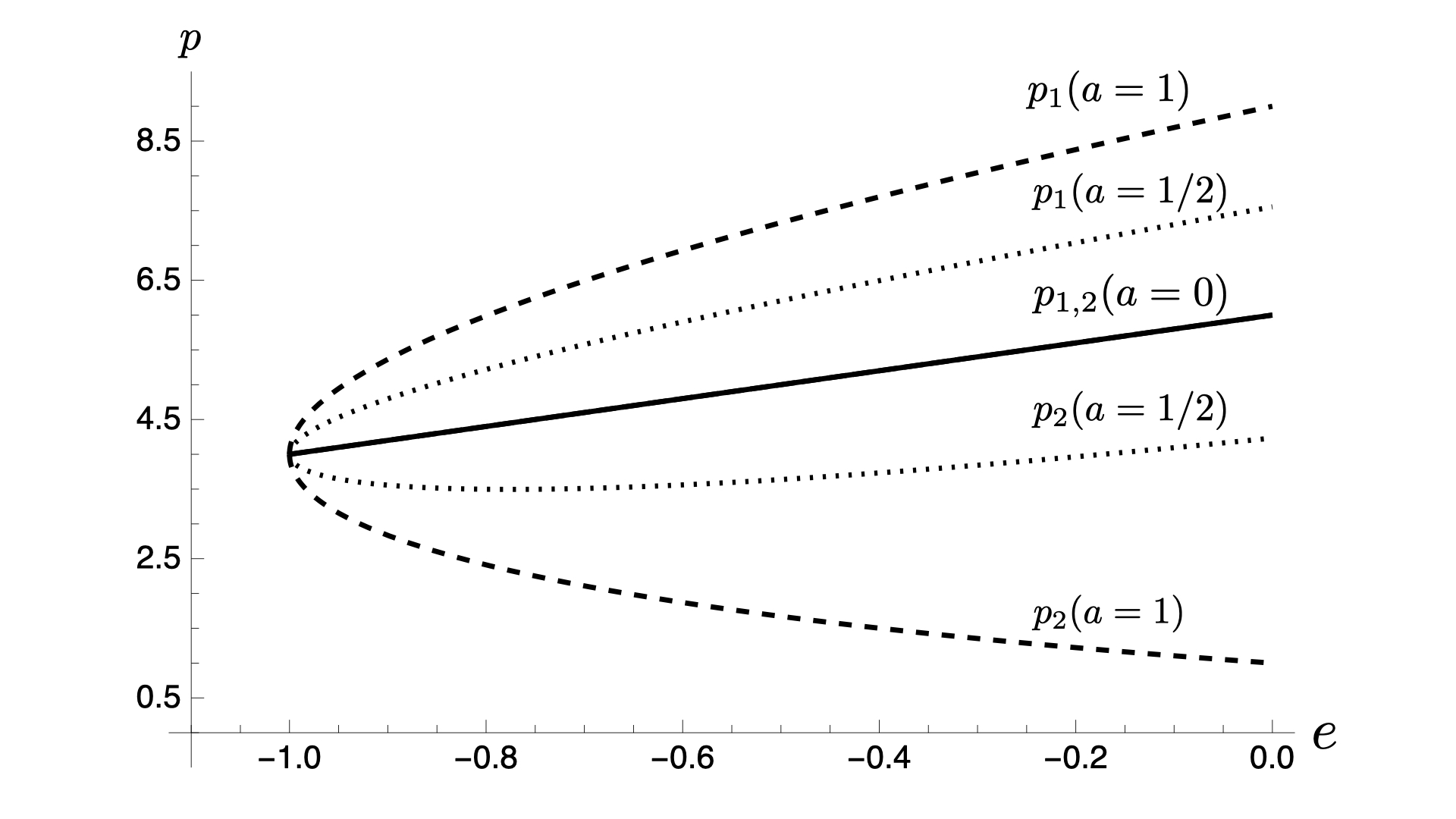}
    \caption{Part of the separatrix containing a stable double root: $\vert + \bullet - \bullet \hspace{-5pt} \bullet - \rangle$. This occurs in the range $E_{ISCO^+} \leq  E < 1$. The bound $Q\ge 0$ is obeyed in the region $p_2\le p\le p_1$ for $-1 < e \leq  0$. The allowed region expands with the spin $a$ of the black hole.}
    \label{upper}
\end{figure}

\begin{itemize}
    \item For $a=0$, the shaded region degenerates to a line 
    \be
    p_1(e;a=0)=p_2(e;a=0)=6+2e.
    \ee 
    \item For $a=1$, we have exactly 
    \bea
    p_1(e;a=1)&=&5+e+4\sqrt{1+e},\\
    p_2(e;a=1)&=&5+e-4\sqrt{1+e}.
    \eea 
    When $e=0$ the orbit lies in the NHEK region, and we have $E=1/\sqrt{2}$, $\ell=\sqrt{2}$, $Q=1/2$. Since $\ell=\ell_*\equiv \frac{2}{\sqrt{3}}\sqrt{1+Q}$, such orbits are critical in the sense of \cite{Compere:2020eat}.  
    \item For $e=0$, we have 
    \bea
    p_1(e=0;a)&=&r_{ISCO-},\\
    p_2(e=0;a)&=&r_{ISCO+}.
    \eea 
    \item For $e\to -1$, we have 
    \bea
   p_1(e=-1;a)=p_2(e=-1;a)=4,
    \eea which is independent of $a$. In that limit, $E \rightarrow 1$, $\ell \rightarrow -\frac{2}{\sqrt{1+e}}+O(1)$ and $Q \rightarrow \frac{16}{a\sqrt{1+e}}+O(1)$. 
\end{itemize}

\subsection{Unbounded orbits: Turning point becoming a double root}

Unbounded motion occurs for $E\ge 1$. The phase space is depicted in Figure \ref{fig:Egtr1}. Generic unbound motion has either no turning point (which corresponds to the root structure $\vert + \rangle$) or one turning point (which corresponds to the root structure ${\vert + \bullet - \bullet  \hspace{2pt}+  \rangle}$). The separatrix between these two classes of orbits is given by the root structure $\vert + \bullet \hspace{-6pt}\bullet+\rangle$. In this case, there must be a double root and one real root which is less than 0. We parametrize the three roots as 
\bea
r_{\text{double}}&=&\frac{p}{1+e},\quad r_{\text{negative}}=\frac{p}{1-e},\quad e>1. 
\eea 
The interpretation of $e$ is now the inverse of the relative distance between the absolute value of the negative root and the double root: $e=(\vert r_{\text{negative}} \vert+r_{\text{double}})/(\vert r_{\text{negative}} \vert-r_{\text{double}})$. The deflecting point is located at $r=r_{\text{double}}$. At the deflecting point, we have 
\be
R(r_{\text{double}})=R'(r_{\text{double}})=0,
\ee 
while we also have 
\be
R(r_{\text{negative}})=0.
\ee 
The solution to these equations is exactly \eqref{solQLE} but with now $e>1$. One can check that $\Delta>0$ and $E$ is real. The condition $Q \geq 0$ amounts to the bound for $p$,
\be
p_2(e ; a)\le p\le p_1(e ;a),\quad e>1,
\ee     
where $p_1$ and $p_2$ are the two solutions to the same quartic equation \eqref{quartic}, as derived previously.

\begin{figure}[!htbp]\vspace{-1cm}
    \centering
    \includegraphics[width=0.8\textwidth]{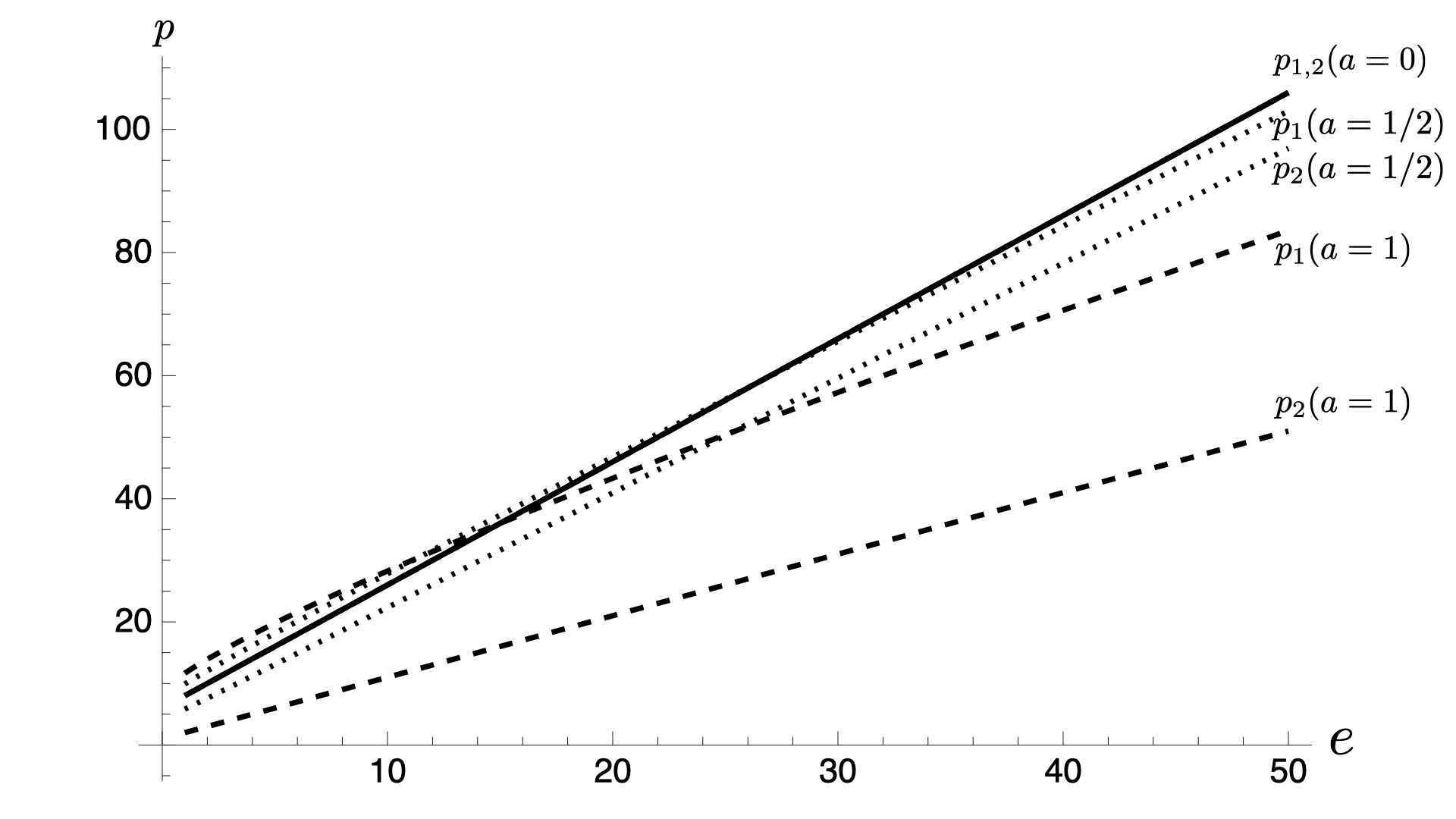}
    \caption{Part of the separatrix with root structure $\vert + \bullet \hspace{-6pt}\bullet+\rangle$ occurring for $E \geq 1$. The parameters $(e,p)$ range in the intervals $e > 1$, $p_2(e ; a)\le p\le p_1(e;a)$. The region grows with increasing spin $a$. In the large $e$ limit, $p_{1,2}(e ;a) \rightarrow r_+ e + O(e^0)$.}
    \label{perelation2}
\end{figure}

\begin{figure}[!htbp]
    \centering
    \includegraphics[width=0.9\textwidth]{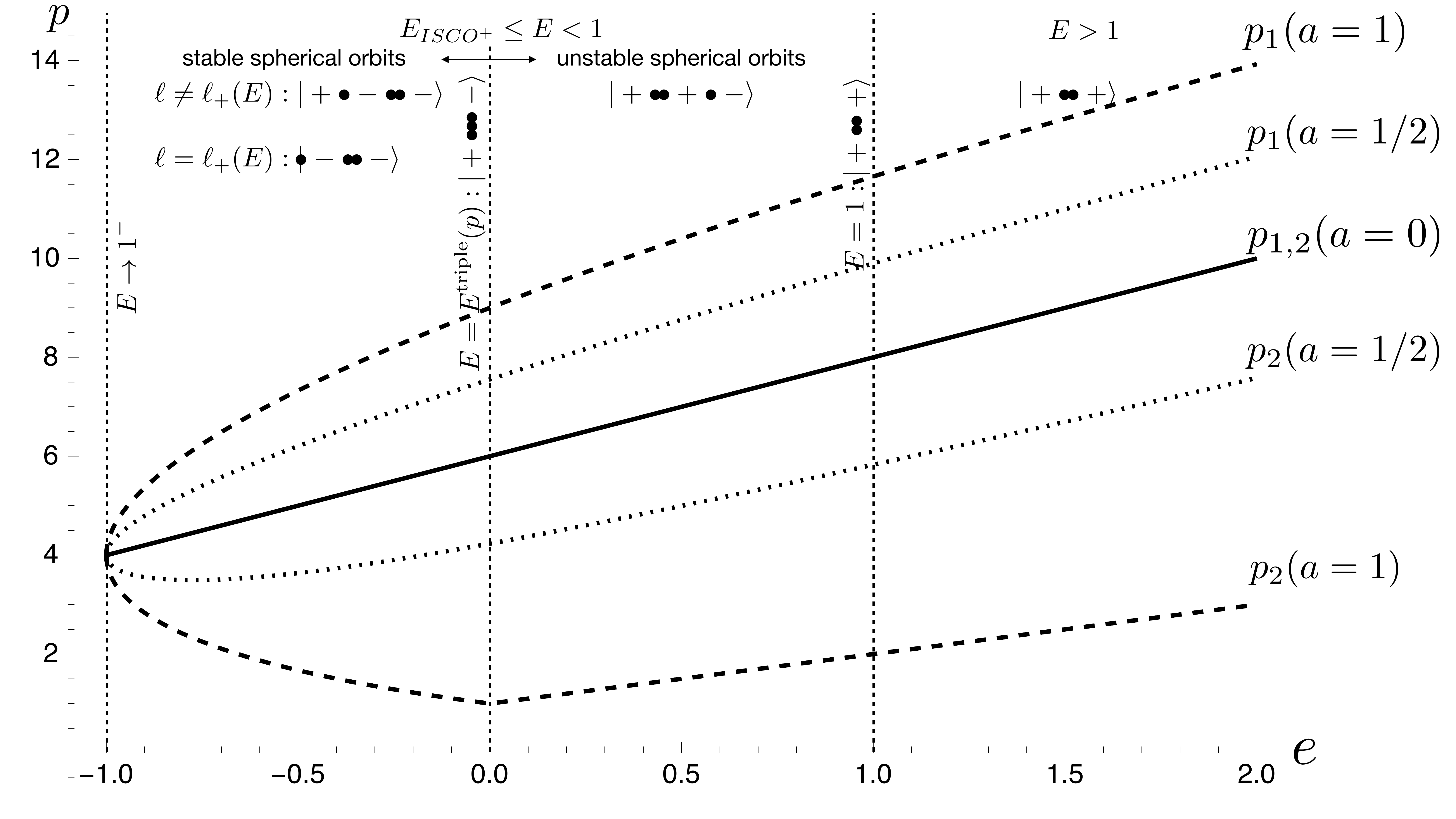}
    \caption{Complete Kerr separatrix where a double root occurs. It is the union of the root structures $\vert + \bullet-\bullet \hspace{-4pt}\bullet- \rangle$, $\vert +\bullet \hspace{-4pt}\bullet + \bullet- \rangle$,  
 $\vert +\bullet \hspace{-4pt}\bullet \hspace{-4pt}\bullet \hspace{2pt} - \rangle$, ${\vert \hspace{-7pt}\bullet - \bullet \hspace{-4pt}\bullet \hspace{2pt} - \rangle}$, and
 $\vert +\bullet \hspace{-4pt}\bullet + \rangle$. The allowed range is $-1<e<\infty$ and $p_2(e ;a ) \leq p \leq p_1(e ;a)$, where $p_1$, $p_2$ are the two real solutions of the quartic \eqref{quartic}. The cusp for the lower bound $p_2$ occurs in the near-horizon extremal Kerr (NHEK) region $a=r=1$, which can be resolved after introducing the NHEK radius $(r-1)/\sqrt{1-a^2}$. It leads to the NHEK separatrix, which requires a distinct treatment \cite{Compere:2020eat}. The large $e$ behavior is depicted in Figure \ref{perelation2}.}\label{summary}\vspace{-1cm}
\end{figure}

When $e\to\infty$, the two bounds approach $p_{1,2}(e ;a) \rightarrow r_+ e + O(e^0)$ and the orbit approaches to $r \rightarrow r_+$. In the parabolic limit $e \rightarrow 1$, one recovers the values \eqref{p1p2}. The values $p_{1,2}(e;a)$ are depicted in Figure \ref{perelation2}. The summary of the three distinct regions of the separatrix is given in Figure \ref{summary}. 

\clearpage

\section{Conclusion}
\label{sec:ccl}

We performed the taxonomy of inequivalent root structures of the quartic potential determining the radial Kerr geodesic motion using  the reality of polar motion as constraints. Distinct generic root structures are separated by codimension 1 boundaries in phase space that are of three types: (1) the complete separatrix, i.e., the root structures containing a double root whose geodesic classes contain, correspondingly, spherical orbits and ``whirling orbits'' that asymptotically approach or leave spherical orbits, (2) the region where one root coincides with the outer horizon and (3) the marginal case where the energy is $\vert E\vert =1$ such that the order of the radial potential degenerates to three since one root disappears. The classification was achieved by establishing the phase space for these degenerate cases, taking into account the bound on Carter's constant arising from the reality of polar motion. We further established which radial motion is allowed due to the constraints in the ergoregion on the existence of time and azimuthal motion.

The result reads as follows. For $0 \leq  E <1$, the eight inequivalent root structures are summarized in Table \ref{table:GeoClasses2} and their phase space in $(E,Q,\ell)$ basis is given in Figure \ref{fig:classEminus1}. For $ E \geq 1$, the four inequivalent root structures are summarized in Table \ref{table:GeoClasses1} and the corresponding phase diagram in $(E,Q,\ell)$ basis is given in Figures \ref{marginalRoots} and \ref{fig:Egtr1}. The large $E$ limit reproduces the null case \cite{Gralla:2019ceu}. The near-horizon near-extremal limit reproduces the (near-)NHEK classification of \cite{Compere:2020eat} up to one correction (the range of existence of near-NHEK deflecting orbits has to be extended), see Figure \ref{fig:nNHEK0}. The resulting 11 distinct classes of radial motion for $E\geq 0$ (which are coinciding between Kerr and Schwarzschild) are explicitly listed in Table \ref{table:Kerrfinal} and Tables \ref{table:marginallyboundorbits}-\ref{table:boundorbits}-\ref{table:unboundorbits}. We distinguished generic orbital classes where both radial endpoints are either a turning point, the horizon, or infinity from nongeneric orbital classes, where at least one endpoint is a double, or triple root, or a root at the horizon. Negative energy orbits only exist within the ergoregion and  are trapped orbits, consistently with \cite{1984GReGr..16...43C}. Explicitly real, fully explicit, initial data-dependent analytical solutions, in terms of elliptic functions, are known for specific radial motion, such as bounded radial motion, see \cite{vandeMeent:2019cam}. The derivation of such a solution for all types of radial motion is also beyond the scope of this paper. 

We further classified the inequivalent root structures on the equator strictly within the ergoregion by explicitly evaluating the position of the ergosphere with respect to the radial roots, see Figures \ref{fig:Ep0} and \ref{fig:Em0}. This led to the identification of six distinguished values of the angular momentum of the black hole listed in Table \ref{valuesa}. 
We also provided a qualitative description of nonequatorial orbits. 

Furthermore, we obtained an explicit description of the complete separatrix. We showed that it can be algebraically  described in terms of a single quartic in appropriate variables $(e,p)$ as the union of three distinct regions respectively obtained when (1) the pericenter of bound motion becomes a double root (this occurs in the range $E_{ISCO^+} \leq E<1$ and the associated spherical orbits are unstable), (2) the eccentricity of bound motion becomes zero (this also occurs in the range $E_{ISCO^+} \leq E<1$ but the associated spherical orbits are stable), (3) the turning point of unbound motion becomes a double root (this occurs in the range $E \geq 1$ and the associated spherical orbits are unstable). The phase space of the separatrix is summarized in Figure \ref{summary}. In the range $0 \leq e<1$, the parameters $(e,p)$ are interpreted as usual as the eccentricity and semilatus rectum of corresponding bounded orbits, while for $-1<e \leq 0$ or $e > 1$ the interpretation is given in terms of relative distance between roots of the corresponding root system. This completes the result of \cite{Stein:2019buj} to the complete separatrix using a single fourth-order polynomial in $(e,p)$. 

The classification, tables, and figures provided in this work may lead to further insights on the properties of matter surrounding astrophysical black holes. In particular, the distinguished values of $a$ listed in Table \ref{valuesa} might be of astrophysical significance. For example, for highly spinning black holes with $a> a_c^{(2)} \approx 0.83$, equatorial bounded motion starts to occur within the ergoregion, while for $a \geq a_c^{(4)}\approx 0.94$, equatorial bounded motion completely occurs within the ergoregion, see Figure \ref{fig:Ep0}. Accretion disks then contribute to a collisional Penrose process due to collision of particles within the disk even in the absence of magnetic fields  \cite{1975ApJ...196L.107P,PhysRevD.16.1615,Banados:2009pr,Berti:2009bk,Harada:2010yv,Leiderschneider:2015kwa,Schnittman:2018ccg,Tursunov:2019oiq,Okabayashi:2019wjs}. On a different note, the thermodynamic bound $\ell \leq E/\Omega_+$ is always obeyed for trapped geodesics. This illustrates that geodesics can be used to test the fundamental laws of thermodynamics of black holes.

A natural continuation of this paper would be to obtain an explicitly real, fully explicit, initial condition-dependent solutions in terms of elliptic integrals and Jacobi elliptic functions for all classes of allowed radial motion. This would complete the analysis of Fujita and Hikida \cite{Fujita:2009bp} for bounded timelike motion and the one of Gralla and Lupsasca \cite{Gralla:2019ceu} for null motion. Given the length of such an analysis, it is differed for further endeavor.

\vspace{10pt}
{\noindent \bf Acknowledgments.}

G.C. is Senior Research Associate of the F.R.S.-FNRS and acknowledges support from the FNRS research credit J.0036.20F, bilateral Czech convention PINT-Bilat-M/PGY R.M005.19 and the IISN convention 4.4503.15. Y.L. is financially supported by the China Scholarship Council. The work of J.L. is supported by NSFC Grant No. 12005069.

\appendix

\section{Discriminant and roots of a polynomial}
\label{rootsec}

The roots of a polynomial can be characterized by the discriminant of the polynomial. For a polynomial with degree $n$, 
\be
P_n(x)=x^n+a_{n-1}x^{n-1}+\cdots+a_0,
\ee
we define the discriminant 
\be
\Delta_n=\prod_{i<j}(x_i-x_j)^2
\ee 
where $x_i,i=1,\cdots,n$ are the $n$ roots of the equation $P(x)=0$. 
\begin{enumerate}
    \item $n=2$. The discriminant is 
    \be
    \Delta_2=a_1^2-4a_0.
    \ee 
    \begin{enumerate}
        \item $\Delta_2>0$, two distinct real roots. 
        \item $\Delta_2<0$, two distinct complex roots.
        \item $\Delta_2=0$, two equal real roots.
    \end{enumerate}
    \item $n=3$. The discriminant is 
    \bea
    \Delta_3&=&-27 a_0^2+18 a_0 a_1 a_2-4 a_0 a_2^3-4 a_1^3+a_1^2a_2^2,\\
    &=& -108(\frac{q^2}{4}+\frac{p^3}{27}),
    \eea
    where $p=a_1-\frac{a_2^2}{3},q=a_0-\frac{a_1a_2}{3}+\frac{2a_2^3}{27}$.
    \begin{enumerate}
        \item $\Delta_3>0$, three distinct real roots. We define 
        \be
        \upsilon=\sqrt{-\frac{p^3}{27}}>0,\quad \theta=\frac{1}{3}\arccos\left( -\frac{q}{2\upsilon}\right).
        \ee 
        Then the three roots are 
        \bea
        x_j&=&-\frac{a_2}{3}+2\sqrt[3]{\upsilon}\cos(\theta+\frac{2\pi j}{3}),\quad j=0,1,2.
        \eea 
        \item $\Delta_3<0$, one real root and two complex roots. The real root is 
        \be
        x_1=-\frac{a_2}{3}+\sqrt[3]{-\frac{q}{2}+\sqrt{-\frac{\Delta_3}{108}}}+\sqrt[3]{-\frac{q}{2}-\sqrt{-\frac{\Delta_3}{108}}}.
        \ee 
        \item $\Delta_3=0$, three real roots with at least two equal. 
        \begin{enumerate}
            \item $p=q=0$, three equal real roots. The three real roots are 
            \be
            x_1=x_2=x_3=-\frac{a_2}{3}.
            \ee 
            \item $q\not=0$, two and only two equal. The three real roots are
\bea
x_1&=&-\frac{a_2}{3}+2\sqrt[3]{-\frac{q}{2}},\\
x_2&=&x_3=-\frac{a_2}{3}-\sqrt[3]{-\frac{q}{2}}.
\eea 
        \end{enumerate}
    \end{enumerate}
    \item $n=4$. The discriminant is 
    \bea
    \Delta_4&=&-27 a_1^4-4 a_3^3 a_1^3+18 a_2 a_3 a_1^3-4 a_2^3 a_1^2+a_2^2 a_3^2 a_1^2-6 a_0 a_3^2 a_1^2+144 a_0 a_2 a_1^2+18 a_0 a_2 a_3^3 a_1\nn\\&&\hspace{-40pt}-192 a_0^2 a_3 a_1-80 a_0 a_2^2 a_3 a_1+16 a_0 a_2^4-27 a_0^2 a_3^4+256 a_0^3-128 a_0^2 a_2^2-4 a_0 a_2^3 a_3^2+144 a_0^2 a_2 a_3^2.\nn\\
    \eea 
    The reduced form of the polynomial is 
    \be
    P_4(x)=y^4+p y^2+q y+s,\quad y=x+\frac{a_3}{4}.
    \ee 
    where 
    \bea
    p&=&a_2-\frac{3 a_3^2}{8},\\
    q&=&\frac{1}{8} \left(a_3^3-4 a_2 a_3+8 a_1\right),\\
    s&=&a_0-\frac{1}{256} a_3 \left(3 a_3^3-16 a_2 a_3+64 a_1\right).
    \eea 
    Using $p,q,s$, we find 
    \be
    \Delta_4=16 p^4 s-4 p^3 q^2-128 p^2 s^2+144 p q^2 s-27 q^4+256 s^3.
    \ee 
    We define the resolvent cubic equation as 
    \be
   Res(y)= y^3-p y^2-4s y+4ps-q^2=0.
    \ee 
    The discriminant of the resolvent cubic equation is 
    \be
    \Delta_{3,y}=\Delta_4.
    \ee 
    Then the general four roots are 
    \bea
    x_1&=&-\frac{a_3}{4}+\frac{1}{2}\sqrt{y_1-p}+\frac{1}{2}\sqrt{-y_1-p-\frac{2q}{\sqrt{y_1-p}}},\\
    x_2&=&-\frac{a_3}{4}+\frac{1}{2}\sqrt{y_1-p}-\frac{1}{2}\sqrt{-y_1-p-\frac{2q}{\sqrt{y_1-p}}},\\
    x_3&=&-\frac{a_3}{4}-\frac{1}{2}\sqrt{y_1-p}+\frac{1}{2}\sqrt{-y_1-p+\frac{2q}{\sqrt{y_1-p}}},\\
    x_4&=&-\frac{a_3}{4}-\frac{1}{2}\sqrt{y_1-p}-\frac{1}{2}\sqrt{-y_1-p+\frac{2q}{\sqrt{y_1-p}}},
    \eea 
    for $y_1\not=p$.
    We always choose $y_1$ to be the largest real root of the resolvent cubic equation. Since 
    \be
    Res(p)=-q^2\le0.
    \ee 
    By definition, we have 
    \be
    y_1\ge p.
    \ee 
    When $y_1=p$, we have $q=0$. The four roots can then be written as 
    \bea
    x_1&=&-\frac{a_3}{4}+\sqrt{\frac{-p+\sqrt{p^2-4s}}{2}},\\
    x_2&=&-\frac{a_3}{4}-\sqrt{\frac{-p+\sqrt{p^2-4s}}{2}},\\
    x_3&=&-\frac{a_3}{4}+\sqrt{\frac{-p-\sqrt{p^2-4s}}{2}},\\
    x_4&=&-\frac{a_3}{4}-\sqrt{\frac{-p-\sqrt{p^2-4s}}{2}}.
    \eea The following discussion is borrowed from \cite{Rees:1922}.
    \begin{enumerate}
        \item $\Delta_4>0$, roots distinct, all real or all complex.
        \begin{enumerate}
            \item $p<0$ and $s>\frac{p^2}{4}$, roots complex.
            \item $p<0$ and $s<\frac{p^2}{4}$, roots real.
            \item $p\ge0$, roots complex. 
        \end{enumerate}
        \item $\Delta_4<0$, roots distinct, two real and two complex. 
        \begin{itemize}
            \item $q>0$. It is clear that $x_3,x_4$ are real and $x_1,x_2$ are complex, $x_3>x_4$.
            \item $q=0$. Since $\Delta_4<0$, it leads to $s<0$. Therefore, $x_1,x_2$ are real and $x_3,x_4$ are complex, $x_1>x_2$.
            \item $q<0$. It is clear that $x_1,x_2$ are real and $x_3,x_4$ are complex, $x_1>x_2$.
        \end{itemize}
        \item $\Delta_4=0$, at least two equal roots. 
        \begin{enumerate}
            \item $p<0$ and $s>\frac{p^2}{4}$, two equal real roots, two complex.
            \item $p<0$ and $-\frac{p^2}{12}<s<\frac{p^2}{4}$, roots real, two and only two equal.
            \item $p<0$ and $s=\frac{p^2}{4}$, two pairs of equal real roots. The four real roots are 
            \bea
            x_1&=&x_3=-\frac{a_3}{4}+\sqrt{-\frac{p}{2}},\\
            x_2&=&x_4=-\frac{a_3}{4}-\sqrt{-\frac{p}{2}}.
            \eea 
            It is clear that $x_1>x_3$.
            \item $p<0$ and $s=-\frac{p^2}{12}$, triple roots and one real root.\begin{itemize}
                \item $q>0$,
               \bea
            x_1&=&x_2=x_3=-\frac{a_3}{4}+\frac{\sqrt[3]{q}}{2},\\
            x_4&=&-\frac{a_3}{4}-\frac{3\sqrt[3]{q}}{2}.
            \eea
            It is clear that 
            \be
            x_1>x_4.
            \ee 
            \item $q<0$, 
            \bea
            x_1&=&-\frac{a_3}{4}-\frac{3}{2}\sqrt[3]{q},\\
            x_2&=&x_3=x_4=-\frac{a_3}{4}+\frac{1}{2}\sqrt[3]{q}.
            \eea 
            It is clear that $x_1>x_4$.
            \end{itemize} 
             
            \item $p=0$ and $s>0$, two equal real roots and two complex. 
            \begin{itemize}
                \item $q>0$,
                 \be
            x_3=x_4=-\frac{a_3}{4}+\sqrt[3]{-\frac{q}{4}},
            \ee while $x_1, x_2$ are complex.
            \item $q<0$,
            \be
            x_1=x_2=-\frac{a_3}{4}+\sqrt[3]{-\frac{q}{4}},
            \ee while $x_3,x_4$ are complex.
            \end{itemize}
           
            \item $p=0$ and $s=0$, four equal real roots. The four real roots are 
            \be
            x_1=x_2=x_3=x_4=-\frac{a_3}{4}.
            \ee 
            \item $p>0$ and $s>0$ and $q\not=0$, two equal real roots and two complex.
            \item $p>0$ and $s=\frac{p^2}{4}$ and $q=0$, two pairs of equal complex roots.
            \begin{itemize}
                \item $q>0$. $x_3=x_4$ are real, while $x_1,x_2$ are complex.
                \item $q<0$. $x_1=x_2$ are real, while $x_3,x_4$ are complex.
            \end{itemize}
            \item $p>0$ and $s=0$, two equal real roots and two complex. The two real roots are 
            \be
            x_1=x_2=-\frac{a_3}{4}.
            \ee 
       
    \end{enumerate}
\end{enumerate}

\end{enumerate}

\section{Schwarzschild geodesics}
\label{sec:schw}

The radial function is 
\bea
R(r) = r \left(  (E^2-1)r^3+2r^2 -k r +2 k  \right),
\eea
where $k \equiv Q+\ell^2$ is the square of the angular momentum along the direction orthogonal to the plane of motion. The reality of polar motion only requires $k \geq 0$. The horizon is located at $r= r_+ \equiv 2$. There is no ergosphere and, consequently, $E \geq 0$. 


\subsection*{Marginal orbits}

Marginal orbits are by definition orbits such that $E = 1$. This condition reduces $R(r)$ to a cubic polynomial with roots $r_0=0$ and 
\bea 
r_{1,2}&=&\frac{1}{4}\left(k\pm \sqrt{k(k-16)}\right).
\eea
The orbits are classified according to the sign of the discriminant $k(k-16)$. We distinguish
\begin{itemize}
    \item $0 \leq k< k_{ibco} \equiv 16$. Then $r_{1,2}$ are generically complex, while $r_1=r_2=0$ for $k=0$. The orbit class is denoted as $\vert + \rangle$. There are only two types of orbits both with $r_+\le r<\infty$. The orbit is either plunging to the horizon or escaping to infinity,  depending upon the initial velocity. They are unbounded with respect to the black hole. We denote them as $\mathcal P(k,E_{ibco})$. 
    \item $k= 16$. This leads to a double root $r_1=r_2 \equiv r_{ibco} \equiv 4$ and the root structure $\vert +\bullet \hspace{-4pt}\bullet + \rangle$. There are three types of orbits:
    \begin{itemize}
        \item $r_+\le r<r_{ibco}$. For positive initial velocity, the orbit originates from the white hole and reaches the radius $r_{ibco}$ asymptotically. For negative initial velocity, the orbit originates asymptotically from $r_{ibco}$ and reaches the black hole in finite affine time. We label these two trapped orbits that are whirling in spacetime around $r_{ibco}$ as $\mathcal W \mathcal T^u(E_{ibco})$. 
        
        \item $r=r_{\text{ibco}}$. This is a circular orbit, which we label as $\mathcal C^u(E_{ibco})$.  Since $R''(r_{ibco})>0$ the orbit is unstable.
        \item $r_{\text{ibco}}< r < \infty$.  The radius $r_{ibco}$ is reached asymptotically, either in the infinite past or future affine time. Both types of whirling deflecting orbits are labeled as $\mathcal W \mathcal D^u(E_{ibco})$.
    \end{itemize}
    \item $16 < k < \infty$. We have $r_+ < r_2 < r_1$ and the root structure $\vert + \bullet  - \bullet \hspace{2pt}+ \rangle$. There are two types of orbits:
    \begin{itemize}
        \item $r_+\le r\le r_2$. This orbit is trapped in between the white hole and the black hole while reaching an intermediate turning point at $r_2$. We call it $\mathcal T(k,E_{ibco})$.
        \item $r_1 \leq r < \infty$. This is a deflecting orbit starting and ending at infinity. We call it $ \mathcal D (k,E_{ibco})$.
    \end{itemize}
\end{itemize}
This leads to the taxonomy displayed in Table \ref{table:marginallyboundorbits}, which will smoothly join with the taxonomy of $E>E_{ibco}$ orbits derived below.

\begin{table}[!tbh]    \centering
\begin{tabular}{|c|c|c|c|c|}\hline
\rule{0pt}{13pt}\textbf{Energy} & \textbf{Carter constant}& \textbf{Root structure} & \textbf{Radial range} & \textbf{Name} \\ \hline
$E = E_{ibco}$ & $0 \leq k < k_{ibco}$ & $\vert + \rangle$ & $r_+ \leq  r < \infty$ &  $\mathcal P(k,E_{ibco})$ \\\cline{2-5}
 & $k= k_{ibco}$ & $\vert +\bullet \hspace{-4pt}\bullet + \rangle$ & $r_+ \leq r < r_{ibco}$ & $\mathcal W \mathcal T^u(E_{ibco})$ \\\cline{4-5}
& & & $r  = r_{ibco}$ & $\mathcal C^u(E_{ibco})$ \\ \cline{4-5} 
 & & & $r_{ibco} < r < \infty$ & $\mathcal W \mathcal D^u(E_{ibco})$ \\  \cline{2-5}
 & $k_{ibco} < k< \infty$ & $\vert + \bullet  - \bullet \hspace{2pt}+ \rangle$ & $r_+ \leq r \leq r_2 $ & $ \mathcal T(k,E_{ibco})$ \\ \cline{4-5}
  &&  & $r_1 \leq r < \infty$ & $\mathcal D (k,E_{ibco})$ \\ \hline 
\end{tabular}\caption{Radial taxonomy of marginally bound orbits. Here $E_{ibco} = \mu$, $k_{ibco}=16M^2\mu^2$ and $r_{ibco}=4M$. }\label{table:marginallyboundorbits}
\end{table}

\subsection*{Generic nonmarginal orbits}

For $E \neq 1$, the radial potential $R(r)/r$ is cubic in $r$. 

\paragraph{Discriminant} The discriminant $\Delta_3$ is given by 
\bea
\Delta_3 = -\frac{4k \left[  (1- E^2) k^2+ (27E^4-36 E^2+8)k+16 \right]}{(E^2-1)^4} =\frac{4}{(E^2-1)^3}k (k-k^s)(k-k^u),
\eea
where 
\bea
k = k^{s,u}(E) \equiv  \frac{ 27E^4-36+8 \mp \vert E \vert (9E^2-8)^{3/2}}{2 (E^2-1)}. 
\eea
The discriminant vanishes for three special values of $k$. Note that $k^{s,u}$ are real (and therefore relevant) only for $E \geq E_{isco} \equiv \frac{2\sqrt{2}}{3}$. At $E = E_{isco}$, $k^s=k^u = k_{isco} \equiv 12$. 
 The bound $k^{u} \geq 0$ is obeyed for any $E \geq E_{isco} $ but $k^s \geq 0$ is only obeyed for $E_{isco} \leq E < E_{ibco} \equiv 1$. 
 
The discriminant vanishes for three subcases: $k=0,k^s,k^u$. For $k=0$, the double root of $R(r)/r$ occurs at $r=0$, while the simple root occurs at $r=r_1 \equiv \frac{2}{1-E^2}$. For $E>1$, it is inside the horizon and therefore irrelevant. For $0\leq  E<1$, it is outside the horizon. Since $R'(r_1)<0$, motion is allowed in $r_+ \leq r \leq r_1$. 

For $k=k^{s,u}$ real, the three real roots are given by 
\bea
r^{s,u}_1 = \frac{2\sqrt{9E^2-8}}{3 \cubicroot{2} (E^2-1)} \cubicroot{\pm 9E(3E^2-2 )-(9E^2-2)\sqrt{9E^2-8}}- \frac{2}{3(E^2-1)} , \\
r^{s,u}_2=r^{s,u}_3 = -\frac{\sqrt{9E^2-8}}{3\cubicroot{2} (E^2-1)} \cubicroot{\pm 9E(3E^2-2)-(9E^2-2)\sqrt{9E^2-8}}- \frac{2}{3(E^2-1)} ,
\eea
where $ \cubicroot x$ is the real cube of $x$.

\paragraph{Classification of orbits}

We discuss the orbits according to the range of $E$. When $E^2 < 1$, there is always one positive root since the function $R(r)/r$ evaluated at 0 is positive, while evaluated at $\infty$ is negative.

For $0 \leq E < \frac{2\sqrt{2}}{3}$, we have $\Delta_3 < 0$ and therefore only one real root $r_1$ (not given explicitly here), which is outside the horizon, $r_1 > r_+$. The only allowed motion is $r_+ \leq r \leq r_1$, and the root structure is $\vert + \bullet \hspace{2pt}- \rangle$. We call this trapped orbit as $\mathcal T(k,E)$. 

For $E= E_{isco} \equiv \frac{2\sqrt{2}}{3}$, $k^s=k^u= k_{isco}\equiv 12$ and all three radial roots coincide, $r_1=r_2=r_3 \equiv r_{isco} \equiv 6$, leading to $\vert + \bullet \hspace{-4pt}\bullet  \hspace{-4pt}\bullet \hspace{2pt}- \rangle$. The constant $r = r_{isco} = 6$ orbit, where $R''(r_{isco})=0$ is the outermost unstable circular orbit or the innermost stable circular orbit $\mathcal C_{isco}$. The whirling trapped  orbit $r_+ \leq r < r_{isco}$ called $\mathcal W \mathcal T_{isco}$ is also allowed. 

For $\frac{2\sqrt{2}}{3}<E <1$, we have $k^u < k^s$. Therefore, we have $\Delta_3 < 0$ (one real root) for either $k<k^u$ or $k > k^s$, while $\Delta_3 >0$ (three real roots) for $k^u < k < k^s$, which leads to the root structure $\vert + \bullet -\bullet +\bullet \hspace{2pt} - \rangle$. 

For the branch $k=k^s$ existing in the range $\frac{2\sqrt{2}}{3} \leq E \leq 1$, we have $6  \leq r^s_2=r^s_3 < \infty$ and $2  \leq r^s_1 \leq 6 $  with $r^s_1=r^s_2=r^s_3=6$ at $E = \frac{2\sqrt{2}}{3}$. The potential $R(r)/r$ is positive at $r=0$ and negative at large $r$,  leading to the root structure $\vert + \bullet  - \bullet  \hspace{-4pt}\bullet \hspace{2pt}- \rangle$. Since $(\frac{R}{r})''(r^s_2)<0$ for $\frac{2\sqrt{2}}{3} < E< 1$, the constant radius orbits $r= r^s_2=r^s_3$ are stable circular orbits, which we denote as $\mathcal C^s(E)$. There are trapped orbits $\mathcal T(k^s,E)$ in the range $r_+ \leq r \leq r_1^s$, which continuously join to the class found for $E \leq E_{isco}$. For $k > k^s$, the double root becomes complex, and only the trapped orbit $\mathcal T(k,E)$ exists, corresponding to the root structure $\vert + \bullet \hspace{2pt}- \rangle$, which again continuously join to the class found for $E \leq E_{isco}$. 

For the branch $k=k^u$ valid in the range $\frac{2\sqrt{2}}{3}\leq E \leq \infty$, we have $3  < r^u_2=r^u_3 \leq 6 $. In  the range $\frac{8}{9} \leq E^2 < 1$ , we have $6  \leq r^u_1 \leq \infty$. This leads to the root structure $\vert + \bullet \hspace{-4pt} \bullet  + \bullet \hspace{2pt}- \rangle$. Also $R''(r^u_2)>0$ and therefore constant $r=r^u_2$ orbits denoted as $\mathcal C^u(E)$ are unstable circular orbits. The orbits $r_+ \leq e < r_2^u$ are whirling trapped orbits $\mathcal W \mathcal T^u(E)$, while the orbits $r_2^u < r \leq r_1^u$ are homoclinic orbits $\mathcal H^u(E)$. For $0 \leq k<k^u$, the double root becomes complex, and there are only the trapped orbits  $\mathcal T(k,E)$ corresponding to $\vert + \bullet  \hspace{2pt}- \rangle$. 

For $E > 1$, we have $k^s <0$ and $k^u > k_{ibco}=16$. Therefore, we have $\Delta_3 < 0$ for $0< k < k^u$ and $\Delta_3 > 0$ for $k>k^u$. 
In the range $E^2 > 1$, we have $r^u_1 < 0$. Since $R''(r^u_2)>0$, constant $r=r^u_2$ are unstable circular orbits $\mathcal C^u(E)$. For $k = k^u$, we have the root structure $\vert + \bullet \hspace{-4pt} \bullet  \hspace{2pt}+ \rangle$. There are whirling trapped orbits $\mathcal W \mathcal T^u(E)$, $r_+ \leq r < r^u_2$ and whirling deflecting orbits $\mathcal W \mathcal D^u(E)$, $r^u_2 < r < \infty$. For $k > k^u$, the double root becomes two simple roots with, as a convention, $r_3 < r_2$ (which can be obtained numerically), and there are two types of orbits: trapped $\mathcal T(k,E)$ with $r_+ \leq r \leq r_3$ or deflecting $\mathcal D(k,E)$ with $r_2 \leq r < \infty$. This corresponds to  $\vert + \bullet - \bullet  \hspace{2pt}+ \rangle$. For $k<k^u$, the double root becomes imaginary. Since $r_1$ remains below the horizon, we have  $\vert + \rangle$. There is a single plunging orbit $\mathcal P(k,E)$ (which is instead moving outwards in case of positive initial velocity) with $r_+ \leq r < \infty$. 

This leads to the taxonomy displayed in Tables \ref{table:boundorbits} and \ref{table:unboundorbits}. There are 11 distinct geodesic orbit classes, counting as separate classes all orbits with qualitatively distinct endpoints (simple, double, or triple root, horizon or infinity). 

\begin{table}[!tbh]    \centering
\begin{tabular}{|c|c|c|c|c|}\hline
\rule{0pt}{13pt}\textbf{Energy} & \textbf{Carter constant} & \textbf{Root structure} & \textbf{Radial range} & \textbf{Name} \\ \hline
$E < E_{isco}$ & $0 \leq k <\infty$ & $\vert + \bullet \hspace{2pt}- \rangle$ & $r_+ \leq  r \leq r_1$ & $\mathcal T(k,E)$ \\\cline{1-5}
 $E =E_{isco}$ & $0 \leq k <k_{isco}$ & $\vert + \bullet \hspace{2pt}- \rangle$ & $r_+ \leq  r \leq r_{1}$ & $\mathcal T(k,E_{isco})$ \\\cline{2-5}
 & $ k =k_{isco}$ & $\vert + \bullet \hspace{-4pt}\bullet  \hspace{-4pt}\bullet \hspace{2pt}- \rangle$ & $r_+ \leq  r < r_{isco}$ & $\mathcal W \mathcal T_{isco}$ \\\cline{4-5}
  && & $r = r_{isco}$ & $\mathcal C_{isco}$ \\\cline{2-5}
 & $  k_{isco} < k <\infty $ &$\vert + \bullet \hspace{2pt}- \rangle$ & $r_+ \leq  r \leq r_{1}$ & $\mathcal T(k,E_{isco})$ \\ \cline{1-5}
  $E_{isco} < E < E_{ibco} $ & $0 \leq k <k^u$ & $\vert + \bullet  \hspace{2pt}- \rangle$ & $r_+ \leq  r \leq r_{1}$ & $\mathcal T(k,E)$ \\ \cline{2-5}
  & $ k =k^u$ & $\vert + \bullet \hspace{-4pt} \bullet  + \bullet \hspace{2pt}- \rangle$ & $r_+\leq  r < r^u_{2}$ & $\mathcal W \mathcal T^u(E)$   \\ \cline{4-5}
 &  &  & $r= r^u_{2}$ & $\mathcal C^u(E)$ \\ \cline{4-5}
    &  &  &  $r_2^u<  r \leq r^u_{1}$ & $\mathcal H^u(E)$  \\ \cline{2-5}
  & $k^u < k < k^s$ & $\vert + \bullet -\bullet +\bullet \hspace{2pt} - \rangle$ & $r_3 \leq  r \leq r_{1}$ & $\mathcal B(k,E)$  \\ \cline{4-5}
  &&  & $r_+ \leq  r \leq r_{2}$ & $\mathcal T(k,E)$ \\ \cline{2-5}
    & $k=k^s$ & $\vert + \bullet  - \bullet  \hspace{-4pt}\bullet \hspace{2pt}- \rangle$ & $r=r_2^s$ & $\mathcal C^s(E)$  \\ \cline{4-5}
  &&  & $r_+ \leq  r \leq r^s_{1}$ & $\mathcal T^s(k^s,E)$  \\ \cline{2-5}
    & $k^s < k < \infty$ & $\vert + \bullet \hspace{2pt}- \rangle$ & $r_+ \leq  r \leq r_{1}$ & $\mathcal T(k,E)$ \\ 
     \hline
  \end{tabular}\caption{Radial taxonomy of bound orbits. Here $E_{isco} = \frac{2\sqrt{2}}{3}\mu$, $k_{isco} = 12M^2\mu^2$, $r_{isco}=6M$.}\label{table:boundorbits}
\end{table}

\begin{table}[!tbh]    \centering
\begin{tabular}{|c|c|c|c|c|}\hline
\rule{0pt}{13pt}\textbf{Energy} & \textbf{Carter constant} & \textbf{Radial range} & \textbf{Name} \\ \hline
$E > E_{ibco}$ & $0 \leq k < k^u$ & $\vert + \rangle$ & $r_+ \leq  r < \infty$ & $\mathcal P(k,E)$\\\cline{2-5}
 & $k= k^u$ & $\vert + \bullet \hspace{-4pt} \bullet  \hspace{2pt}+ \rangle$ & $r_+ \leq r < r^u_2$ & $\mathcal W \mathcal T^u(E)$ \\\cline{4-5}
 & && $r=r^u_2$ &  $\mathcal C^u(E)$\\\cline{4-5}
 & && $r^u_2 <  r < \infty$ & $\mathcal W \mathcal D^u(E)$,  \\\cline{2-5}
  & $k> k^u$ & $\vert + \bullet - \bullet  \hspace{2pt}+ \rangle$ & $r_2 \leq r < \infty $ &  $\mathcal D(k,E)$\\ \cline{4-5}
 &  & & $r_+ \leq r \leq r_3 $ & $\mathcal T(k,E)$ \\ \cline{1-5}
\end{tabular}\caption{Radial taxonomy of unbound orbits. It reduces for $E=E_{ibco}$ to the taxonomy of marginal orbits, see Table \ref{table:marginallyboundorbits}.}\label{table:unboundorbits}
\end{table}

\begin{figure}
    \centering

    \includegraphics[width=0.9\textwidth]{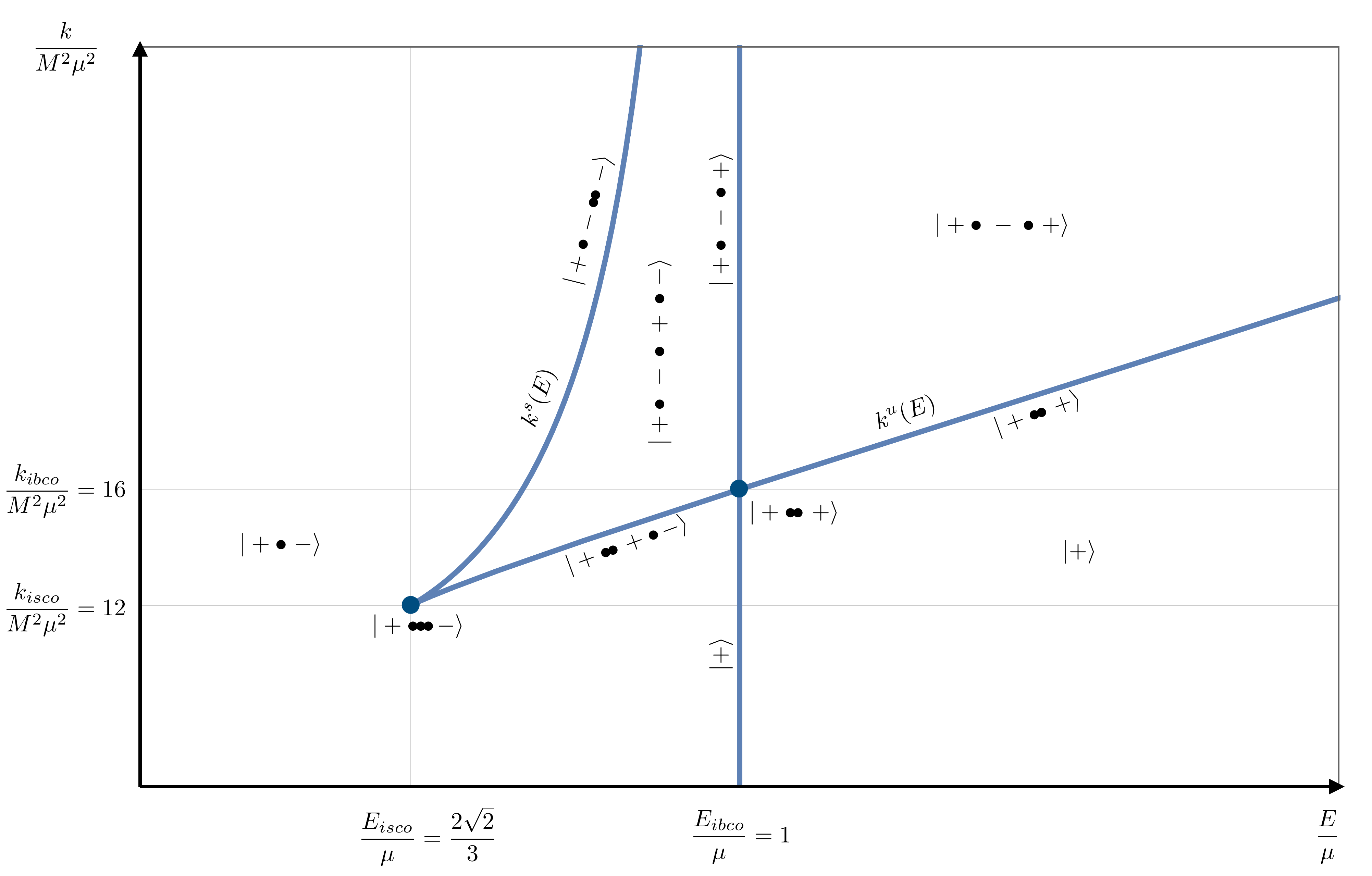}
    \caption{Phase space diagram representing the eight distinct root diagrams (four generic and four nongeneric) for Schwarzschild with $E \geq 0$ and $k \geq 0$. Here $|$ represents the horizon, $\rangle$ spatial infinity, $+$ an allowed radial range, $-$ a disallowed radial range, $\bullet$ a root (turning point/no velocity), ${\bullet \hspace{-3pt} \bullet}$ a double root (attractor point/no velocity nor acceleration), and ${\bullet \hspace{-4pt} \bullet\hspace{-4pt} \bullet}$ a triple root.}
    \label{PlotSchwarzschild_classes}
    
        \includegraphics[width=0.9\textwidth]{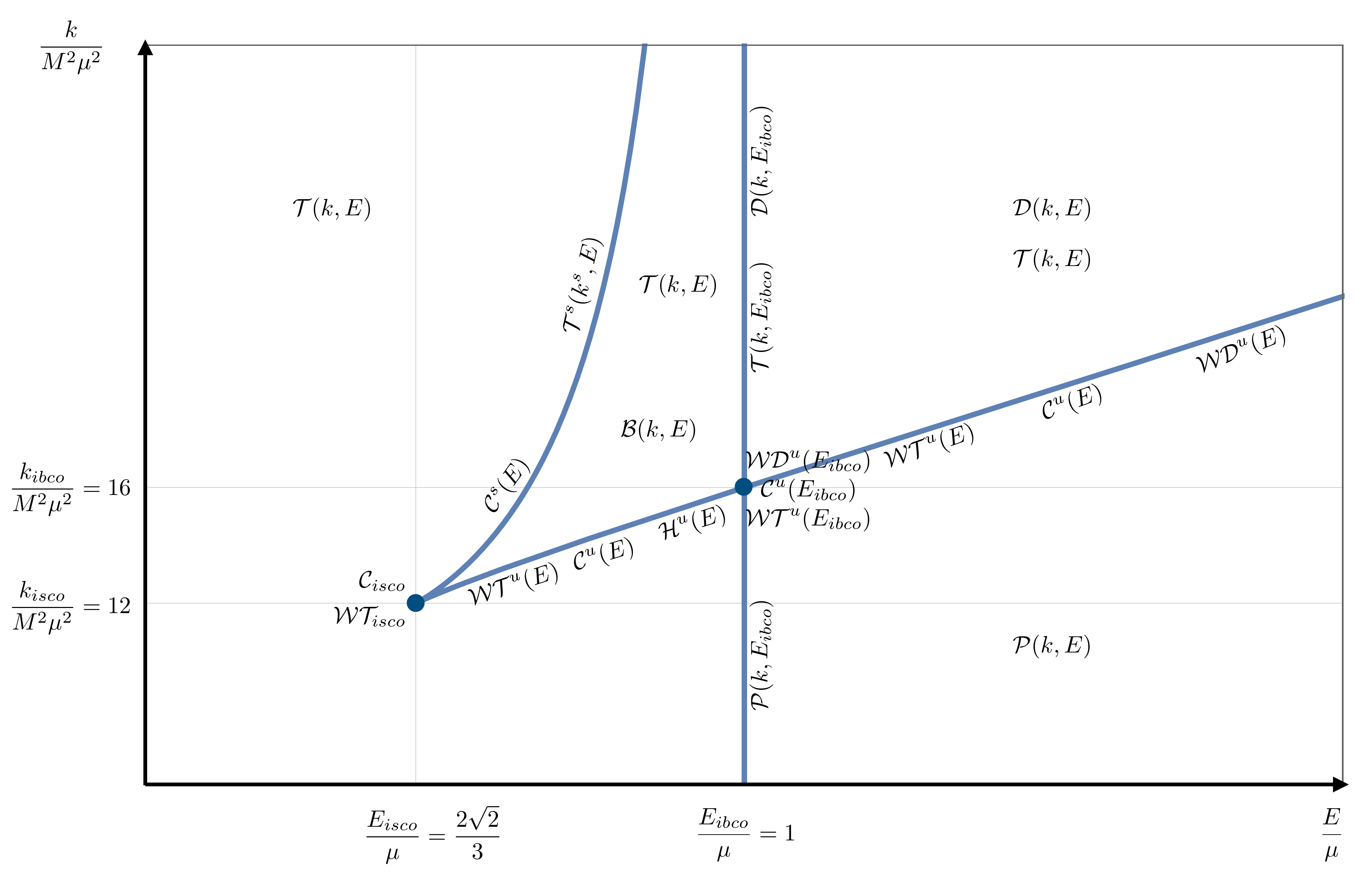}
    \caption{Phase space diagram representing all 11 distinct classes of Schwarzschild orbits. The 11 distinct classes of orbits are listed in Tables \ref{table:marginallyboundorbits}, \ref{table:boundorbits}, and \ref{table:unboundorbits}.}
    \label{SchwarzschildPlot}
\end{figure}

The classification of the root structures is displayed in Figure \ref{PlotSchwarzschild_classes}. The full classification is displayed in Figure \ref{SchwarzschildPlot}.

\bibliography{refs}

\end{document}